\documentclass[a4paper,11pt]{article}
\pdfoutput=1 

\usepackage{jheppub} 
\usepackage[T1]{fontenc} 
\usepackage{xcolor}
\usepackage{subfigure}
\usepackage[export]{adjustbox}

\newcommand{\ket}[1]{\left|#1\right\rangle}
\newcommand{\bra}[1]{\left\langle #1\right|}
\newcommand{\tr}{\text{Tr}}

\title{\boldmath Effective entropy of quantum fields coupled with gravity}

\author[a]{Xi Dong,}
\author[b]{Xiao-Liang Qi,}
\author[b]{Zhou Shangnan,}
\author[b]{and Zhenbin Yang}

\affiliation[a]{Department of Physics, University of California, Santa Barbara, California 93106, USA}
\affiliation[b]{Stanford Institute for Theoretical Physics, Stanford University, Stanford, California 94305, USA}

\abstract{Entanglement entropy, or von Neumann entropy, quantifies the amount of uncertainty of a quantum state. For quantum fields in curved space, entanglement entropy of the quantum field theory degrees of freedom is well-defined for a fixed background geometry. In this paper, we propose a generalization of the quantum field theory entanglement entropy by including dynamical gravity. The generalized quantity named effective entropy, and its Renyi entropy generalizations, are defined by analytic continuation of a replica calculation. The replicated theory is defined as a gravitational path integral with multiple copies of the original boundary conditions, with a co-dimension-$2$ brane at the boundary of region we are studying. We discuss different approaches to define the region in a gauge invariant way, and show that the effective entropy satisfies the quantum extremal surface formula. When the quantum fields carry a significant amount of entanglement, the quantum extremal surface can have a topology transition, after which an entanglement island region appears. Our result generalizes the Hubeny-Rangamani-Takayanagi formula of holographic entropy (with quantum corrections) to general geometries without asymptotic AdS boundary, and provides a more solid framework for addressing problems such as the Page curve of evaporating black holes in asymptotic flat spacetime. We apply the formula to two example systems, a closed two-dimensional universe and a four-dimensional maximally extended Schwarzchild black hole. We discuss the analog of the effective entropy in random tensor network models, which provides more concrete understanding of quantum information properties in general dynamical geometries. We show that, in absence of a large boundary like in AdS space case, it is essential to introduce ancilla that couples to the original system, in order for correctly characterizing quantum states and correlation functions in the random tensor network. Using the superdensity operator formalism, we study the system with ancilla and show how quantum information in the entanglement island can be reconstructed in a state-dependent and observer-dependent map. We study the closed universe (without spatial boundary) case and discuss how it is related to open universe. }

\begin{document} 
\maketitle
\flushbottom

\section{Introduction}

Holographic duality \cite{maldacena1999large} points out a deep connection between quantum gravity theory and quantum field theory. A gravity theory in $d+1$-dimensional anti-de Sitter (AdS) space is the holographic dual of a $d$-dimensional quantum field theory living on the asymptotic boundary of the hyperbolic space. If we believe that these two theories have a one-to-one correspondence (which can be used as a definition of the bulk gravity theory), the gravity theory can be considered as a reorganization of the quantum field theory degrees of freedom. The Ryu-Takayanagi (RT) formula \cite{ryu2006holographic} and its generalizations \cite{Hubeny:2007xt,dong2016gravity,jafferis2016relative} provide important clues about how the bulk degrees of freedom corresponds to the boundary ones. With quantum corrections \cite{faulkner2013quantum,Engelhardt:2014gca,Dong:2017xht}, the Hubeny-Rangamani-Takayanagi (HRT) formula tells us that the von Neumann entropy of a boundary region $A$ is given by the dominant saddle point of $S_A=S_{\rm bulk}(\Sigma)+\frac{|\gamma|}{4G_N}$, with $|\gamma|$ a surface that is homologous to the boundary region, and $\Sigma$ is a Cauchy surface bounded by $\gamma$ and $A$. 
The bulk long-wavelength degrees of freedom in a given geometry are mapped to a subspace of boundary Hilbert space, where each bulk operator can be reconstructed at the boundary. Operators in the causal diamond of bulk region $\Sigma$, known as the entanglement wedge of $A$, can be reconstructed in region $A$ of the boundary. 
The emergent locality in the bulk is related to quantum error correction  \cite{almheiri2015bulk}. 
The quantum HRT formula seems to generate physically meaningful results even when the bulk field theory contribution is not subleading to the area law term. In particular, in a series of recent work \cite{Penington:2019npb,Almheiri_2019,Almheiri_2020} (see also \cite{Akers:2019nfi,Rozali:2019day,Chen:2019uhq, Bousso:2019ykv, Almheiri:2019psy}), the quantum HRT formula has been used to obtain the entropy change of an evaporating black hole, {\it i.e.} the Page curve.

Although a lot of efforts have been made to generalize the holographic duality beyond asymptotic AdS geometries, a lot of fundamental questions remain unclear, such as what the Hilbert space of the theory is. In this work, we propose a framework for computing the quantum field theory (QFT) entanglement entropy of a spatial region in the bulk. In a curved space quantum field theory without dynamical gravity, the quantum field theory entanglement entropy of a given spatial region $A$ is well-defined, although it is UV divergent. We would like to find a generalization of this quantity in systems with dynamical gravity. The intuition is that such entanglement entropy should be well-defined, because in our real world there is dynamical gravity, yet an experimentalist can identify a spatial region in her lab, and measure its entanglement entropy in the particular given state that is prepared. Although one cannot directly specify the region in term of coordinates, since it is not diffeomorphism invariant, one can set the initial state of the universe such that there is a planet that is identified as earth, and then define the region by its relative position to earth. This is schematically how we think about defining a spatial region in the gauge invariant way. Even though we cannot guarantee that such approach can work with arbitrary quantum gravitational systems, it can at least apply to states with semiclassical geometry.

\begin{figure}
    \centering
    \includegraphics[width=4.7in]{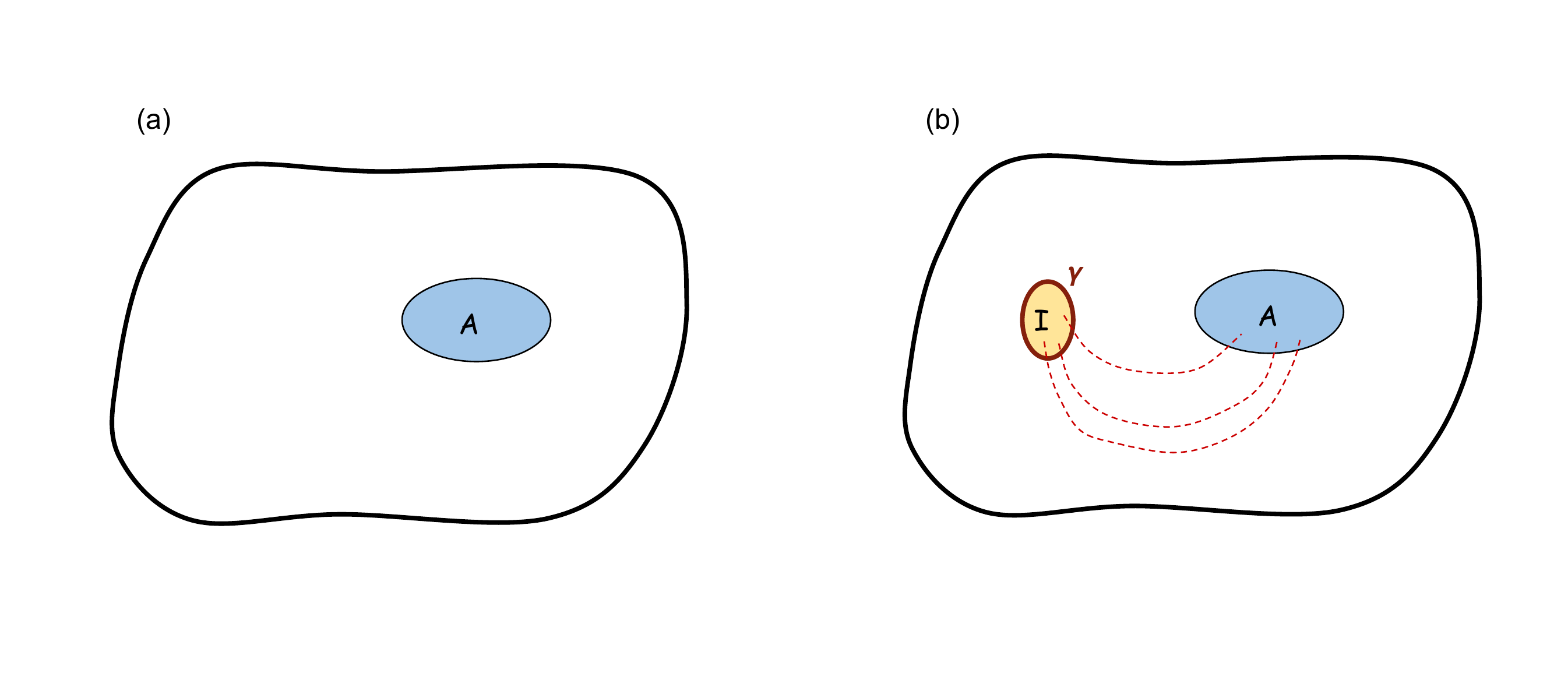}
    \caption{Illustration of the QES formula in Eq. (\ref{eq:QES}) for two different situations. The curved black lines represent a Cauchy surface. (a) $\Sigma=A$ when the formula reduced to the ordinary quantum field theory entropy. (b) $\Sigma=A\cup I$ when a new quantum extremal surface $\gamma=\partial I$ appears, contributing an area law entropy $\frac{|\gamma|}{4G_N}$. The quantum field theory entropy becomes $S^{\rm qft}_{AI}$ instead of $S^{\rm qft}_{A}$, which can reduce the entropy when there is entanglement between $I$ and $A$, as is indicated by the red dashed lines. }
    \label{fig:QES}
\end{figure}

One example of such computation is to compute the entropy of early Hawking radiation of a Schwarzchild black hole in asymptotically flat space. By choosing the spatial region $A$ to be the exterior of a sphere around the black hole, the entropy as a function of the radius of the sphere is expected to follow the Page curve. It is natural to expect that a nontrivial quantum extremal surface is responsible for the Page curve, similar to the case in asymptotic AdS spaces, except that the entropy that is computed here is for a region in the spacetime with dynamical gravity, rather than in a bath system with fixed background. In the limit that the region is far away from the black hole and the gravity is semiclassical, this difference does not matter much \cite{Gautason:2020tmk, Anegawa:2020ezn, Hashimoto:2020cas, Hartman:2020swn}.  The main goal of this paper is to set up a framework where the generalization of QFT entropy from fixed background to dynamical background is well-defined, and the quantum extremal surface formula of such entropy can be justified in a way similar to the proof of quantum HRT formula in the asymptotic AdS case \cite{faulkner2013quantum,Dong:2017xht}. Our result is based on a replica calculation of Renyi entropy. In quantum field theory, the $n$-th Renyi entropy of region $A$ is computed by a replica geometry with a branching surface at the boundary of the region $A$. On the contrary, when gravity is dynamical, the branching surface has a conical singularity and thus violates Einstein's equation. The actual geometry with the same boundary condition (if there is a boundary) should be smooth, without the conical singularity. As we will discuss in more details in Sec. \ref{sec:gravity}, we consider a replicated geometry with an extra brane at the boundary of region $A$. The brane introduces a conical singularity with the angle $2n\pi$, which stabilizes the geometry that computes the quantum field theory entropy in the fixed background case. If this is the dominant saddle point, in the weakly coupled limit the gravitational calculation will result in an entropy that is the same as the quantum field theory entropy. However, in general there are other saddle point geometries with other topology, which are the replica wormholes, very similar to the case of AdS evaporating black hole \cite{Penington:2019kki,EastCoast}. If we assume the dominant saddle point does not break replica symmetry $Z_n$, in the limit $n\rightarrow 1$ we obtain the quantum extremal surface formula:
\begin{align}
    S_A^g={\rm Ext}_{\gamma=\partial I}\left[\frac{\left|\gamma\right|}{4G_N}+S^{\rm qft}(\Sigma=I\cup A)\right]\label{eq:QES}
\end{align}
$S^{\rm qft}(\Sigma=I\cup A)$ is the entropy of a spatial region $I$ union the original region $A$ in the quantum field theory with fixed background curved space, and $\gamma$ is the boundary of $I$. ${\rm Ext}$ refers to taking extremal value of this quantity. If there are multiple saddle points, the one with the smallest entropy should be chosen. It is interesting to note that the area law term only contains the extra quantum extremal surface $\gamma$, and excludes the boundary of $A$. This is consistent with the fact that in the trivial case $\gamma=\emptyset$, $\Sigma=A$, our result reduces to the quantum field theory entropy, without the area law term. Physically, the entropy we define is that carried by quantum field theory degrees of freedom with length scale above certain UV cutoff scale, in a dynamical spacetime. Therefore we name this quantity {\it effective entropy}.

We apply the quantum extremal surface formula (\ref{eq:QES}) to two examples systems in Sec. \ref{sec:examples}. The first example is a one-dimensional region in a two-dimensional closed universe, with a matter conformal field theory coupled with two-dimensional Jackiew-Teitelboim gravity \cite{jackiw1985lower,teitelboim1983gravitation}. This example illustrates how the current proposal can apply to close universe. The second example is the case of asymptotically flat space evaporating black hole, where $A$ is a region that includes all Hawking radiation until time $t$. The entanglement island when $t$ reaches the Page time, similar to the AdS black hole case \cite{Penington:2019kki,EastCoast}. In Sec. \ref{sec:schwarzchild} we study a particular state in the maximally extended four-dimensional Schwarzchild black hole geometry. It is similar to the eternal geometry studied in two-dimensional models, both in AdS and flat geometries \cite{Almheiri:2019yqk,Gautason:2020tmk, Anegawa:2020ezn, Hashimoto:2020cas, Hartman:2020swn}. However, we choose a different state such that the space far from black hole is in the vacuum, rather than in thermal equilibrium with the black hole. This avoids the problem of having a finite energy density at an infinite flat space region.

To obtain further intuition of the effective entropy, and gain further understanding of quantum information properties in general geometries, in Sec. \ref{sec:tensornetwork} we study the random tensor network (RTN) model proposed in Ref.  \cite{hayden2016holographic}. RTN models are defined on generic graphs, which are viewed as a discrete analog of the spatial geometry. The previous results on RTN models have been mainly about graphs with a large boundary, which is the analog of asymptotic AdS geometries. In the current work we study more general geometries where bulk degrees of freedom may not be able to be encoded in the boundary. We discuss the structure of correlation functions and show that it is helpful to define quantum state in an ``observer-dependent" Hilbert space using the formalism of superdensity operators \cite{cotler2018superdensity}.  A similar formula to Eq. \ref{eq:QES} for Renyi entropy appears in this model. The tensor network model helps us understand how quantum information in the entanglement island is reconstructed, which is a generalization of the entanglement wedge reconstruction in AdS/CFT. Using ancilla introduced in the superdensity operator formalism, we can also explicitly study the quantum information recovery process. Compared with the AdS/CFT case, the main feature is that observers ({\it i.e.} ancilla systems coupled with the original system) play an essential role in determining the quantum information structure in the system. The quantum information recovery from the entanglement island is state-dependent and observer-dependent. We also discuss special properties of a closed universe and how it is related to the open universe case. Finally, we conclude our paper and provide further discussion and outlook in 
Sec. \ref{sec:conclusion}.

\section{Effective entropy in gravitational system}
\label{sec:gravity}

\subsection{Overview of entropy in quantum field theory}\label{sec:QFTEE}

We consider a quantum field theory with a fixed background metric $g_{\mu\nu}$.
Denoting the field as $\phi$, a quantum state $|\Psi\rangle$ can be defined as a path integral of a manifold up to some Cauchy slice $\mathcal{S}$:
\begin{equation}
    \langle \phi_b|\Psi\rangle=\int_{\phi |_{\mathcal{S}}=\phi_b} \mathcal{D}\phi~ e^{-S_{QFT}(\phi,g_{\mu\nu})}
\end{equation}
$|\Psi\rangle$ defined this way is not necessary normalized and its normalization can be calculated from the path integral over the whole (time reflected symmetric) manifold $\mathcal{ M}$:
\begin{equation}
    Z_{\mathcal{M}}\equiv \langle \Psi|\Psi\rangle=\int_{\mathcal{ M}} \mathcal{D}\phi~ e^{-S_{QFT}(\phi,g_{\mu\nu})}
\end{equation}
There is implicitly a UV cutoff $\epsilon$. The form of the cutoff is not important, as long as it is finite so that the entropy is finite.
The density matrix of a spatial region $A$ on the Cauchy slice $\mathcal{S}$ is obtained by tracing out the fields in the complement of $A$ and is given by the path integral on $\mathcal{M}$ with a slit $A_{\pm}$ open:
\begin{equation}
    \langle \phi_+|\rho|\phi_-\rangle={\frac 1 {Z_{\mathcal{M}}}}\int_{\phi|_{A_+}=\phi_+,\phi|_{A_-}=\phi_-}\mathcal{D}\phi~ e^{-S_{QFT}(\phi,g_{\mu\nu})}
\end{equation}
In the limit where $A$ shrinks to zero, the denominator is equal to the numerator and the whole expression equals to one.
The $n$-th Renyi entropy of the density matrix $\rho_A$ can be computed by
\begin{align}
    e^{-(n-1)S^{(n)}_A}\equiv\tr \rho_A^n=\frac{1}{Z_\mathcal{M}^n}\bra{\Psi}^{\otimes n}X_{An}\ket{\Psi}^{\otimes n}
\end{align}
with $X_{An}$ a cyclic permutation operator that acts in region $A$ and permutes the $n$ replica cyclically. Denoting the $n$ copies of fields as $\phi^{(a)},a=1,2,...,n$, we have $X_{An}\phi^{(a)}X_{An}^\dagger=\phi^{(a+1)}$ with $\phi^{(n+1)}\equiv \phi^{(1)}$.

\begin{figure}[htbp]
    \centering
    \includegraphics[width=4in]{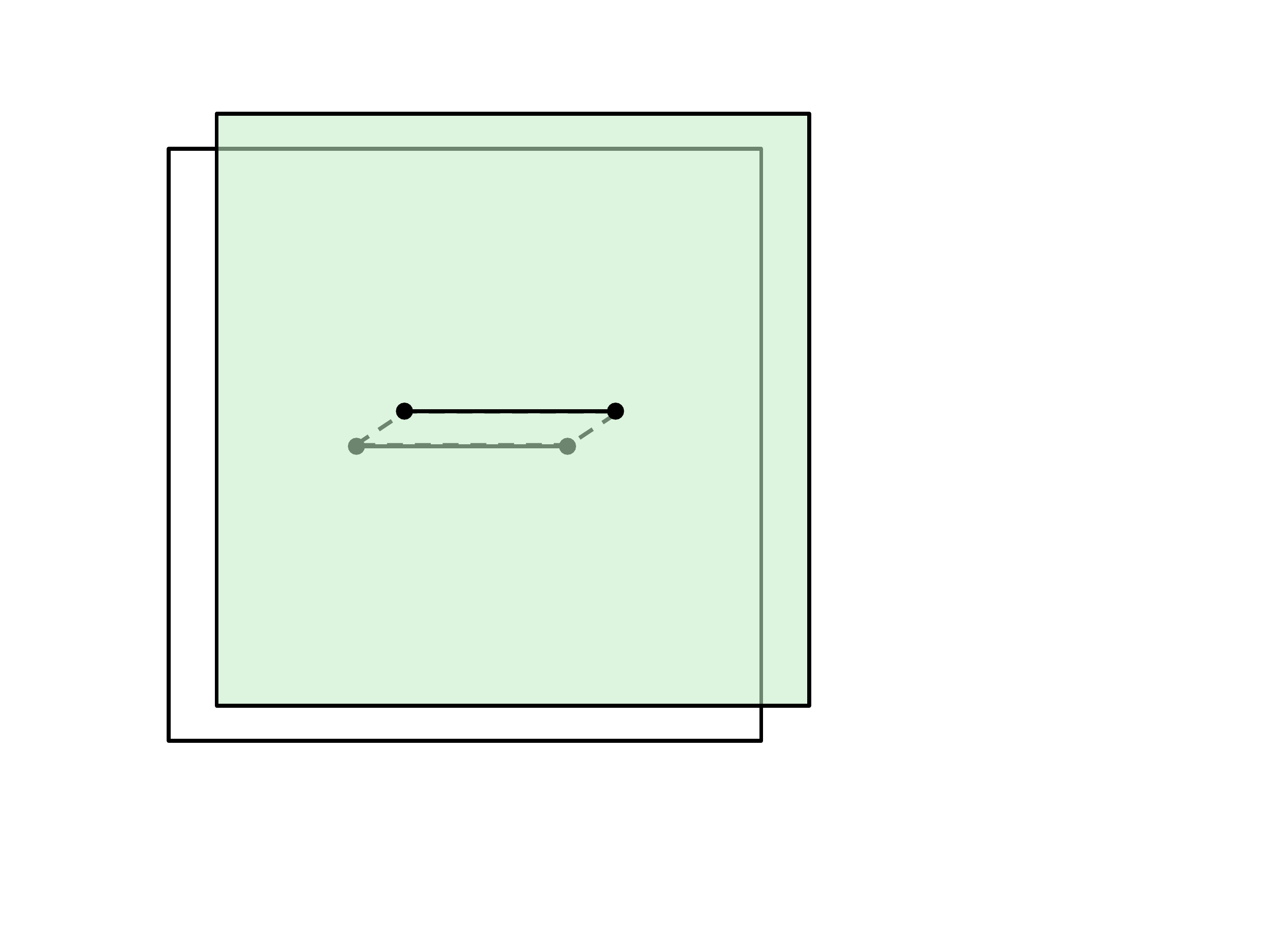}
    \caption{Illustration of the replica calculation of Renyi entropy in fixed background (Eq. (\ref{eq:QFTentropy})) for $n=2$. The replica geometry has conical singularity at the branch surfaces.}
    \label{fig:QFTentropy}
\end{figure}

In the path integral language, 
this is computed by a replica geometry, obtained by taking an $n$-fold branched cover space $\mathcal{M}_n(A)$ of the original geometry $\mathcal{M}$, with the boundary of $A$ (which has co-dimension $2$ in spacetime) being the branching surface. 
The metric of $\mathcal{M}_n(A)$, which we denote as $\tilde{g}_n$, has the same curvature locally as the original geometry, except for the conical singularity at $\partial A$ with a conical angle of $2n\pi$. (See Fig. \ref{fig:QFTentropy}.) The Renyi entropy is determined as
\begin{align}
    e^{-(n-1)S_A^{(n)}}=\frac{~Z_{\mathcal{M}_n(A)}}{Z_\mathcal{M}^n};~~~~~Z_{\mathcal{M}_n(A)}\equiv\int_{\mathcal{M}_n(A)} \mathcal{D}\phi~ e^{-S_{QFT}\left[\tilde{g}_{\mu\nu},\phi\right]}\label{eq:QFTentropy}
\end{align}
with $Z_{\mathcal{M}_n(A)}$ the quantum field theory path integral over the branched cover space. Here we do not need to explicitly write the replica index of $\phi^{(i)}$ any more, since we can view it as one single field living on the branched cover manifold.

\subsection{Generalization to systems with dynamical gravity}

When we include dynamical gravity in the system, we need to generalize the quantity (\ref{eq:QFTentropy}) by allowing the geometry to fluctuate. As a preparation, we first rewrite Eq. (\ref{eq:QFTentropy}) using another replica trick:
\begin{align}
    e^{-(n-1)S_A^{(n)}}= Z_{\mathcal{M}_n}(A)Z_{\mathcal{M}}^{m}\Big|_{m\rightarrow -n}\equiv Z_{\mathcal{M}_{n,m}(A)}\Big|_{m\rightarrow -n}\label{eq:QFTentropy2}
\end{align}
In the second equality, we view the product $Z_{\mathcal{M}_n}(A)Z_{\mathcal{M}}^{m}$ as the partition function of the QFT on a manifold 
\begin{align}
    \mathcal{M}_{n,m}(A)=\mathcal{M}_n(A)\otimes \mathcal{M}^{\otimes m}
\end{align}
which is $n+m$ copies of the original manifold $\mathcal{M}$, with a branch covering over the first $n$ of them at the boundary of $A$. When considering geometry fluctuation, this replica trick avoids the complication of treating numerator and denominator in Eq. (\ref{eq:QFTentropy}) separately. The natural way to include dynamical gravity is to replace Eq. (\ref{eq:QFTentropy2}) by a path integral over geometries with the same boundary condition as $\mathcal{M}_{n,m}(A)$, weighted by a certain gravitational action:
\begin{align}
    e^{-(n-1)S_A^{(n)}}=\left.\int_{n,m} D\tilde{g} D\phi e^{-S_{\rm grav}[\tilde{g}]-S_{\rm QFT}[\tilde{g},\phi]}\right|_{m\rightarrow -n}\label{eq:graventropy1}
\end{align}
To be consistent with the previous subsection, we denote the metric by $\tilde{g}$. The subscript $n,m$ refers to the fact that the boundary condition is given by $n+m$ copies of the original geometry. The action $S_{\rm grav}[\tilde{g}]$ should include the information about $A$ in some proper way, as will be discussed below. 

The key questions are: 1) what gravitational action $S_{\rm grav}[\tilde{g}]$ should be used here; 2) how to define region $A$ in a gauge invariant way and include the information about $A$ in $S_{\rm grav}[\tilde{g}]$. The most natural choice for $S_{\rm grav}[\tilde{g}]$ appears to be the Einstein-Hilbert action for metric $\tilde{g}$. However, this choice leads to physically incorrect results. In particular, in the limit that gravitational fluctuations are weak and the quantum field theory entropy is small, we expect that the saddle point of path integral (\ref{eq:graventropy1}) should reproduce the QFT entropy in Eq. (\ref{eq:QFTentropy2}), which means that the saddle point should be the branch covering manifold $\mathcal{M}_{n,m}(A)$. However, this manifold has conical singularity and cannot be a saddle point of the Einstein-Hilbert action. In other words, if $S_{\rm grav}[\tilde{g}]$ is the Einstein-Hilbert action, the entropy we obtain will be quite different from the QFT value even in the limit of weak gravity and low QFT entropy.

To find out a physically reasonable action, it is helpful to consider a system with two quantum fields $\phi$ and $\eta$. The two fields are coupled, but have independent degrees of freedom. The QFT Hilbert space of the system is a direct product of them: $\mathbb{H}=\mathbb{H}_{\phi}\otimes\mathbb{H}_\eta$. Therefore it is well-defined to consider $S_{A\phi}^{(n)}$, the Renyi entropy of $\phi$ field in region $A$, while $\eta$ field is traced out. Now if we assume $\eta$ is very massive, we can integrate over $\eta$ field, which will lead to a correction to the gravitational action $S_{\rm grav}[\tilde{g}]$. Since $\eta$ field is not acted by the twist operator in the entropy computation, the action contributed by $\eta$ will be an Einstein-Hilbert term of the original manifold without branch covering:
\begin{align}
    \delta S_{\rm grav}[\tilde{g}]\propto S_{\rm EH}[g]
\end{align}
where $g$ represents the metric of the original manifold without branch covering. As a generalization of the quantum field theory entropy, the entropy we are defining for a low energy field $\phi$ is a characterization of its correlation properties and should not be sensitive to the difference between bare gravitational dynamics and induced action by integrating over high energy fields. Therefore this reasoning suggests that a natural gravitational action for the replica system is the action of the untwisted manifold $g$ rather than that of $\tilde{g}$. It should be noted that the additional quantum field $\eta$ is only introduced as a tool to clarify the argument. Even if there is only one field $\phi$, integrating over the high energy degrees of freedom of the $\phi$ field have the same effect. 

In summary, we propose that the generalized notion of entropy, which we name as effective entropy, of a region $A$ is computed by Eq. (\ref{eq:graventropy1}) with the gravitational action being the Einstein-Hilbert action of the untwisted manifold, while the quantum field $\phi$ lives on the twisted manifold. More explicitly, there are two equivalent ways to write down the entropy formula. The first one is in term of the untwisted metric $g$:
\begin{equation}
     e^{-(n-1)S_A^{(n)}}=\left.\int_{n,m} Dg \int D\phi e^{-S_{EH}(g)-S_{QFT}(\phi,g)} X_{An}\right|_{m\rightarrow -n}\label{eq:graventropy2}
\end{equation}
where $n,m$ in the integral represents the boundary condition that is $n+m$ copies of the original geometry, with the twist operator $X_{An}$ inserted in the first $n$ copies. Here $S_{EH}(g)$ is the Einstein-Hilbert action $S_{EH}(g)= -\frac 1{16\pi G}\int  R(g)\sqrt{|g|}d^dx$ for the untwisted metric\footnote{This action may also be viewed as the action for the twisted manifold $\tilde{g}$ with a fixed conical angle of $2\pi n$ on $\partial A$ (as a boundary condition), which is defined (e.g.\ in \cite{Dong:2019piw}) by excluding all localized contributions from the conical defect itself as required to do so by a well-defined variational principle.}, and $X_{An}$ is the twist operator on the first $n$ copies that only acts on the quantum field $\phi$, without affecting the geometry.

Alternatively, one can use the twisted geometry with metric $\tilde{g}$, and rewrite the Einstein-Hilbert action $S_{EH}(g)$ in term of $\tilde{g}$:
\begin{equation}
         e^{-(n-1)S_A^{(n)}}=\left.\int_{n,m}D \tilde{g}\int D\phi e^{-S_{EH}(\tilde{g})-\frac{1-n}{4G}|\partial A|-S_{QFT}(\phi,\tilde{g})}\right|_{m\rightarrow -n} \label{eq:gravitationalentropy}
\end{equation}
where $|\partial A|$ is the area of $\partial A$, and we have used the relation 
\begin{align}
    S_{EH}(g)=S_{EH}(\tilde{g})+\frac{1-n}{4G}\left|\partial A\right|.
\end{align}
This expression makes the role of region $A$ more manifest: computing the Renyi entropy of region $A$ corresponds to inserting a brane at the co-dimension-$2$ surface $\partial A$ with a particular brane tension. This brane sources a conical singularity with angle $2\pi n$. As a consequence, in the limit that the back-reaction induced by $S_{QFT}(\phi,g)$ is negligible, the action in Eq. (\ref{eq:gravitationalentropy}) has the branch covering manifold $\mathcal{M}_{n,m}(A)$ as a saddle point. The effect of branch covering in causing the conical singularity is compensated by the brane term in the action and does not violate Einstein's equations. 

Obviously, our prescription is only meaningful if $A$ (or at least $\partial A$) is defined in a gauge invariant way. We now discuss how this is done. In the spirit of Mach's Principle, a bulk region is defined with reference to a gauge-invariant object such as a distant star. In geometries with asymptotic boundaries, it is convenient to use the boundary as a reference to define a bulk region using diffeomorphism invariant quantities. There can be multiple ways to choose such a region. For instance, one approach could be shooting light rays from past and future from the boundary and define $A$ as the region between the intersection of the light rays and the boundary, or one can define such a region using the proper distance away from the boundary.
In general these different ways of defining a bulk region can disagree with each other in replica geometries due to gravitational backreactions. 

\begin{figure}
    \centering
    \includegraphics[width=4in]{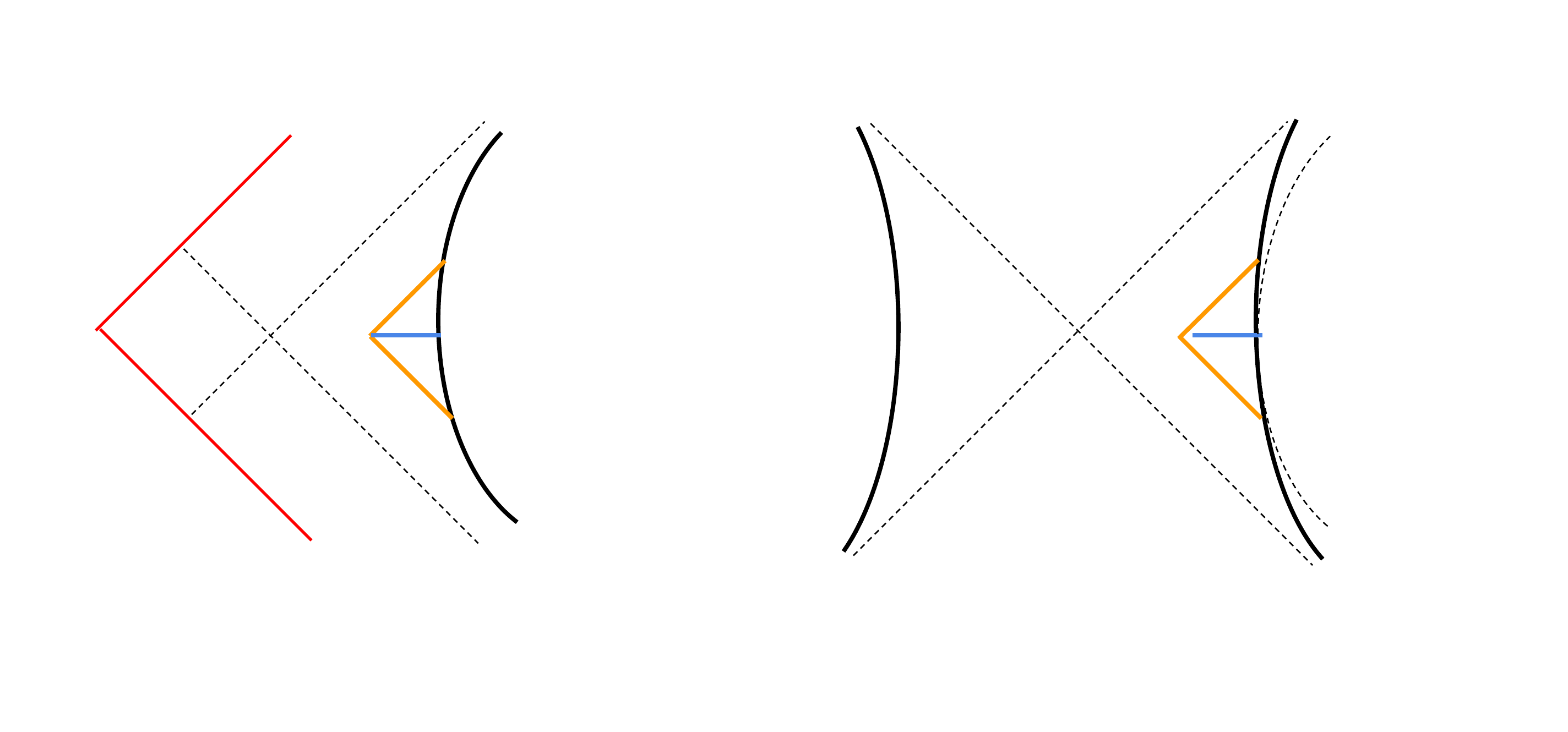}
    \caption{The left panel is the original black hole geometry, with a bulk region defined either by sending two light rays from the boundary (the orange lines) or by fixing the proper distance from the boundary (the blue line). 
    The right panel is the two replica wormhole geometry, which has a temperature lower than the original black hole temperature. The dashed curve is the original boundary and the solid curve is that in the two replica geometry. Using the same light rays will thus define a bulk region that is larger than the region defined by fixing proper distance.}
    \label{fig:AinReplicaGeo}
\end{figure}

A simple example is the connected two-replica geometry of brane states in JT gravity where the black hole temperature is lower than its original temperature in the disconnected geometry (Figure \ref{fig:AinReplicaGeo}).
The consequence is that if we choose the light rays that define a region with fixed distance from the boundary in the disconnected geometry, the same light rays will define a region with different distance from the boundary in the connected geometry which means that the Renyi entropies can have strong dependence on the method used to choose a bulk region. Fortunately in the limit of von Neumann entropy ($n\rightarrow 1$), such discrepancies vanish since they only contribute to higher orders in $n-1$.

In a geometry without asymptotic boundary, it is more tricky to place a distant star and we will not make such attempts. Instead, we adopt the philosophy advocated by Hawking and Ellis, "we shall take the local physical laws that have been experimentally determined, and shall see what these laws imply about the large scale structure of the universe." \cite{hawking1973large}. For a bulk observer living inside a closed universe, we can think of the condition of no spatial boundary as her ignorance of the global structure of the universe. Therefore, instead of fixing any data at the boundary, one'd better fix it to be nothing. This is similar to the no-boundary proposal for the initial condition of the universe. For such a bulk observer, the only data he/she can fix is the observed geometry on $A$. The gravitational path integral with this boundary condition describes a density matrix for the observer \cite{PhysRevD.34.2267,Hawking:1986vj,Barvinsky:2008vz,Maldacena:2019cbz,maldacenaStrings2019}:
\begin{equation}
    \rho= \includegraphics[scale = .2,valign = c]{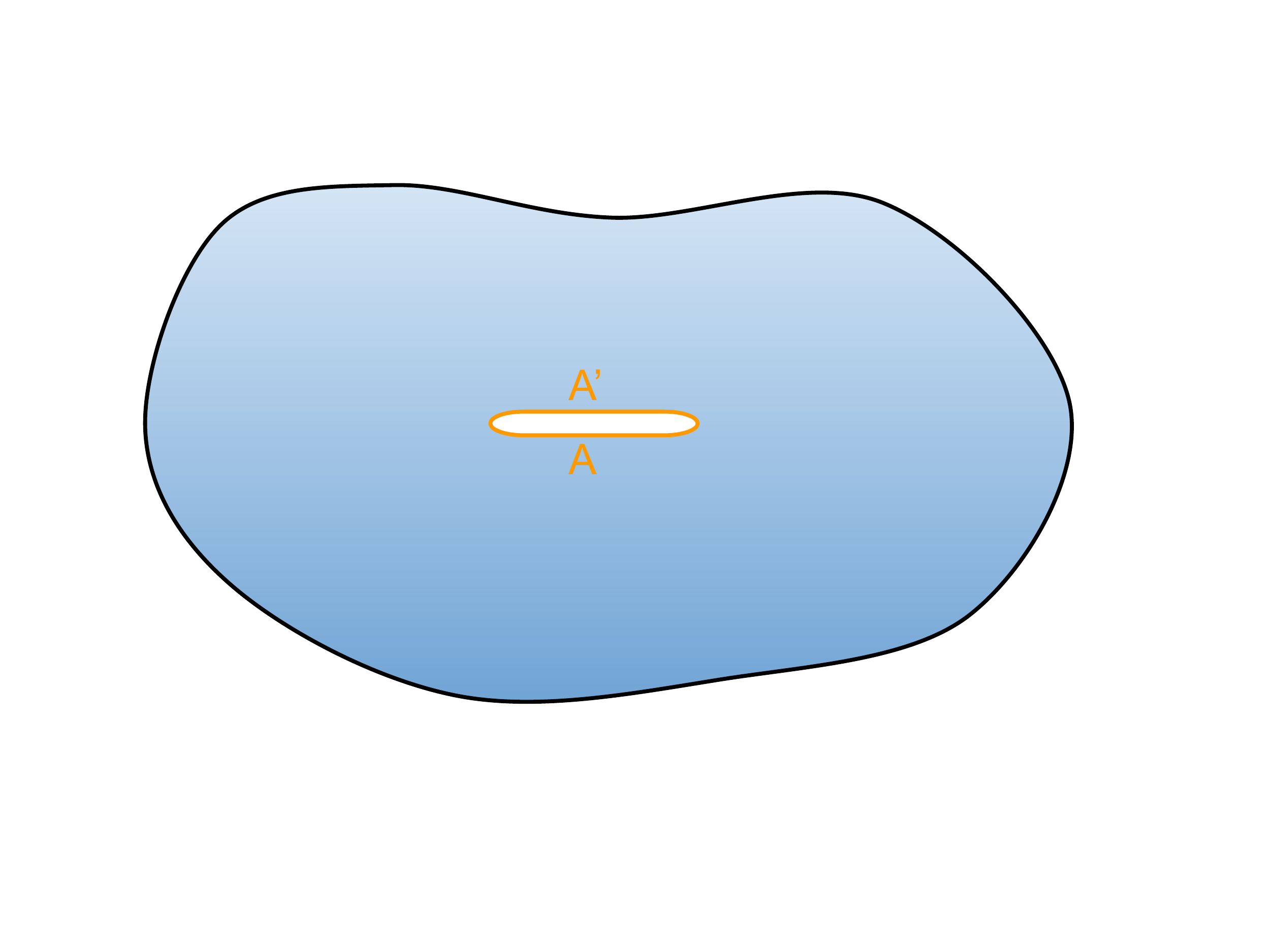}\label{fig:gravityDM}
\end{equation}
We will talk more about such types of gravitational density matrix in section \ref{sec:GDM}.
Notice that this description of closed universe is different from the closed universe in the fully evaporated black hole.
Such a difference is due to the different location of the observer.

Compared to the QFT calculation, a key difference introduced by dynamical gravity is the possibility of different topology. Starting from the disconnected geometry in the QFT calculation as a reference, other geometries can be considered replica wormholes connecting different copies. A special situation that needs some further discussion is the closed universe case where the boundary conditions are the euclidean preparation in the past. $n+m$ copies of such closed universes can have $(n+m)!$ fully disconnected geometries by permutation of the boundary conditions: 
\begin{equation}
    \includegraphics[scale = .4,valign = c]{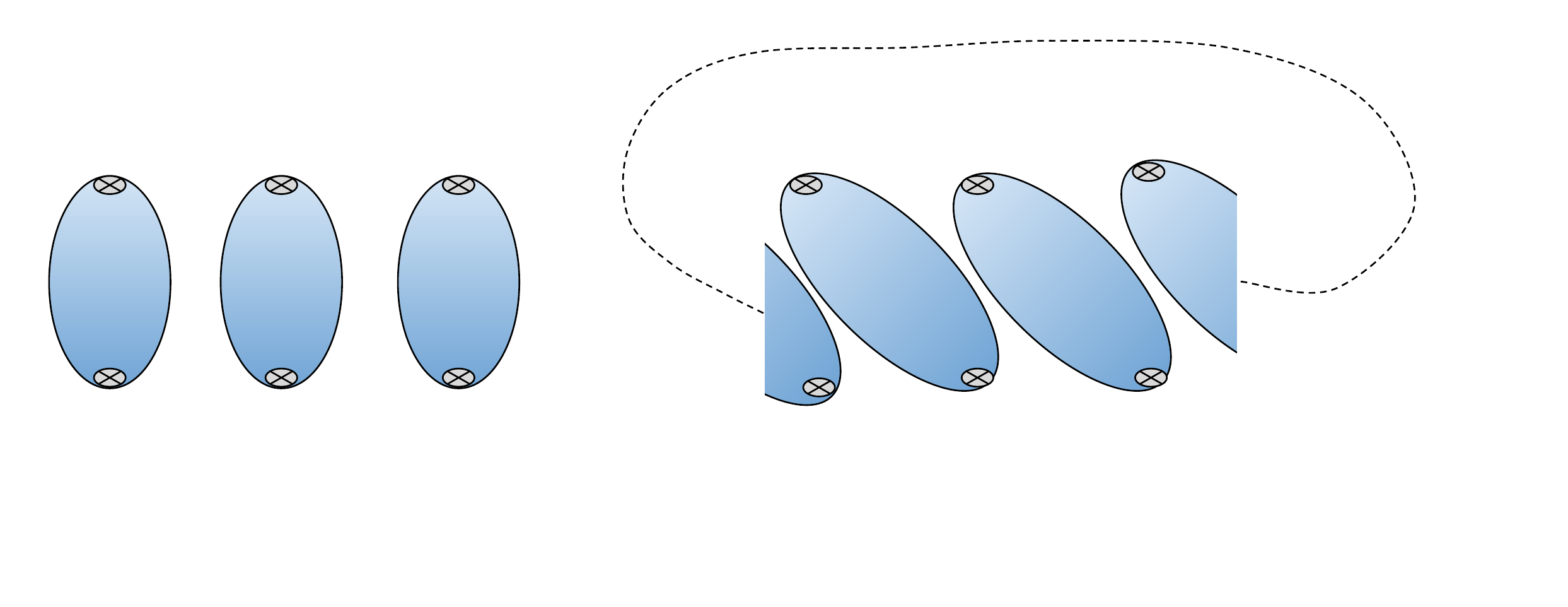}
\end{equation}
In a fully evaporated black hole case, one can think of the interior of the black hole as such a closed universe and the euclidean boundary is the physical process used to create the closed universe, namely the formation and evaporation of the black hole. 
In the analytic continuation $m\rightarrow -n$, such additional permutation ambiguity does not lead to a contribution to the Renyi entropy.

If we restrict ourselves to only consider geometry with asymptotic boundaries or closed universes with a bulk observer, then the gravitational path integral should not introduce strong correlation between $Z_{\mathcal{M}}$ and $Z_{\mathcal{M}_n(A)}$. This means we can interchange the integration over metric and analytic continuation to reduce the gravitational path integral over the numerator and denominator separately:
\begin{equation}
     e^{-(n-1)S^{(n)}_A}\simeq  \frac{\int_{n} \mathcal{D} g e^{-S_{EH}(g)-S_{QFT}(\phi,g)-\frac{n-1}{4G}|\partial A|}}{\int_n \mathcal{D} g e^{-S_{EH}(g)-S_{QFT}(\phi,g)}}
\end{equation}
For example, for $n=2$ the purity can be pictorially illustrated as follows:
\begin{equation}
    \tr \rho^2=\includegraphics[scale = .4,valign = c]{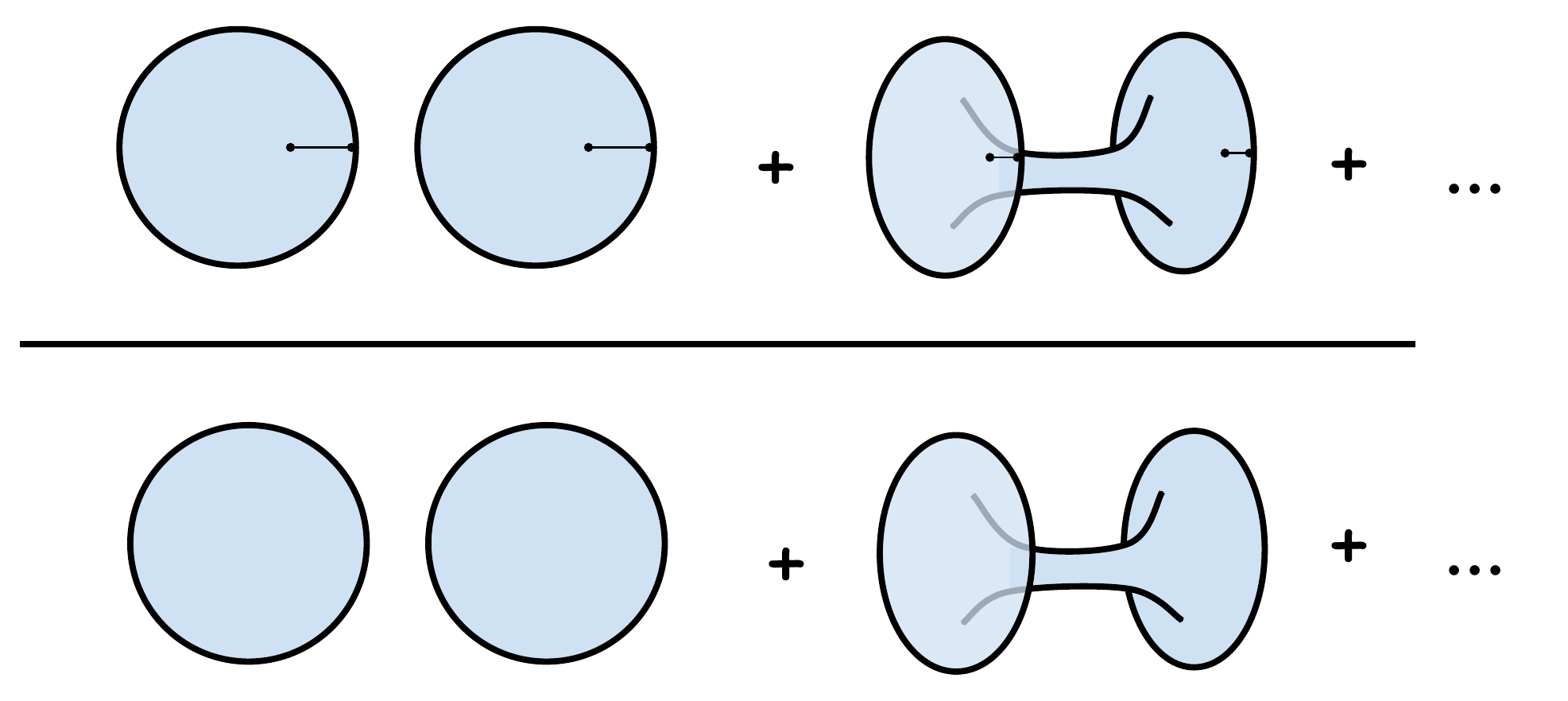}~~~\label{fig:graventropy}
\end{equation}
where the line segments with black dots at both ends indicate the region $A$ which is the branch cuts in the numerator. 

We expect the denominator to be dominated by $n$ copies of the original manifold, which is the first term in (\ref{fig:graventropy}). As we discussed earlier, the branch covering manifold $\tilde{M}_n(A)$ (first term in the numerator of (\ref{fig:graventropy})) is a saddle point of the path integral in the numerator, if back-reaction is negligible. However, it may or may not be the dominant saddle point. If the quantum field theory entropy is comparable with gravitational entropy, 
there could be non-perturbative effect caused by other saddles, such as the second term in the numerator of Fig. \ref{fig:graventropy} with additional wormholes. This situation is the same as the ``replica wormhole" discussed in asymptotically AdS geometries \cite{Penington:2019kki,EastCoast}, except that $A$ is now a bulk region. When such a nontrivial saddle is dominant, the effective entropy is different from the QFT value, which is the situation for the Hawking radiation of an evaporating black hole after Page time.

\begin{figure}
    \centering
    \includegraphics[width=4.2in]{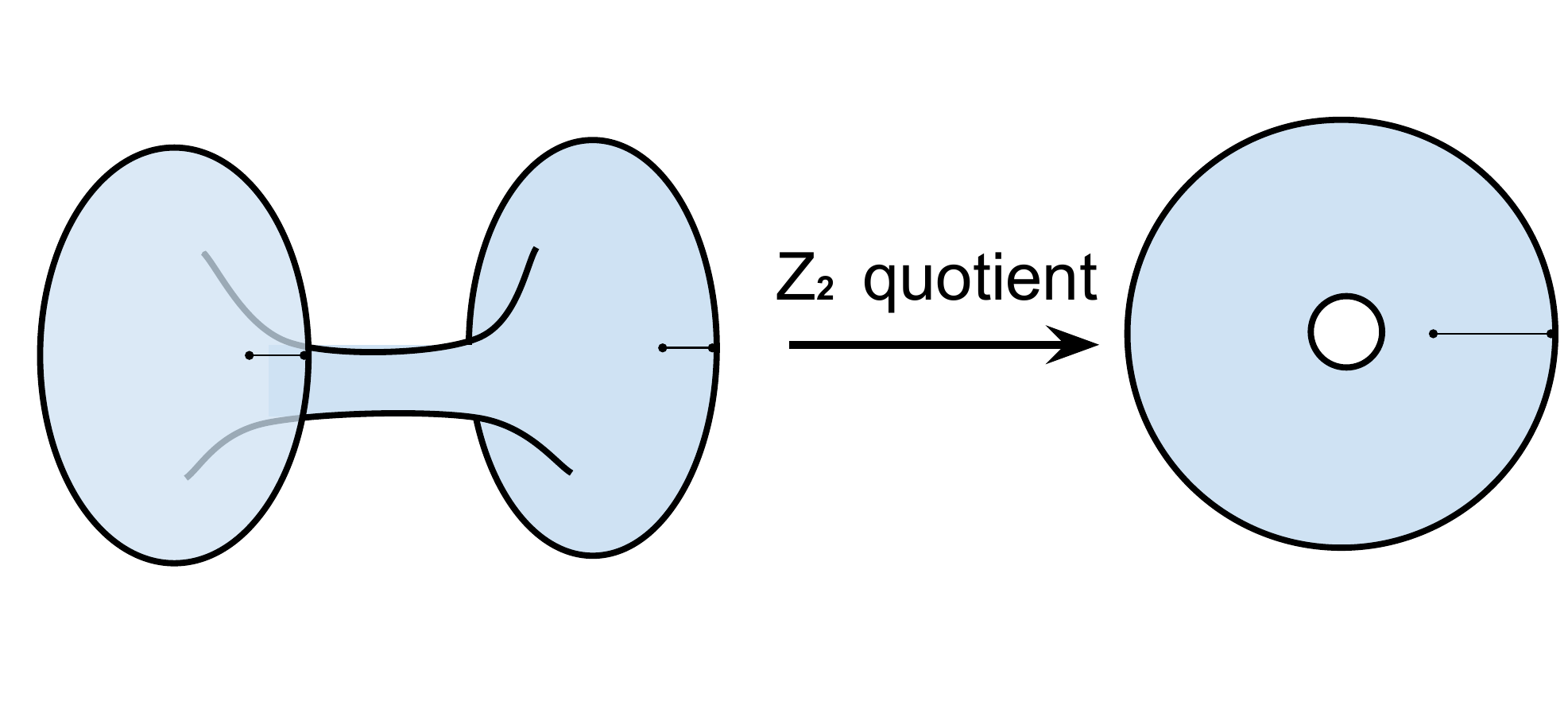}
    \caption{The $n$-replica geometry $\mathcal{M}_n$ with a possible replica wormhole (left), and the $Z_n$ quotient geometry $\mathcal{M}_n/Z_n$. The quotient geometry has no conical singularity at the boundary of $A$, but has a conical singularity with angle $\frac{2\pi}n$ at the boundary of the extra branching surface ({\it i.e.} the ``island") $I$.}
    \label{fig:quotient}
\end{figure}

In general, the replica symmetry may or may not be broken. If it is broken, we have to deal with the entire new geometry and there is no generic calculation to the partition function $Z_A$. If we assume that the dominant saddle $\mathcal{M}_n$ is still replica symmetric, even if it contains extra replica wormhole, the computation can be simplified in a similar way as the Renyi entropy calculation in AdS/CFT \cite{lewkowycz2013generalized,faulkner2013quantum,dong2016gravity}. The key is to consider a $Z_n$ quotient geometry $\bar{\mathcal{M}}=\mathcal{M}_n/Z_n$, illustrated in Fig. \ref{fig:quotient}. Due to $Z_n$ symmetry, the saddle point action satisfies
\begin{align}
-\log Z_A\simeq  S\left(\mathcal{M}_n\right)=nS\left(\bar{\mathcal{M}}\right)
\end{align}
The new geometry $\bar{\mathcal{M}}$ has no conical singularity at boundary of $A$ (since it is removed by the quotient), but if there is an additional replica wormhole, there will be extra $Z_n$ fix points, which are the boundary of another region $I$ (blue region in Fig. \ref{fig:quotient}). Since the geometry is smooth before the quotient, after quotient the boundary of $I$ becomes singular with a conical angle $\frac{2\pi}n$ and the action $S(\bar{\mathcal{M}})$ is evaluated without including the contribution from the conical angle.

In the limit $n\rightarrow 1$, the bulk geometry has order $n-1$ back-reaction caused by the brane and the change of gravitational action is equal to the area law contribution from the conical singularity with angle $\frac{2\pi}{n}$ by equation of motion. (The derivation is the same as in AdS/CFT \cite{lewkowycz2013generalized}.) Therefore
\begin{align}
    S\left(\bar{\mathcal{M}}\right)\simeq \left(1-\frac1n\right)\left[\frac{\left|\partial I\right|}{4G_N}+S^{\rm QFT(n)}_{A\cup I}\right]
\end{align}
with $S^{\rm QFT(n)}_{A\cup I}$ the $n$-th Renyi entropy of the quantum field theory in the original geometry $\mathcal{M}$. Taking the $n\rightarrow 1$ limit gives the quantum extremal surface formula of von Neumann entropy (Eq. (\ref{eq:QES})):

\begin{align}
    S_A={\rm ext}_I\left[\frac{\left|\partial I\right|}{4G_N}+S^{\rm QFT}_{A\cup I}\right]\label{eq:QES2}
\end{align}
with ${\rm ext}_I$ representing taking extremal value of this quantity by varying $I$. If there are multiple saddle points, the one with lowest entropy should be taken. 

This discussion is completely in parallel with the AdS/CFT case, with $A$ the analog of a boundary region in AdS/CFT, and $I$ the analog of a spatial slice of the entanglement wedge of $A$. The main difference is that in the current case the region $A$ has the same dimension as $I$. 

An important point we would like to comment about Eq. (\ref{eq:QES2}) is the UV cutoff dependence. It should be noted that the area law entropy only contains the area of extra region $I$, and does not contain the boundary area of $A$, which is the consequence that we have inserted a source brane at the boundary of $A$ but no source for $I$. When we change the cutoff of the QFT, say lowering the cutoff scale by integrating over some high energy modes, the gravitational coupling $G_N$ should correspondingly be renormalized. If this change of cutoff happens in region $A$, it will change the value of entropy $S_A$, just like what happens in a QFT. In contrast, the choice of UV cutoff in region $I$ does not affect $S(A)$ since the sum of the two terms in Eq. (\ref{eq:QES2}) remain invariant. This is similar to the AdS/CFT case, where entropy of a boundary region should depend on the boundary UV cutoff but not that of the bulk
 QFT. 
 
We would like to discuss a bit more about the physical interpretation of the effective entropy. When the geometry is fluctuating, $S_A^{(n)}$ defined by the path integral in Eq. (\ref{eq:graventropy2}) or Eq. (\ref{eq:gravitationalentropy}) is generically not the Renyi entropy of a density operator. To understand its physical meaning, we can use a relation between Renyi entropy and correlation functions. For a system with finite Hilbert space dimension, and a Hilbert space that factorizes to $\mathbb{H}=\mathbb{H}_A\otimes\mathbb{H}_{\overline{A}}$, one can choose an orthonormal basis of operators $T_a,~a=1,2,...,D_A^2$ in region $A$ which satisfies 
 \begin{align}
     {\rm tr}\left(T_aT_b\right)=\delta_{ab},~\sum_{a}T_a^{\alpha\beta}T_a^{\gamma\delta}=\delta^{\alpha\delta}\delta^{\beta\gamma}
 \end{align}
 with $T_a^{\alpha\beta}$ the $\alpha\beta$ matrix element of $T_a$ in a certain basis. By decomposing the cyclic permutation operator one can show (more details in appendix \ref{app:renyi})
 \begin{align}
    e^{-S^{(n)}_A}&\equiv\sum_{a_1,a_2,...,a_{n-1}}\langle T_{a_1}\rangle\langle T_{a_1}T_{a_2}\rangle...\langle T_{a_{n-1}}T_{a_{n-2}}\rangle\langle T_{a_{n-1}}\rangle\label{eq:correlationdecom}
\end{align}
Here $\left\langle T_{a}\right\rangle={\rm tr}\left(\rho_AT_{a}\right)$ is evaluated in the original quantum state. Now we can change the point of view and view this as the definition of the $n$-th Renyi entropy. For a quantum field theory, the Hilbert space dimension is infinite, but one can imagine generalizing Eq. (\ref{eq:correlationdecom}) and define $T_a$ to be an orthonormal basis of operators which creates excitations below certain cutoff scale. (Note that the definition only requires $T_a$ to form an orthonormal basis. They do not necessarily generate a closed algebra under multiplication.) The generalization of the above correlation function to dynamical gravity case is
\begin{align}
    e^{-S_A^{(n)}}=\lim_{m\rightarrow -n}\int_{n,m}Dg e^{-S_{\rm EH}(g)}\left\langle T_{a_1}^{(1)}T_{a_1}^{(2)}T_{a_2}^{(2)}...T_{a_{n-1}}^{(n-1)}T_{a_{n-2}}^{(n-1)}T_{a_{n-1}}^{(n)}\right\rangle_g\label{eq:correlationcecom_grav}
\end{align}
Here $T_{a}^{(s)}$ labels the operator defined on $s$-th copy of $a$, and the expectation value is computed in the QFT with the background metric of $g$. In summary, as long as there is a gauge invariant definition of $A$ in the $n+m$ copied geometry and the QFT correlation functions are well-defined, $S_A^{(n)}$ can be defined using Eq. (\ref{eq:correlationcecom_grav}). 

\section{Examples}
\label{sec:examples}
\subsection{Euclidean partition function as a density matrix}\label{sec:GDM}

In this section, we will study a simple example of the gravitational no boundary density matrix. 
Considering the partition function of a Euclidean AdS (EAdS) gravity theory coupled to a QFT with Dirichlet boundary condition, a partition function can be viewed as a wavefunction of the QFT on the boundary of EAdS which describes the Hartle-Hawking state in a dS space under analytical continuation \cite{PhysRevD.28.2960,Maldacena_2003, Hertog_2012}.
However, if the boundary geometry is itself reflection symmetric, one can alternatively view the partition function as a density matrix on half of the boundary geometry. Such construction gives a general class of density matrices.
Such density matrix gives an example of the gravitational no boundary density matrix where we only fix the geometry on a spatial slice and sum over all possible geometries that are compatible with the boundary condition \cite{PhysRevD.34.2267,Hawking:1986vj,Barvinsky:2008vz,Maldacena:2019cbz,maldacenaStrings2019}.
This is a generalization of the  Hartle-Hawking no boundary wavefunction.

\begin{figure}
\center
\includegraphics[width=0.9\textwidth]{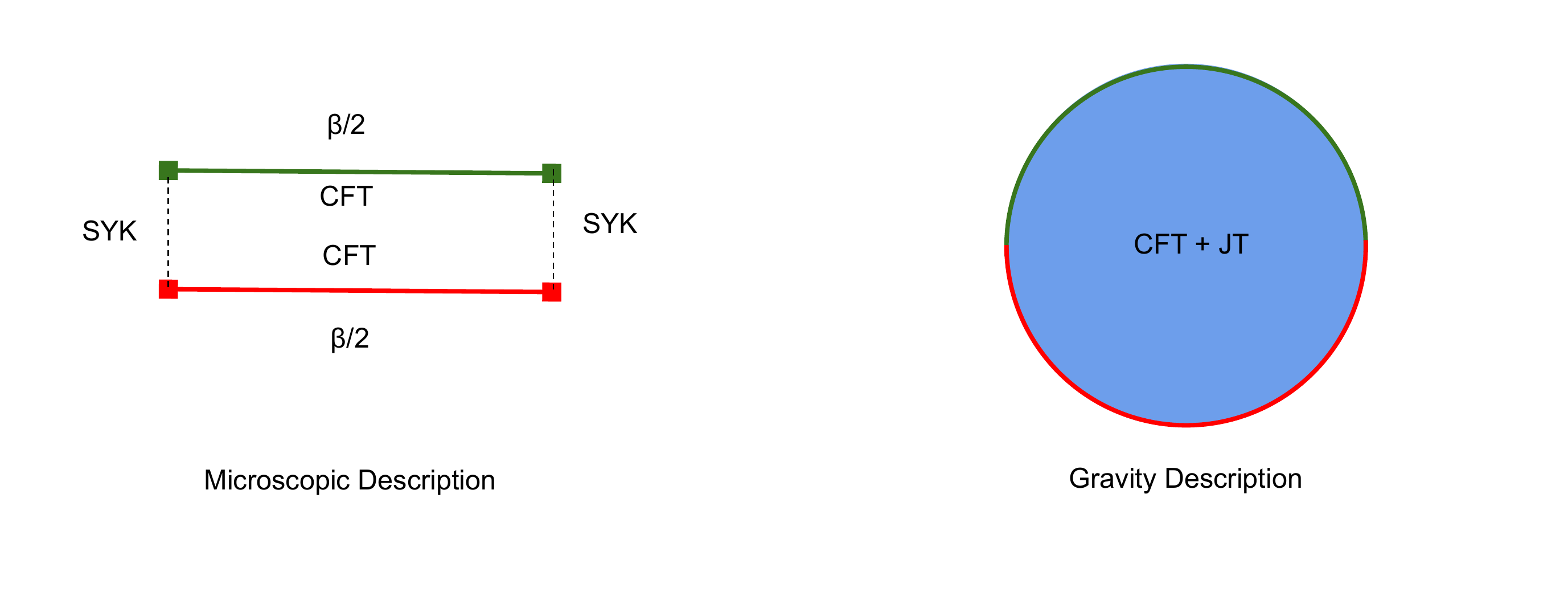}
\label{fig:DMdescription}
\caption{The microscopic (left panel) andgGravitational (right panel) description of the CFT density matrix.}
\end{figure}

Concretely, we can consider the partition function of JT gravity coupled to a 2d CFT  \cite{Almheiri:2014cka,Engels_y_2016,Maldacena:2018lmt,Yang_2019,Almheiri_2019}.\footnote{A closely related model will be discussed in  \cite{CGM} and we thank Juan Maldacena for discussion on this model.}
The boundary geometry is a Euclidean circle of length $\beta$, which can be split into two semicircles of length $\frac{\beta}{2}$.
With respect to the matter variables along the two semicircles, the partition function is a hermitian function and therefore defines a (unnormalized) density matrix of the 2d CFT on an interval of length $\frac{\beta}{2}$.\footnote{The reader should not confuse this density matrix with the TFD state of the dual CFT, which has a lower dimension.}
The path integral representation of the  (unnormalized) density matrix is the following:
\begin{equation}
    \rho(\psi_+,\psi_-)=\int\mathcal{D}\psi \mathcal{D}\phi \mathcal{D}g e^{S_0\chi(M)+\int_{M} \phi (R+2)+2\int_{\partial M}\phi_b K-S_{CFT}(\psi)}
\end{equation}
where $\psi$ is the CFT field variable. $\psi_+$ is its boundary value in the region $(0,\beta/2)$ (the green semicircle in figure \ref{fig:DMdescription}), and $\psi_-$ is that in the region $(\beta/2,\beta)$ (the red semicircle in figure \ref{fig:DMdescription}). 
We also fix the boundary value of the dilaton to be $\phi_b$.
It is instructive to look at the microscopic description of the density matrix using the duality between SYK system $(\chi)$ and JT gravity.
Suppose the CFT is the 2d free fermion theory, the boundary description of $\rho$ can be written as:
\begin{equation}
    \rho(\psi_+,\psi_-)=\tr \mathcal{P}\left(e^{-\beta/2 H_{SYK}-\int_0^{\beta/2} du \chi(u)\psi_+(u)} e^{-\beta/2 H_{SYK}-\int_{\beta/2}^{\beta} du \chi(u)\psi_-(u)}\right).
\end{equation}
where $\mathcal{P}$ stands for the path-ordering product.
 This can be regarded as a short-range entangled state of a 2d CFT with 2 SYK systems at the two ends of the semicircle prepared by the space evolution. 
$\rho(\psi_+,\psi_-)$ is the reduced density matrix of the 2d CFT after tracing out the SYK fermions.
In the bulk picture, this entanglement is described by the CFT living in a dynamical gravity background.
Clearly, from the microscopic description, the entropy of the CFT is bounded by the maximum entropy of the two SYK system.
In the bulk picture, this becomes the the Bekenstein bound with the maximum entropy of the SYK system replaced by $S_0+2\pi \phi_b$.
Such a bound is due to the emergence of nontrivial quantum extremal surface and below we will give an explicit calculation of the quantum extremal surface.
We will first discuss the classical saddle of the density matrix and the correlation functions, then discuss the von Neumann entropy with and without nontrivial quantum extremal surfaces.

\begin{figure}
\center
\includegraphics[width=0.9\textwidth]{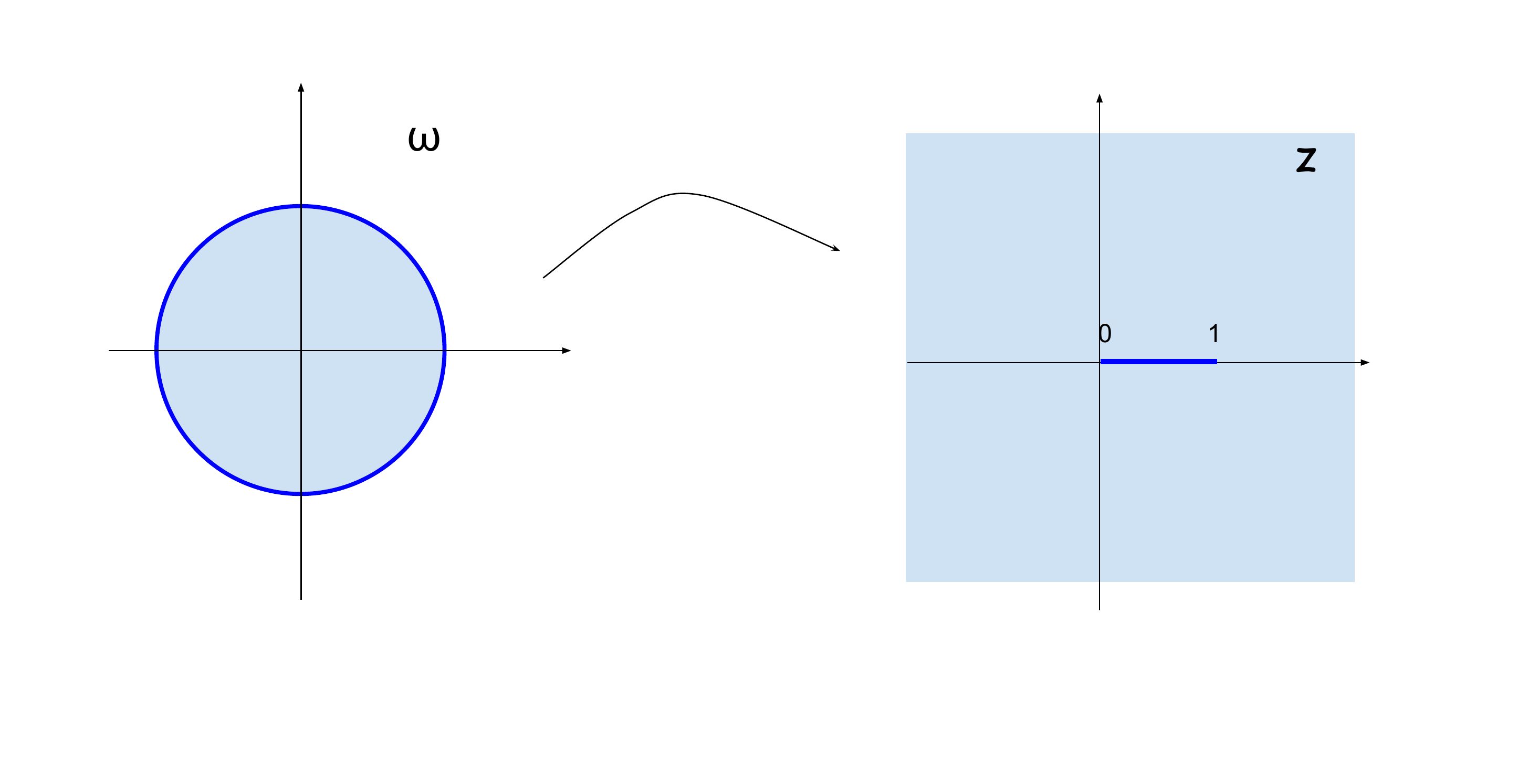}
\label{fig:ConfTrans}
\caption{Conformal transformation of the unit disk to the complex plane with a slit cut.}
\end{figure}

In the leading saddle approximation, the density matrix is a Euclidean path integral on the disk with hyperbolic metric:
\begin{equation}
	ds^2={4d\omega d\bar \omega\over (1-\omega\bar\omega)^2}
\end{equation}
Putting the boundary at $|\omega|=1-\epsilon$, we can determine $\epsilon$ is equal to ${2\pi\over\beta}$.
We can split the boundary along the real axis and treat the semicircle in the upper half plane as bra and the semicircle in the lower half plane as ket for the CFT respectively.
After gluing the bra and ket, the topology of the manifold is a sphere so this density matrix describes a bulk region in a closed universe. 
The complement region is the other time reflection symmetric slice, which is the diameter connecting the two ends of the semicircle. 
Due to the conformal invariance of the state, the density matrix can be equivalently viewed as the density matrix on the whole complex plane with a slit cut from $z_1=(0,0)$ to $z_2=(1,0)$, as is shown in Fig. \ref{fig:ConfTrans}.

The conformal transformation is given by:
\begin{equation}
	\omega={\sqrt{z}-\sqrt{z-1}\over \sqrt{z}+\sqrt{z-1}},~~\text{or}~~~z={(1+\omega)^2\over 4\omega},
\end{equation}
where the branch cut of the square root is taken to be from $0$ to $-\infty$. It can be easily checked that the boundary of the disk $\omega=e^{i\theta}$ is mapped to the slit with the bra in the lower half plane and the ket in the upper half plane:
\begin{equation}
	z=\cos^2{\theta\over 2}-{i \epsilon\over 2}\sin\theta.
\end{equation}
After the conformal transformation, the metric  becomes:
\begin{equation}
	ds^2\equiv e^{2\rho}dzd\bar z={4\omega(z) \bar\omega(z)\over |z||z-1|(1-\omega(z)\bar\omega(z))^2}dz d\bar z,
\end{equation} 
In particular, along the real line, the metric is
\begin{equation}
	ds^2={dz^2\over (1-z)z \epsilon^2}={\beta^2d\theta^2\over 4\pi^2},~~~~z\in(0,1);~~~~ds^2={dz^2\over 4(z-1)^2z^2},~~~z\in(-\infty,0)\cup(1,\infty).
\end{equation}
This geometry has a conical angle $\pi$ at the two ends of the slit, coming from the identification of the original partition function.\footnote{Since we are not integrating the dilaton field along the slit, the curvature on the slit does not need to satisfy the constant curvature constraint. The readers confused about this point may think of an ordinary quantum mechanical particle. If the position of the particle is fixed,  then its momentum can jump.  Notice that this is slightly different from the situation when we determine the boundary location of $A$ without fixing the entire metric of it.
}
From the trace anomaly $T^{\mu}_{\mu}={c\over 24\pi}R$, this indicates local high energy excitations at the two edges and we will regularize the conical angle in a small rage of size $\delta$, using the metric one can relate $\delta$ with the CFT UV cutoff $\tilde\epsilon$;
\begin{equation}
	ds^2={\beta^2\delta\over 4\pi^2}=\tilde \epsilon^2
\end{equation}
The CFT two-point function on the slit is uniquely determined by conformal symmetry and for operators with conformal dimension $\Delta$ it is equal to:
\begin{equation}
	\langle \mathcal{O}(\theta_1)\mathcal{O}(\theta_2)\rangle=\left({4\pi^2\over \beta^2}{\sqrt{(1-z_1)z_1(1-z_2)z_2}\over(z_1-z_2)^2}\right)^{\Delta}=({2\pi\over\beta})^{2\Delta}{\sin^{\Delta}\theta_1\sin^{\Delta}\theta_2\over (\cos\theta_1-\cos\theta_2)^{2\Delta}}.
\end{equation}
The correlator can be well approximated by the vaccum correlator $({2\pi\over \beta})^{2\Delta}{1\over(\theta_1-\theta_2)^{2\Delta}}$ for $\theta_{1,2}$ away from the two end points, which indicates that for most part of the region the density matrix is well approximated by the vacuum state, i.e the Hartle-Hawking no boundary state.
Near the two ends, the correlator vanishes because of the $\sin\theta$ factor in the numerator.
We can also consider the entropy of the density matrix, which is given by the two point function of the twist operators at the two ends $z_{1,2}$:
\begin{equation}
	S_{1}={c\over 3}\log{\beta|z_1-z_2|\over 2\pi\tilde\epsilon  (1-z_1)^{1\over 4}z_1^{1\over 4}(1-z_2)^{1\over 4}z_2^{1\over 4}}={c\over 3}\log{\beta\over 2\pi\tilde \epsilon  \delta^{1\over 2}}={2c\over 3}\log{\beta\over 2\pi \tilde\epsilon}.
\end{equation}
This is the entropy of the CFT density matrix on the fixed disk geometry.
The exact density matrix, on the other hand, is the one given by the gravitational path integral over all geometries with the circular boundary condition.
In general, there are two types of corrections to the density matrix.  
One is perturbative correction coming from the backreaction of the matter and the other is nonperturbative correction from the change of topology.
When the central charge of the matter is small we expect both corrections are small so the exact density matrix should be well approximated by the fixed geometry results.
In the region of large central charge ($c$ is the same order as the gravitational entropy), however, both the perturbative and nonperturbative correction will be important.

For the perturbative corrections, the dilaton field will have large backreaction due to the bulk stress tensor, which may cause the change of the shape of the slit. In order to glue the slit one need to solve the conformal welding problem.
Fortunately, this complication does not occur when calculating the saddle point of $\tr\rho$. 
Using conformal anomaly, we can explicitly write down the bulk stress tensor in the complex $z$ coordinates:
\begin{equation}
	T_{zz}=-{c\over 12\pi}((\partial\rho)^2-\partial^2\rho)={c\over 64 \pi}{1\over (1-z)^2 z^2};
	~~T_{z\bar z}=T_{\bar z z}=-{c\over 12\pi}\partial\bar\partial\rho=-{c\over 48\pi}e^{2\rho};~~~T_{\bar z\bar z}=\bar T_{zz}.
\end{equation}
The Einstein equations in JT gravity are the following:
\begin{eqnarray}
		\partial\bar\partial\phi-{1\over 2}e^{2\rho}\phi &=&{1\over 2}T_{z\bar z}=-{c\over 96\pi}e^{2\rho};\\
		-e^{2\rho}\partial (e^{-2\rho}\partial\phi)={1\over 2}T_{zz};&~&~~-e^{2\rho}\bar\partial (e^{-2\rho}\bar\partial\phi)={1\over 2}T_{\bar z\bar z}.
\end{eqnarray}
If the stress tensor are zero, then we have the vaccum solution which can be easily determined from the solution in the original metric $\omega$:
\begin{equation}
	\phi_{vac}=\phi_h{1+\omega\bar\omega \over 1-\omega \bar \omega}=\phi_h {z+\bar z-1\over\sqrt{(z-1) z}+\sqrt{(\bar z-1)\bar z}},
\end{equation}
where $\phi_h$ is the minimum value of the dilaton field.
With the stress tensor, we have additional inhomogeneous solutions. 
For the $T_{zz}$ and $T_{\bar z \bar z}$ component, the inhomogeneous solution can be easily derived by integrating over the stress tensor:
\begin{equation}
\begin{split}
\phi_{inh}(z,\bar z)&=-{1\over 2}\int^{z} dz_1 e^{2\rho}\int d z_2e^{-2\rho}T_{z_2 z_2}-{1\over 2}\int^{\bar z} d\bar z_1 e^{2\rho}\int d \bar z_2e^{-2\rho}T_{\bar z_2 \bar z_2}\\
&={c\over 32\pi}{(z+\bar z-2z\bar z+2\sqrt{z(z-1)\bar z(\bar z-1)})(\tan^{-1}(1-2z)-\tan^{-1}(1-2\bar z))\over z-\bar z}.
\end{split}
\end{equation}
Later we will use the property of $\phi_{inh}$ along the real axis which takes the form:
\begin{equation}
	\phi_{inh}(z)=-{c\over 16\pi} \theta(1-z)\theta(z).
\end{equation}
It is straightforward to check that the $\phi_{inh}$ piece satisfies both the $T_{zz}$ and $T_{\bar z\bar z}$ equations and in addition it has property:
\begin{equation}
\partial\bar\partial\phi_{inh}-{1\over 2}e^{2\rho}\phi_{inh}={c\over 32\pi} e^{2\rho}.
\end{equation}
This together with the $T_{z\bar z}$ equation determines the backreacted solution of $\phi$:
\begin{equation}
\phi=\phi_{vac}+\phi_{inh}+{c\over 12\pi} .
\end{equation}
The shape of the slit is determined by the boundary condition $\phi=\phi_b$. Since the value of the inhomogeneous solution is a finite constant along the slit, the shape of the slit is undeformed.
The boundary condition of the dilaton field determines $\phi_h$ in a standard way:
\begin{equation}
	\phi_h={2\pi \phi_b\over \beta}.
\end{equation} 
One can also use the following argument from Schwarzian theory to derive the same conclusion. After integrating out the dilaton field, the gravitational action becomes the Schwarzian action along the slit, and we need to consider its backreaction from the CFT partition function.  
The Schwarzian variables can be parametrized by the function $\theta(u)$ where $u$ is the proper length along the boundary. In order to glue the slit along the same boundary location one might need to consider additional conformal map to align the $u$ variables in the bra segment and ket segment. 
This happens when considering the off diagonal elements of the density matrix. 
Fortunately, for the diagonal elements and $\tr\rho$, due to the time reflection symmetry of the state, the saddle point of the two Schwarzian variables along the bra and ket are required to be identical, which means that the matching condition is trivial. Thus, the CFT partition function is independent of these time reflection symmetric schwarzian variables, therefore the classical saddle is again given by the saddle of the Schwarzian action, which is $\theta(u)={2\pi u\over \beta}$.

\begin{figure}
    \centering
    \includegraphics[width=0.8\textwidth]{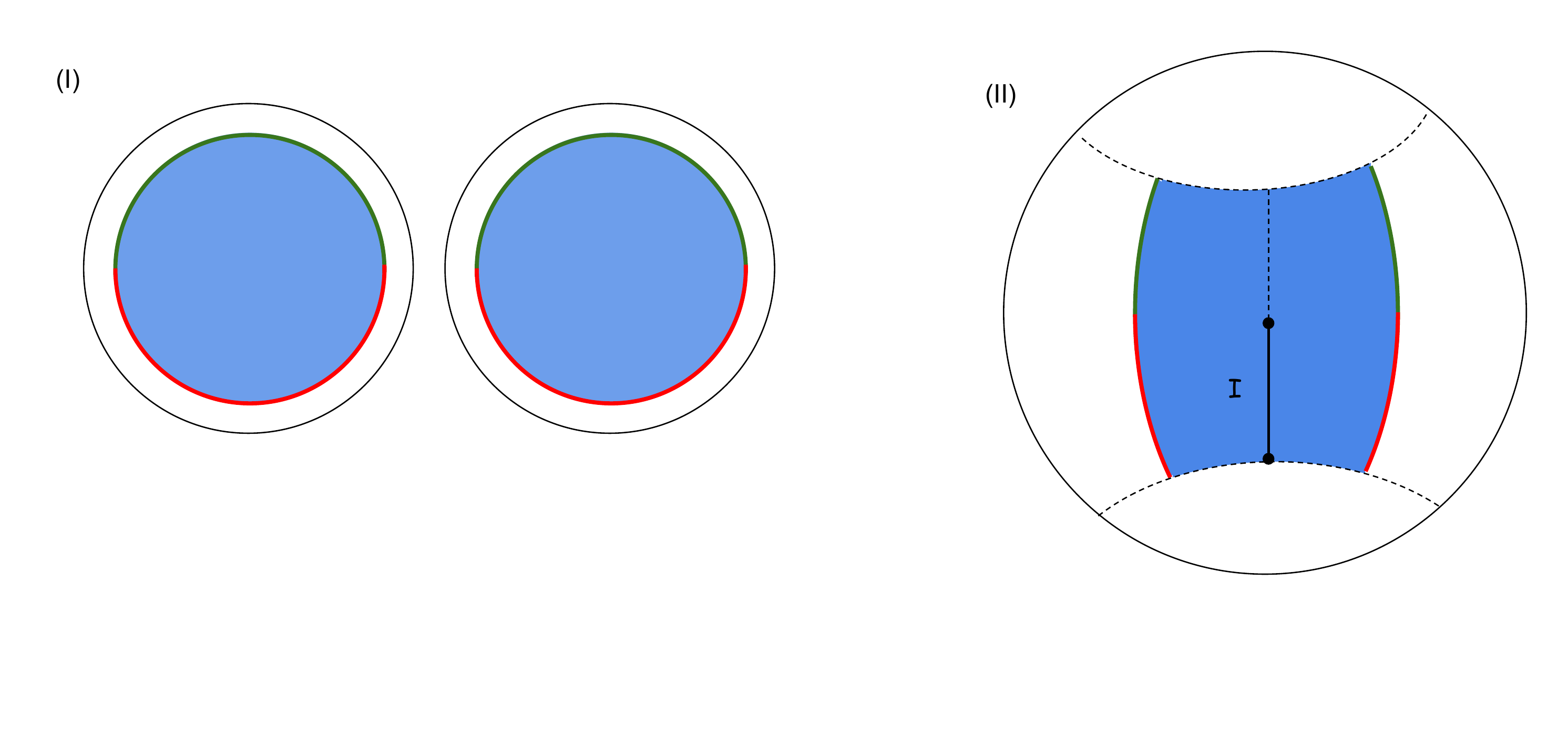}
    \caption{The two replica geometries: (I) is the disconnected saddle, (II) is the connected saddle which is the external traversable wormhole geometry by gluing alond the dashed line. The two red (green) lines are glued together.}
    \label{fig:2RenyiGDM}
\end{figure}

We have finished our discussion on perturbative corrections. Now we talk about the non-perturbative corrections.
For $\tr\rho$ and correlators, the non-perturbative correction are suppressed to order $e^{-S_0}$, which can be ignored.
Nevertheless as discussed in the context of replica wormholes  \cite{Penington:2019kki,EastCoast}, for the Renyi entropies $\tr\rho^n$, the non-perturbative saddles can dominate.
For example, the two-replica geometry, whose boundary condition is given by two coupled euclidean circles, gives the same contribution as the thermal partition function of two coupled SYK systems \cite{Maldacena:2018lmt,Maldacena:2019ufo}, and as a result there are two bulk geometries (figure \ref{fig:2RenyiGDM}): one is a product of two euclidean AdS disks and the other is the eternal traversable wormhole geometry.

\begin{figure}
    \centering
    \includegraphics[width=0.4\textwidth]{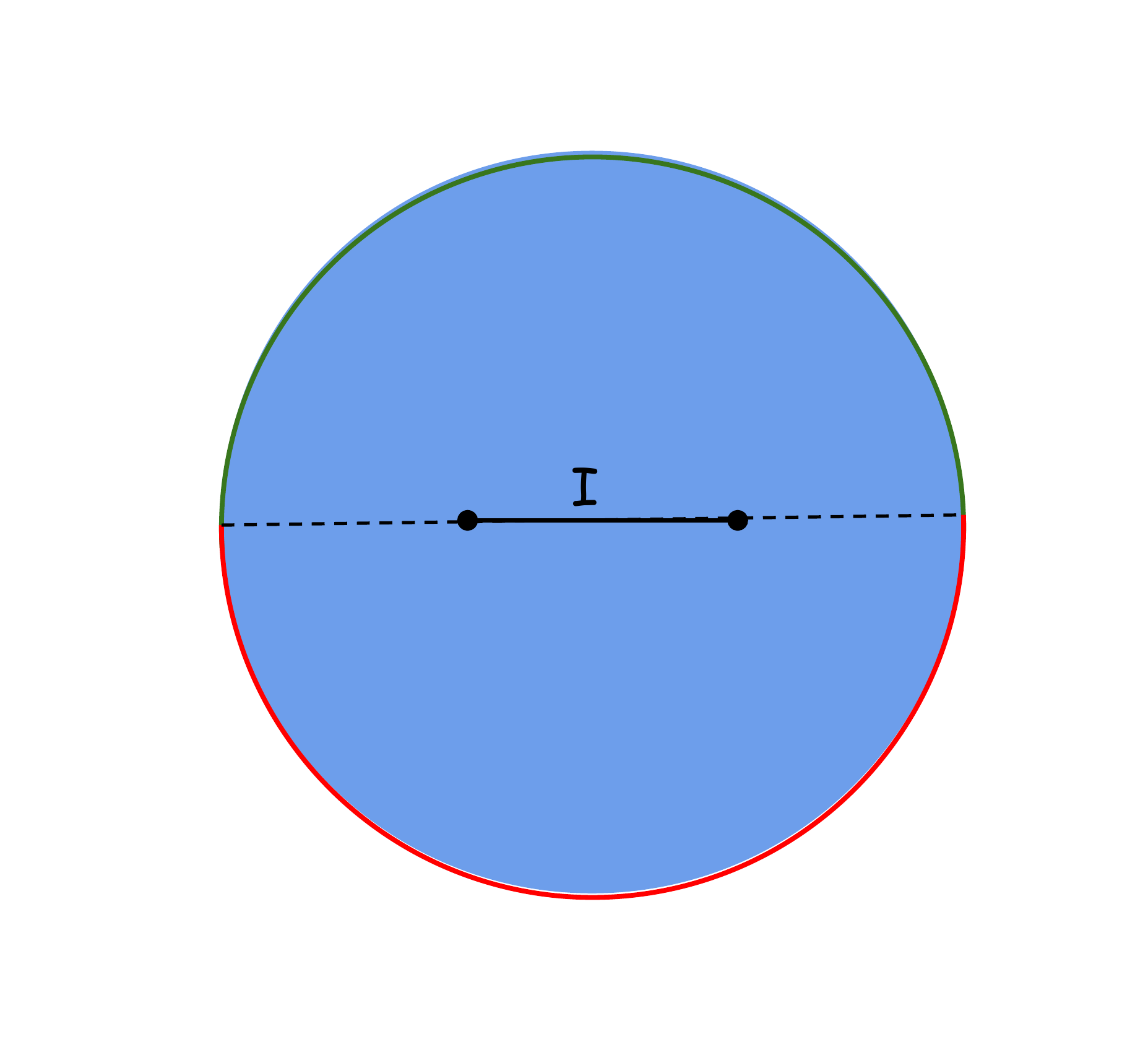}
    \caption{The island region in the original saddle}
    \label{fig:iGDM}
\end{figure}

The parameters of the coupled SYK system are $c,S_0\sim N$, $\phi_b\sim {N\over J}$, and 
the thermal partition function  has first order phase transition around $\beta J \sim O(1) $. 
The eternal traversable wormhole geometry dominates at larger $\beta$, when the free energy becomes order one due to the existence of a gap. As a result, the second Renyi entropy approaches a constant of order $e^{-2S_0}$, which is consistent with our expectation that the Von Neumann entropy should be bounded.

\begin{figure*}
\subfigure[$\tilde\epsilon=0.1$, $S_0=200$, $2\pi \phi_b=20$, $\beta=20$.]{
\includegraphics[width=0.55\textwidth]{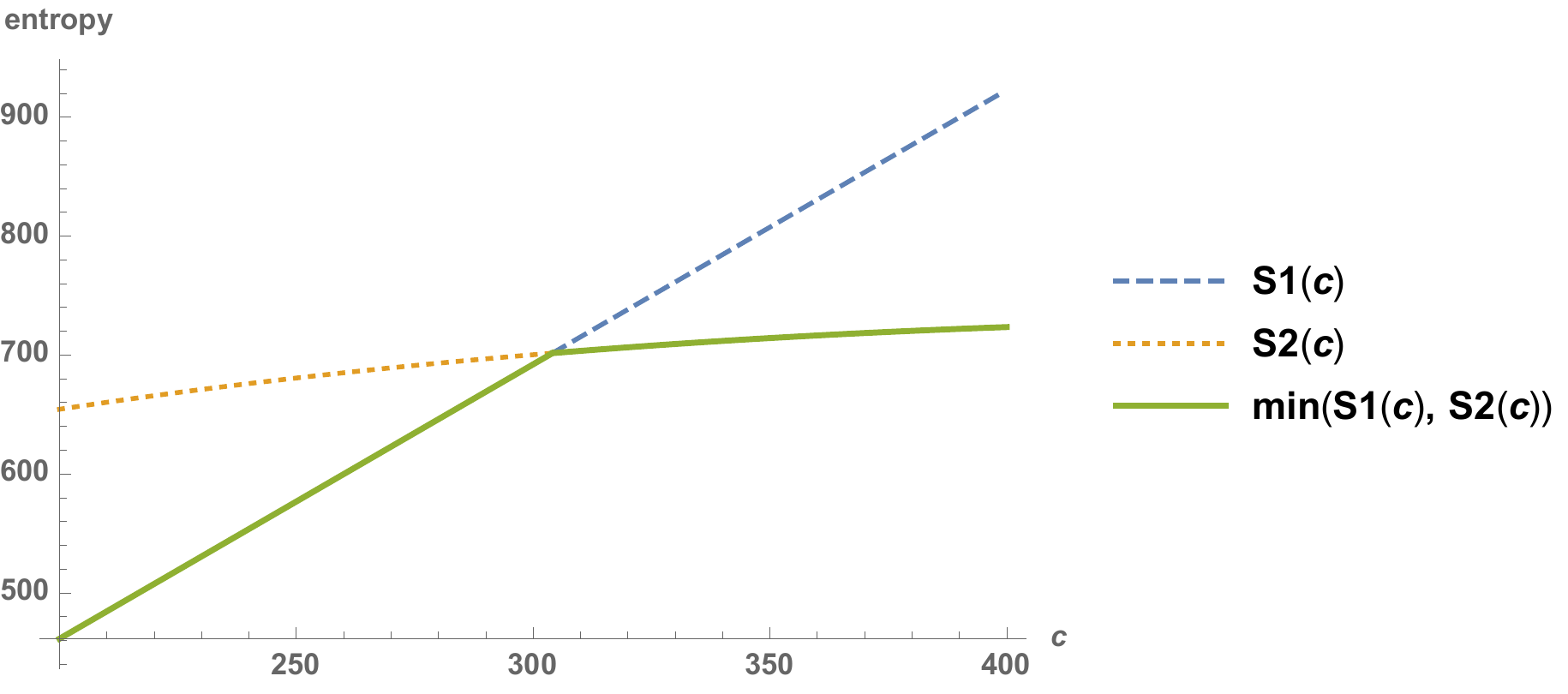}}
\subfigure[$\tilde\epsilon=0.1$, $S_0=200$, $2\pi \phi_b=20$, $c=300$.]{
\includegraphics[width=0.55\textwidth]{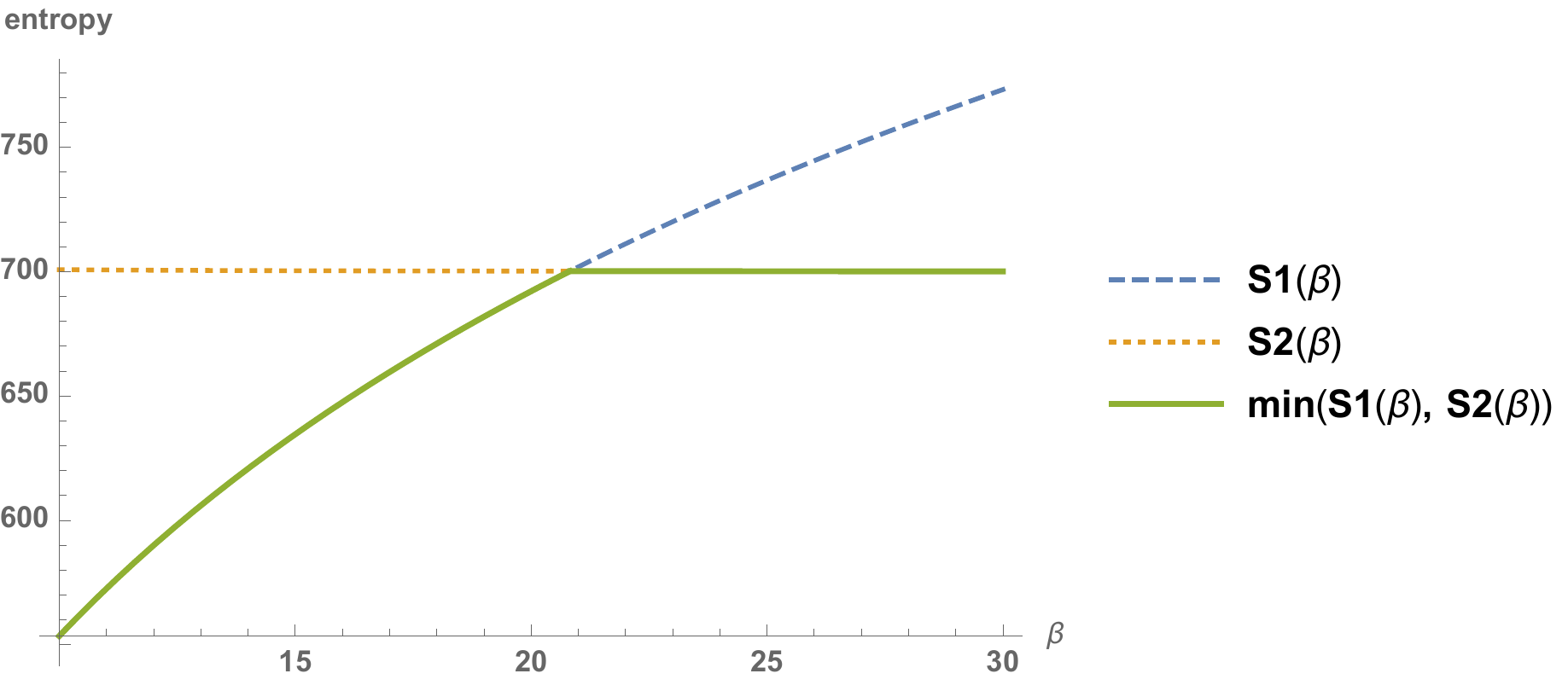}}
\caption{A numerical plot for the entropy as a function of central charge $c$ and proper length $\beta$: $S_1$ is the effective entropy without island and $S_2$ is the effective entropy with island contribution. The von Neumann entropy is the minimum of these two entropies. }
\label{fig:NPlot}
\end{figure*}

In the von Neumann limit this reflects the emergence of a nontrivial quantum extremal surface in the original saddle (figure \ref{fig:iGDM}). The quantum extremal surface consists of pairs of points on the manifold. 
Because of time reflection symmetry, the location of the quantum extremal surface could only be along the real $z$ axis. We consider the simplest case when the QES is a single pair of points at locations $z_3$ and $z_4$. Without loss of generality, we assume that $z_3<0$ and $z_4>1$. The effective entropy is given by the four-point function of the twist operators.
In general such function is dependent on the operator spectrum.
However, for theories with large central charge and small number of low-dimension operators, only the Virasoro block of identity operator will dominate and the four-point function can be approximated by a product of two two-point functions \cite{Hartman:2013mia,Faulkner:2013yia}.
Since we are looking for the saddle that has the smallest effective entropy, the emergent twist operators from the QES should be contracted with the two operators at the two ends of the slit.
This gives the bulk entropy:
\begin{equation}
		S_{bulk}(z_3,z_1=0,z_2=1,z_4)={2c\over 3}\log{\beta\over 2\pi\tilde \epsilon}+{c\over 3}\log\sqrt{-z_3\over 1-z_3}+{c\over 3}\log\sqrt{z_4-1\over z_4} +(\text{UV})
\end{equation}
The CFT UV divergence at the quantum extremal surface can be absorbed into $S_0$.
The first piece is the same as the original entropy $S_{1}$ and the other pieces are negative. As $z_{3,4}$ approaches the two ends, the bulk entropy becomes zero. 
Of course the locations of $z_{3,4}$ are determined by the extremal condition of the effective entropy which is a sum of the QFT entropy and the dilaton field:
\begin{equation}
\begin{split}
	S_{2}&=\min_{\rho_{1,2}} S_{1}-{c\over 3}(\rho_1+ \rho_2)+2\pi \phi_h(\cosh \rho_1+\cosh \rho_2)+2S_0+{c\over 3}\\
	&={2c\over 3}\log{\beta\over 2\pi \tilde \epsilon}+2S_0+{c\over 3}-{2c\over 3}\sinh^{-1}{c\beta\over 12\pi^2\phi_b}+2\sqrt{({4\pi^2 \phi_b\over \beta})^2+{c^2\over 9}}
\end{split}
\end{equation}
where $\rho_1=-\log\sqrt{-z_3\over 1-z_3}\geq 0$ and $\rho_2=-\log\sqrt{z_4-1\over z_4}\geq 0$. The saddle point is given by $\rho_{1,2}=\sinh^{-1}{c\over 6\pi \phi_h}=\sinh^{-1}{c\beta\over 12\pi^2\phi_b}$.
We should keep in mind that the above formula ignores the boundary graviton contributions which requires: ${ \phi_b\over \beta}\gg 1.$
Compared with the entropy without island $S_{1}$, we found the phase transition happens around:
\begin{equation}
  {\beta\over 4\pi^2 \phi_b}\sim {3\over c}\sinh\left(3{S_0\over c}+3/2\right).
\end{equation}
In the region of $c\sim S_0$, the phase transition is around $\beta\sim {\phi_b\over c}$.
When $ {c\beta\over \phi_b}\gg 1$, the effective entropy approaches a constant independent of $\beta$:
\begin{equation}
   S_{2}\sim {2c\over 3}\log{3\pi\phi_b\over c \tilde \epsilon}+2S_0+ c.
\end{equation}
In figure \ref{fig:NPlot}, we show the numerical plot of the behavior of phase transition with respect to $c$ and $\beta$.

The emergence of the island region is an interesting phenomenon. Semiclassically, the global state is the real $z$ axis and our density matrix is given by tracing out everything outside of the region between $(0,1)$.
When the matter entropy of this region is big enough, an island region appears in the spatial region that we traced out. This reduces the matter entropy such that it can never exceed the boundary area of the original region, which is consistent with the Bekenstein bound. Here we consider the density matrix at some bulk region, and the existence of an island region is a direct consequence of the fact that we are not fixing the geometry away from the region of interest. A very important question is then how to determine whether a certain part of bulk geometry should be fixed during the gravitational path integral. This is an observer-dependent question.  The physics probed by a bulk observer (or a group of bulk observers) corresponds to a fixed geometry in the region probed, while the unobserved part of the universe can have a fluctuating geometry. Similar observer dependence will be discussed in tensor network models in Sec. \ref{sec:tensornetwork}. A more quantitative and systematic answer to this question is unknown and requires future work.

\subsection{Four dimensional Schwarzchild black hole in flat spacetime}\label{sec:schwarzchild}

In this section we discuss the bulk field effective entropy of a Schwarzchild black hole in four dimensional asymptotic flat spacetime. The entanglement island in flat spacetime black hole geometry has been discussed in recent works \cite{Gautason:2020tmk, Anegawa:2020ezn, Hashimoto:2020cas, Hartman:2020swn}. Our calculation is qualitatively similar but for a different state, as will be discussed below.

\begin{figure}
    \centering
    \includegraphics[width=4in]{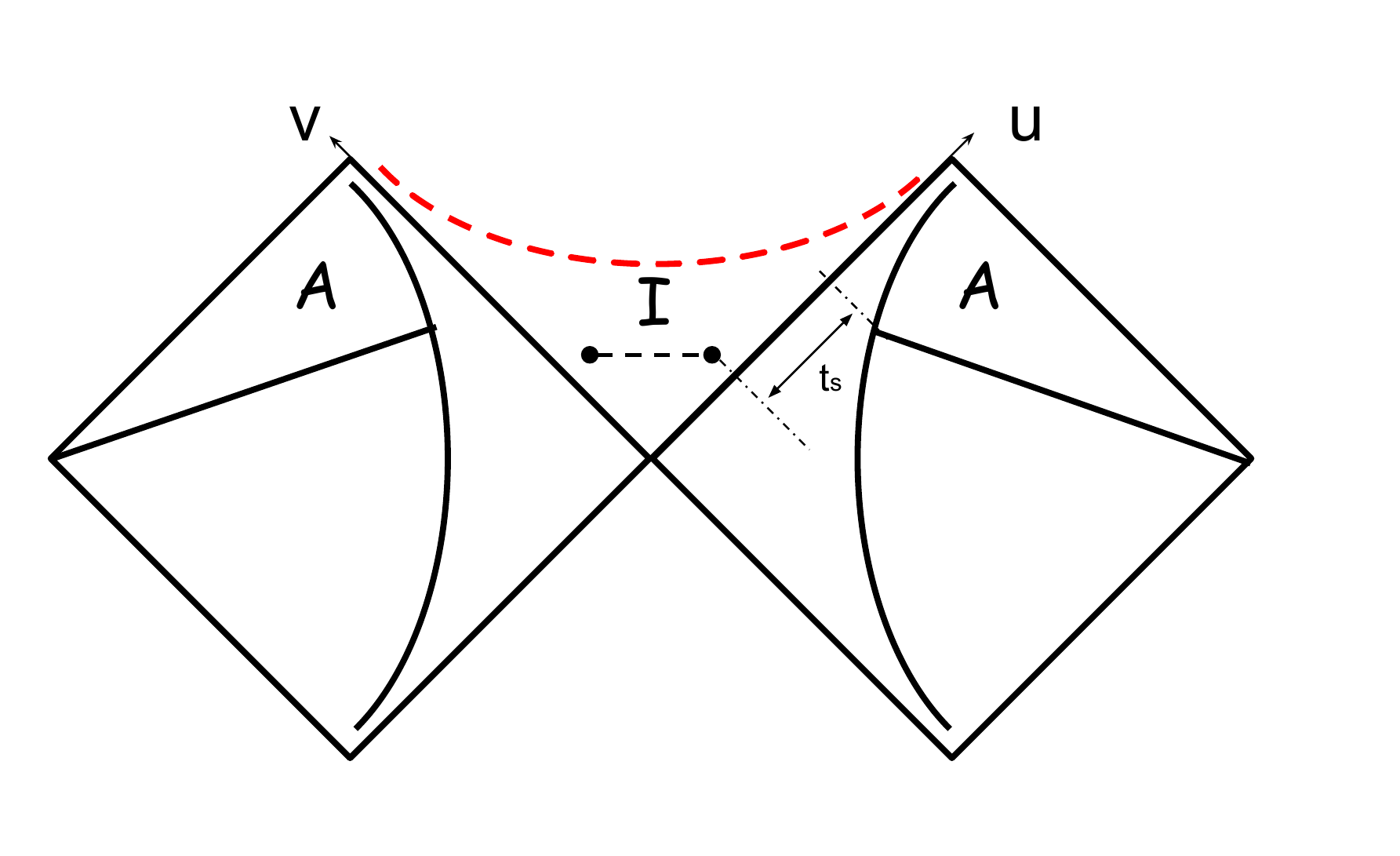}
    \caption{Illustration of the eternal black hole geometry and region $A$ we are considering. }
    \label{fig:kruskal}
\end{figure}

For simplicity we consider the maximally extended Schwarzchild black hole rather than an evaporating one. In Kruskal–Szekeres coordinates, the metric is
\begin{equation}
    ds^2 = - \frac{4r_H^3}{r}\exp\left(-\frac{r}{r_H}\right) dU dV + r^2 d\Omega^2
\end{equation}
where $r$ is the Schwarzchild radius and is related to the Kruskal coordinates as follows:
\begin{equation}
    U V=\left(1-\frac {r}{r_H}\right)e^{r/r_H}.
\end{equation}
and $r_H$ is the radius of the horizon.
The maximally extended geometry describes two entangled black holes. If there are matter fields, the black hole will create particles from vacuum fluctuation and emit Hawking radiations. We are interested in the entropy of the Hawking radiation in the union of two exterior regions, defined by radius $r\geq r_H$ at a Schwarzchild time $t$ (See Fig. \ref{fig:kruskal}). Note that we have taken the time direction in both sides of the eternal black hole to move up, so that the time dependence is nontrivial. In Kruskal coordinate, the boundary of the region in the right-hand-side wedge is defined by
\begin{align}
    U&=Re^t,~V=-Re^{-t}\label{eq:constantr}\\
    \text{with~}R^2&=\left(\frac {r}{r_H}-1\right)e^{r/r_H}\nonumber
\end{align}
Similarly the boundary in the left-hand-side wedge is $U=-Re^{-t},V=Re^{t}$.

We make several simplifications to the problem. Firstly, we decompose the matter field into spherical modes of the transverse direction, only focusing on the massless modes following Polchinski \cite{Polchinski_2016}. We ignore its interaction with other fields. Secondly, we assume the matter field has a small energy-momentum tensor and neglect its back reaction to the metric. (We will later check that this approximation is self-consistent.) Finally, we choose a special state of the massless modes to simplify the entropy calculation. 

The first simplification essentially reduces the setup to a CFT living on a two dimensional geometry in the $U, V$ direction. We choose a particular state of the CFT that is obtained by doing a Weyl transformation on the Minkowski vacuum state. 
Naively, one may Weyl transform the state from $ds^2=-dUdV$ to the Kruskal coordinate. However, because the metric in Kruskal coordinate vanishes exponentially with $r$ near infinity, such transformation results in a large energy momentum near infinity, which is inconsistent with our assumption of small back-reaction. Hence, we instead introduce the following coordinate transformation:
\begin{align}
    U=\sinh u, V=\sinh v
\end{align}
 The metric in $(u,v)$ becomes
\begin{equation}
    ds^2=-g(u,v) du dv;~~~g(u,v)=\frac{4r_H^3}{r}\exp\Big(-\frac{r}{r_H}\Big) \cosh u \cosh v
\end{equation}
The advantage of using $u,v$ coordinates is that at large distance the metric becomes flat:
\begin{equation}
\left.g(u,v)\right|_{|u|\gg 1,|v|\gg 1}\simeq 4 r_H^2
\end{equation}
so that the Weyl transformation brings the flat spacetime vacuum state to some state with nontrivial energy-momentum tensor restricted to the region defined by $|u|<1$ or $|v|<1$.\footnote{It should be noted that this region with nontrivial energy-momentum still extends to null infinity.} For the region near horizon $|u|\ll 1,|v|\ll 1$, $u\simeq U,~v\simeq V$ and the coordinate returns to Kruskal–Szekeres coordinate. Physically, roughly speaking the state we are studying (at Schwarzchild time $t=0$) is like a thermal-field double state for a finite region coupled with a pair of semi-infinite bath at zero temperature. This is the key difference between our result from that of Ref. \cite{Hashimoto:2020cas}.

More generally, the interpolation scale can be tuned by defining $U=a\sinh\frac{u}a,~V=a\sinh\frac{v}a$. Since there is no scale invariance in $U,V$, physics at different $a$'s is not equivalent. We have studied the case with general $a$ and confirmed that the choice of $a$ does not affect the late time behavior that will be discussed below. Therefore for simplicity we will keep $a=1$ in the remainder of the discussion.

\begin{figure}
    \centering
    \includegraphics[width=6.2in]{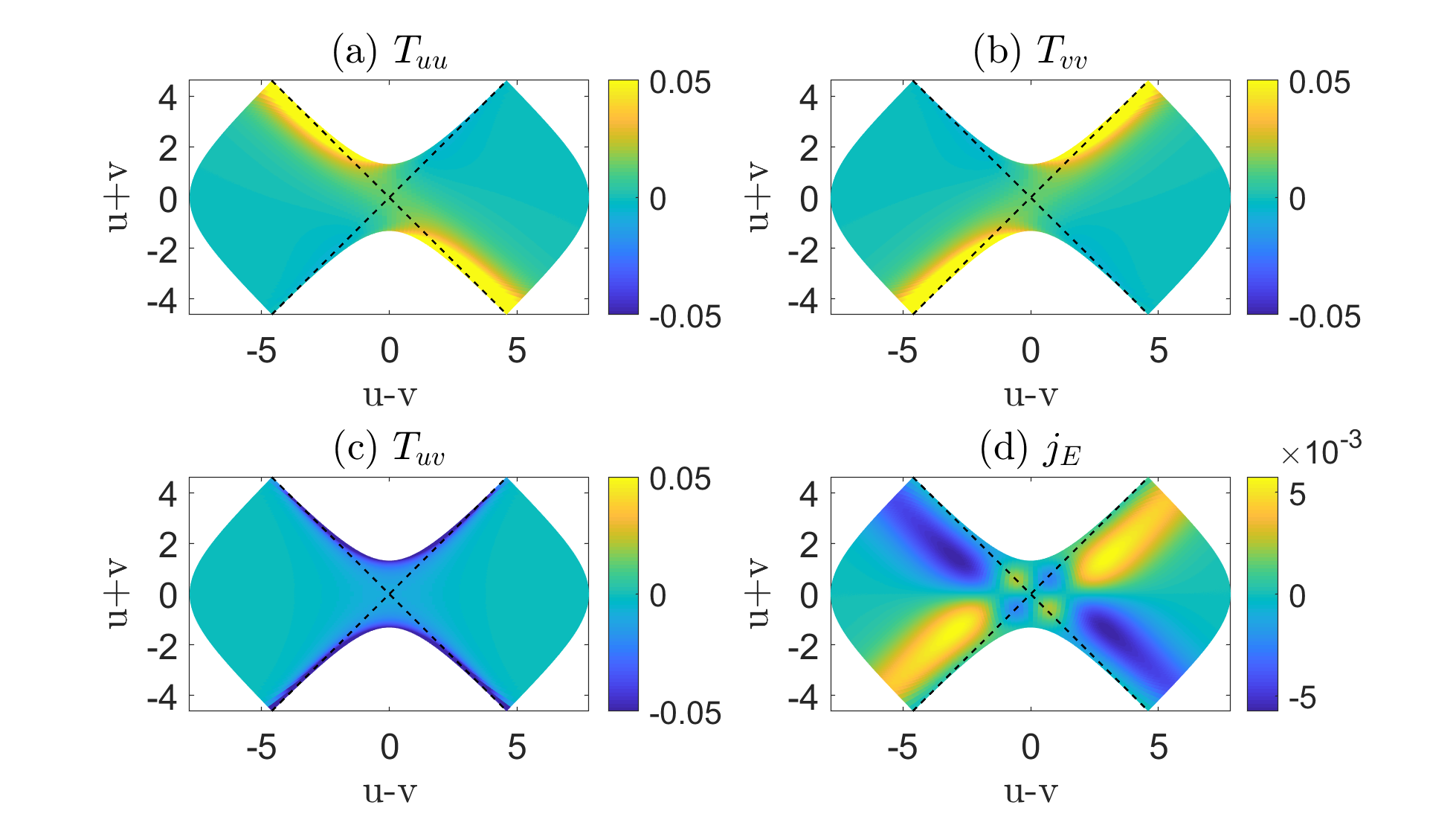}
    \caption{(a)-(c) Numerical results of the three components of energy momentum tensor. $T_{uu}$ and $T_{vv}$ peaks around the horizon, and $T_{uv}$ is slowly varying in most of the places. Divergence occurs near singularity. (d) The outgoing energy current density across a constant $r$ surface, given by Eq. (\ref{eq:energyflux}). (Note that the scale is smaller than Fig. (a)-(c). We plotted this quantity everywhere, although we will only use it in exterior regions.) In (a) (b) and (c), to increase visibility we have replaced data bigger than $0.05$ by $0.05$, and that smaller than $-0.05$ by $-0.05$.}
    \label{fig:energymomentum}
\end{figure}

For this state, the actual stress tensor with the cutoff defined with respect to the physical metric is determined by the conformal anomaly. In general when the metric is transformed as $\tilde{g}_{\mu\nu}\rightarrow g_{\mu\nu}= e^{2\omega}\tilde{g}_{\mu\nu}$, the energy momentum tensor transforms as
\begin{align}
    T_{\mu\nu}^g=T_{\mu\nu}^{\tilde{g}}-\frac c{12\pi}\left[\partial_\mu\omega\partial_\nu\omega-\frac12\tilde{g}_{\mu\nu}\tilde{\nabla}_\sigma\omega\tilde{\nabla}^\sigma\omega-\tilde{\nabla}_\nu\tilde{\nabla}_\mu\omega+\tilde{g}_{\mu\nu}\tilde{\nabla}^2\omega\right]
\end{align}
with $\tilde{\nabla}_\mu$ the covariant derivative in metric $\tilde{g}_{\mu\nu}$, and $c$ the central charge of the CFT. In our case, $\tilde{g}_{\mu\nu}=\eta_{\mu\nu}$ is the flat space, and we consider the vacuum state with $T_{\mu\nu}^{\tilde{g}}=0$. Explicitly, in terms of $g(u,v)$ we obtain
\begin{align}
     T_{uu}&=-\frac{c}{12\pi}\sqrt{g(u,v)}\partial_u^2\frac1{\sqrt{g(u,v)}},~~~
     T_{vv}=-\frac{c}{12\pi}\sqrt{g(u,v)}\partial_v^2\frac1{\sqrt{g(u,v)}},\\
     T_{uv}&=-{c\over 12\pi }\partial_u\partial_v\log\sqrt{g(u,v)}.
\end{align}
The stress tensor vanishes for $|u|\gg 1,|v|\gg 1$ when the metric becomes flat. Physically the state we consider has energy and momentum near the horizon at time $t=0$. We have verified this by numerics, as is shown in Fig. \ref{fig:energymomentum}.

\begin{figure}
    \centering
    \includegraphics[width=5in]{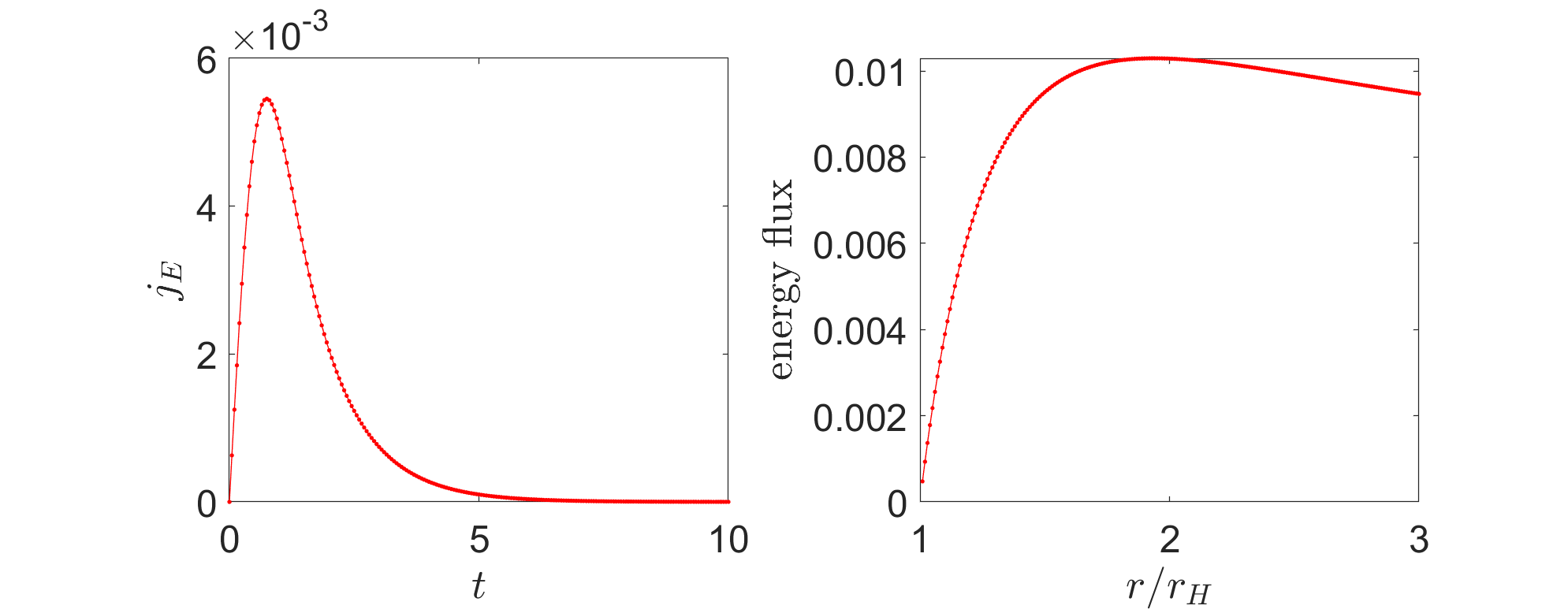}
    \caption{(a) Energy current density as a function of Schwarzchild time. (b) Total energy flux at late time, which is an integration of the energy current density for $t\in[0,+\infty)$, as a function of radius $r$.}
    \label{fig:energyflux}
\end{figure}

It should be noticed that the state is not boost invariant due to our choice of the $u,v$ coordinates, so that the matter field is not in thermal equilibrium with the black hole. In principle one needs to solve the backreaction on the geometry due to the nontrivial stress tensor using the Einstein equations. Such backreaction, for example, will describe the energy loss of the black hole due to the Hawking radiation. We can estimate the amount of energy loss in current state by looking at the energy flux across the constant radius $r$ surface. Using the boost killing vector:
\begin{align}
    \left(\xi^u,\xi^v\right)=\left(\tanh u,-\tanh v\right),
\end{align}
the energy flux across the constant $r$ surface is:
\begin{align}
    I_E=\int_0^tdt'\xi^\mu T_{\mu}^{~\tau}\epsilon_{\tau \nu}\xi^\nu =\int_0^tdt'\frac{2}{g(u,v)}\left(-\tanh^2 u T_{uu}+\tanh^2 v T_{vv}\right).\label{eq:energyflux}
\end{align}
Here $t$ is the Schwarzchild time defined in Eq. (\ref{eq:constantr}). The change of the black hole mass, or equivalently the change of the quasilocal stress tensor along the constant $r$ slice is equal to the total energy flux. As is shown in Fig. \ref{fig:energymomentum} (d) and Fig. \ref{fig:energyflux} (a), the energy current density peaks around the interpolation scale $u\sim 1$ and vanishes in long time, leading to a finite energy flux shown in Fig. \ref{fig:energyflux} (b). Therefore the energy change is of order $c$ and the backreaction can be neglected in the limit $c\ll \frac{r_H^2}{G_N}$.

Now let's look at the entropy calculation. The entropy of a single interval in the state we consider is obtained from a Weyl transformation of that in flat spacetime vacuum of the CFT \cite{Calabrese_2009}\footnote{More precisely the first term should be $\log\left(-\frac{u_{12}v_{12}}{\epsilon^2}\right)$ with a $UV$ cutoff $\epsilon$. We have neglected $\epsilon$ since its effect can be approximately absorbed in a redefinition of $\propto\frac1{G_N}$ in the region we are interested in.}: 
\begin{align}
    S(u_1,v_1;u_2,v_2)&={c\over 6}\log(-u_{12}v_{12})+{c\over 12}\log \left(g(u_1,v_1) g(u_2,v_2)\right)\label{eq:entropygeneral}
\end{align}
with $u_{12}=u_1-u_2$ and $v_{12}=v_1-v_2$. If there is no entanglement island, the boundary of $A$ is given by $(u_1,v_1)=(u_A,v_A)$ and $(u_2,v_2)=(v_A,u_A)$, with 
\begin{align}
    u_A=\sinh^{-1}\left(Re^t\right),~v_A=-\sinh^{-1}\left(Re^{-t}\right)
\end{align}
and $R$ is defined in Eq. (\ref{eq:constantr}). In the late time limit 
\begin{align}
    u_A\simeq t+\log(2R),~v_A\simeq -Re^{-t}
\end{align}
The entropy grows linearly in time, with leading contribution from the Weyl factors:
\begin{align}
    S_{\rm no~Island}(t)&\simeq \frac c6\left(t+\log t+{\rm const.}\right)
\end{align}


The Bekenstein-Hawking entropy of the black hole is given by the transverse area at the horizon:
\begin{equation}
    S_{BH}={\pi r_H^2\over G}.
\end{equation}
We expect that a transition occurs when the entropy of the Hawking radiation is close to $2S_{BH}$ (with factor of $2$ coming from the two-sided geometry), which provides an estimate of the Page time $t_P\simeq 12S_{BH}/c$. 

After the Page time, we expect that the entropy calculation is dominated by a nontrivial quantum extremal surface, so that the entropy is reduced to a value close to the black hole entropy. 
Due to the reflection symmetry of the geometry, the quantum extremal surfaces should locate at $(u_I,v_I)$ and $(v_I,u_I)$. The QFT entropy is then a two-interval entropy which depends on more details of the CFT. However, in the limit we are considering, the two intervals are far away from each other, so that we can approximate the twist operator four-point function by a product of two-point functions. This leads to the effective entropy formula
\begin{equation}
    S_{\rm Island}={\rm Ext}_{u_I,v_I}\left[{c\over 3}\log(-u_{IA}v_{IA})+{c\over 6}\log g(u_A,v_A) g(u_I,v_I)+{2\pi r^2(u_I,v_I) \over G_N}\right]
\end{equation}
with $u_{IA}=u_I-u_A$ and similarly for $v_{IA}$.

Since the effective entropy is expected to be close to twice the black hole entropy, the location of $u_I,v_I$ should be close to the horizon, which requires $\sinh u_I \sinh v_I\ll 1$. In addition, we expect $S_{\rm Island}$ to saturate to a finite value at late time, which requires $u_I-u_A$ to approach a constant value. This in turn requires $u_I\propto t$ and $v_I\propto e^{-t}$. We will take these assumptions and verify that they are self-consistent.  In this region we obtain
\begin{align}
\frac{r(u_I,v_I)}{r_H}&\simeq 1-e^{-1}U_IV_I=1-e^{-1} v_I\sinh u_I\nonumber\\
    g(u_I,v_I)&\simeq \frac 4e r_H^2 \cosh u_I
\end{align}
so that the entropy becomes:
\begin{equation}
    S(u_I,v_I)\sim {c\over 3}\left[\log (-u_{IA} v_{IA})+{1\over 2}\log \cosh u_I +{t\over 2}\right]+2S_{BH}-{4\pi r_H^2\over {eG}} v_I \sinh u_I  +\lbrace\text{constant order c}\rbrace
\end{equation}
Variation with respect to $u_I, v_I$ gives us a pair of equations:
\begin{equation}
    {c\over 3}{1\over u_I-u_A}+{c\over 6}\tanh u_I-{4\pi r_H^2\over {eG}} v_I \cosh u_I =0;~~
    {c\over 3}{1\over v_I-v_A}-{4\pi r_H^2\over {eG}} \sinh u_I=0
\end{equation}
In the limit of $t\gtrsim {1\over G}$ and $G\ll 1$, the solution is:
\begin{align}
    v_I&\simeq e^{-t}\simeq -v_A\nonumber\\
    u_I&\sim t+\log {ceG\over 12 \pi r_H^2}\simeq u_A-\left(\log\frac{12\pi r_H^2}{cG}-1+\log(2R)\right)
\end{align}
This corresponds to $r(u_I,v_I)\simeq r_H\left(1-\frac{cG}{24\pi r_H^2}\right)$. Thus, the quantum extremal surface is close to the horizon, at the interior side. (As an interesting contrast, the island for an eternal geometry is outside the horizon \cite{Almheiri:2019yqk,Hashimoto:2020cas}.) The effective entropy is approximately $S_{\rm Island}\simeq 2S_{BH}$, which justifies the estimation of Page time $t_{P}\simeq 12S_{BH}/c$. The $u_I$ coordinate is of order the scrambling time earlier than the boundary location of $A$, which is consistent with the Hayden-Preskill decoding criterion, that a small diary thrown into the black hole after the Page time, should be reconstructable from the Hawking radiations after waiting for scrambling time. From the geometric perspective, the diary is now in the entanglement wedge of the Hawking radiations \cite{Hayden_2007}, \cite{ Penington:2019npb, Almheiri_2019}.

\section{Random tensor network models}
\label{sec:tensornetwork}

To gain more physical intuition, here we generalize the random tensor network (RTN) models, proposed as toy models for holographic duality in Ref. \cite{hayden2016holographic}, to generic universes. Our main goal is to understand how similar quantum extremal surface formula appears in (Renyi entropy calculation of) RTN models, and gain a more explicit understanding of quantum information recovery. Moreover, we would like to understand the structure of quantum states in general geometry beyond AdS. We first review the original RTN model and then discuss its generalization in the current context.

\subsection{Random tensor network model for holographic duality}

\begin{figure}
    \centering
    \includegraphics[width=5in]{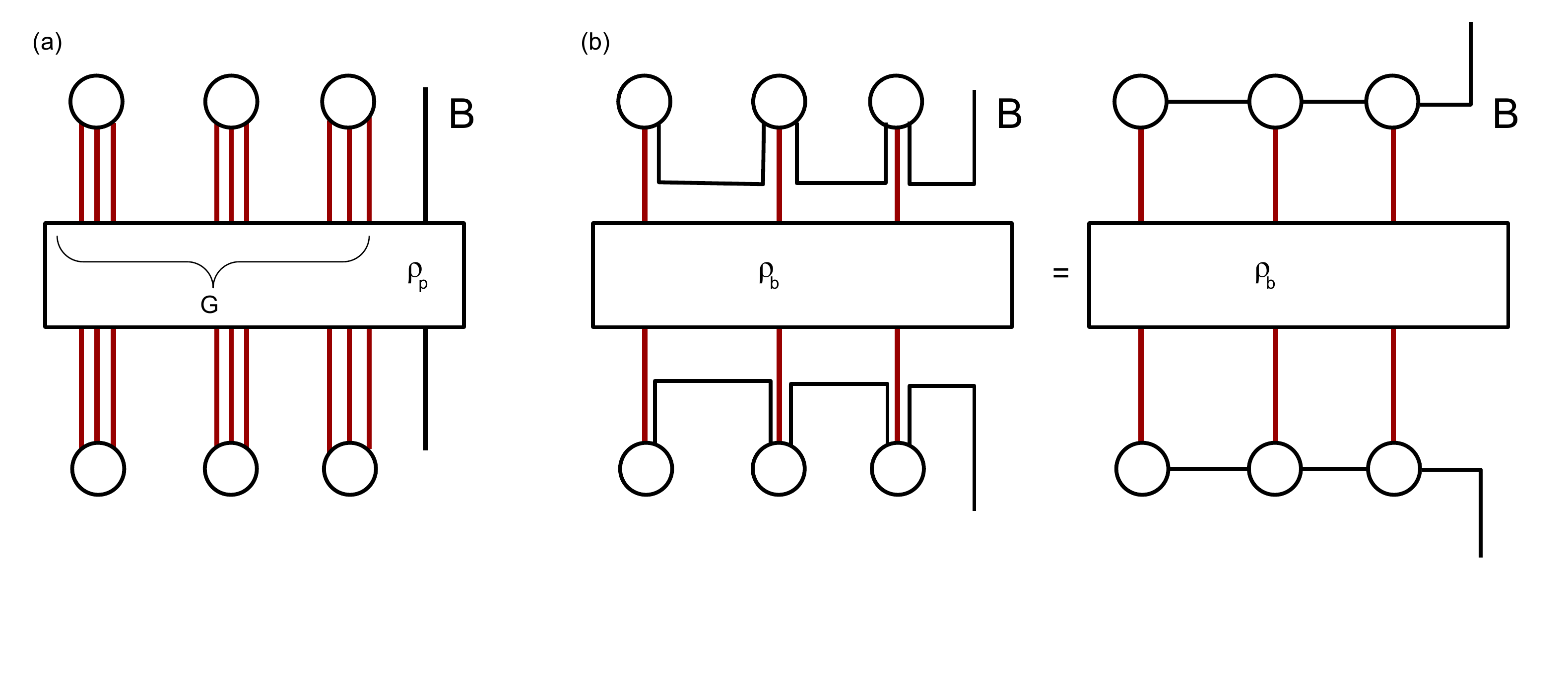}
    \caption{Illustration of (a) a most general random tensor network state defined in Eq. (\ref{eq:RTNdef}); (b) an ``ordinary" random tensor network state in which $\rho_P$ has the tensor factorization structure (\ref{eq:geometryQFTseparation}). }
    \label{fig:RTNintro}
\end{figure}

In its most general form, a random tensor network model is defined by the following elements:
\begin{enumerate}
    \item A quantum state $\rho_P$ in a Hilbert space with a tensor factorization $\mathbb{H}_P=\otimes_{x\in B\cup G}\mathbb{H}_x$. $\rho_P$ is called the parent state. The set of vertices $x$ is divided into two subsets: bulk $G$ (for ``gravitational") and boundary $B$. 
    \item A random pure state on each vertex $\ket{V_x}\in\mathbb{H}_x$ ($x\in G$).  $\ket{V_x}=U_x\ket{0_x}$ with $U_x$ a Haar random unitary in $\mathbb{H}_x$ and $\ket{0_x}\in\mathbb{H}_x$ an arbitrary reference state.
    \item RTN defines an ensemble of physical states in the Hilbert space $\mathbb{H}_B$ by taking a projection on $\mathbb{H}_G$: 
    \begin{align}
        \rho_B={\rm tr}_G\left(\rho_P\otimes_{x\in G}\ket{V_x}\bra{V_x}\right)\label{eq:RTNdef}
    \end{align} (not yet normalized). Physically, one can think this as the state obtained by measuring all qubits at $x\in G$ in a random basis and post-selecting on a particular output state $\otimes_{x\in G}\ket{V_x}$. 
\end{enumerate}
This definition is illustrated in Fig. \ref{fig:RTNintro} (a). Usually, we take $\rho_P$ as a simple state, such as EPR pairs or the ground state of a free field theory. The role of random projection is to generate a state $\rho_B$ which has much richer entanglement structure. 

The RTN models are useful because even if $\rho_B$ for a particular realization of $\ket{V_x}$ is quite complicated, the computation can be greatly simplified by taking the ensemble average. For any operator $\hat{O}$ defined in $k$ copies of the boundary Hilbert space $\mathbb{H}_B^{\otimes k}$, one can consider the expectation value in the product state $\rho_B^{\otimes k}$:
\begin{align}
    \langle \hat{O}\rangle&=\frac{{\rm tr}\left(\rho_B^{\otimes k}\hat{O}\right)}{{\rm tr}(\rho_B)^k}=\frac{{\rm tr}_{B\cup G}\left(\rho_P^{\otimes k}\left[\hat{O}\otimes\left(\otimes_x\ket{V_x}\bra{V_x}^{\otimes k}\right)\right]\right)}{{\rm tr}(\rho_B)^k}
\end{align}
In the cases we are interested in, the correlation between denominator and numerator is not important, so that we can approximate the ensemble average $\overline{\langle \hat{O}\rangle}$ by the separate average:
\begin{align}
    \langle \hat{O}\rangle&\simeq \frac{{\rm tr}_{B\cup G}\left(\rho_P^{\otimes k}\left[\hat{O}\otimes\left(\otimes_x\overline{\ket{V_x}\bra{V_x}^{\otimes k}}\right)\right]\right)}{\overline{{\rm tr}(\rho_B)^k}}
\end{align}
(A more rigorous discussion about the normalization can be found in Ref.\cite{hayden2016holographic}.) The average over Haar random ensemble is known:
\begin{align}
    \overline{\ket{V_x}\bra{V_x}^{\otimes k}}=\frac1{C_{kx}}\sum_{g\in S_k}g_x
\end{align}
Here, $g_x$ is an element that permutes different copies of Hilbert spaces. $C_{kx}$ is a normalization constant that can be determined by requiring $\tr\ket{V_x}\bra{V_x}^{\otimes k} =1$.  Since $\frac1{k!}\sum_{g\in S_k}g_x$ simply symmetrizes any state it acts on, $\frac{C_{kx}}{k!}$ is then equal to the dimension of permutation symmetric states in the $k$ copied Hilbert space. Therefore
\begin{align}
    \overline{\langle \hat{O}\rangle}&\simeq C^{-1}\sum_{\left\{g_x\in S_k\right\}}{\rm tr}_{B\cup G}\left(\rho_P^{\otimes k}\left[\hat{O}\otimes (\otimes_xg_x)\right]\right)\label{eq:randomaverage}
\end{align}
with $C=\prod_xC_{kx}$. It is helpful to define ${\rm tr}_{B\cup G}\left(\rho_P^{\otimes k}\left[\hat{O}\otimes (\otimes_xg_x)\right]\right)\equiv e^{-\mathcal{A}\left[g_x\right]}$, which maps $\overline{\langle \hat{O}\rangle}$ to the partition function of a discrete spin model with the action $\mathcal{A}\left[g_x\right]$.

Eq. (\ref{eq:randomaverage}) relates the expectation value of $\hat{O}$ in $k$ copies of boundary state $\rho_B^{\otimes k}$ to a sum over similar quantities in the (simpler) state $\rho_P^{\otimes k}$, for operators of the form $\hat{O}\otimes \otimes_xg_x$. In particular, if we choose $\hat{O}$ itself to be a permutation acting on some subsystem of the boundary, then each term on the right-side of Eq. (\ref{eq:randomaverage}) is a local unitary invariant. A simple example is the second Renyi entropy, which is computed by taking $k=2$ and $\hat{O}=X_A$, which swaps the two copies of qubits in a subsystem $A\subseteq B$. For $k=2$ there are only two permutation elements, so that Eq. (\ref{eq:randomaverage}) reduces to
\begin{align}
\overline{e^{-S^{(2)}_A}}&=\overline{\hat{X}_A}\simeq C^{-1}\sum_{\Sigma\subseteq G}{\rm tr_{B\cup G}}\left(\rho_P^{\otimes 2}X_{A}X_{\Sigma}\right)=C^{-1}\sum_{\Sigma\subseteq G}e^{-S^{(2)}_{\Sigma A}(\rho_P)}
\end{align}
In other words, the purity $e^{-S^{(2)}_A}$ for a subsystem $A$ is related to a weighted sum of that of the parent states for different regions ($\Sigma A$). 

\begin{figure}
    \centering
    \includegraphics[width=4in]{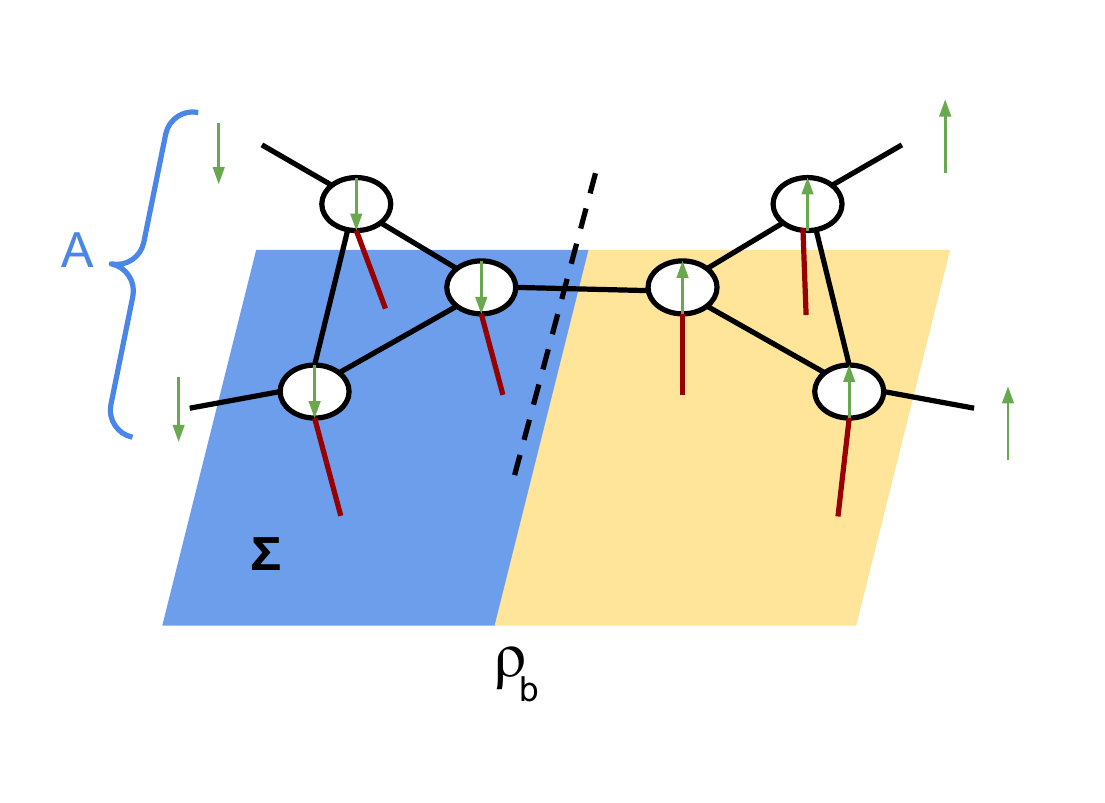}
    \caption{Illustration of the Ising spin configuration for the purity calculation. The boundary condition of the Ising model is defined as spin down ($s_x=-1$) for sites in $A$ and up ($+1$) elsewhere. The link states $\ket{\overline{xy}}$ contribute a ferromagnetic coupling, and the bulk QFT state $\rho_b$ contributes an extra term to the action, given by the second Renyi entropy of the spin down region $\Sigma$. When a single Ising configuration dominates, the boundary of the spin down domain is the quantum extremal surface (\ref{eq:RTNQES}). }
    \label{fig:Ising}
\end{figure}

We usually consider a simple case when the parent state $\rho_P$ is a direct product of EPR pairs and a quantum field theory state $\rho_b$, when the former has much higher Hilbert space dimension:
\begin{align}
    \rho_P=\otimes_{\overline{xy}}|xy\rangle\langle xy|\otimes\rho_b\label{eq:geometryQFTseparation}
\end{align}
Here $\overline{xy}$ denotes a link in the network that connects vertices $x$ and $y$, and $\ket{xy}$ is a entangled state defined on this link. This is illustrated in Fig. \ref{fig:RTNintro} (b). If there are $k$ links connecting to the same vertex $x$, there is a separate qudit for each of them, and the Hilbert space at $x$ is their direct product: $\mathbb{H}_x=\otimes_{y}\mathbb{H}_{\overline{xy}}$. 
Due to the direct product structure in Eq. (\ref{eq:geometryQFTseparation}), the Renyi entropy of $\rho_P$ is a sum of that of the link states and the remaining bulk QFT state $\rho_b$. If for simplicity we take all $\ket{xy}$ to be maximally entangled EPR pairs with entanglement entropy $\log D$, the purity becomes the following Ising model partition function:
\begin{align}
    e^{-S^{(2)}_A}&=C^{-1}\sum_{s_x=\pm 1}\exp\left[-\sum_{\overline{xy}}\frac{1-s_xs_y}{2}\log D-S^{(2)}_{s_x=-1}(\rho_b)\right]
    \nonumber\\
    &=C^{-1}\sum_{\Sigma\subseteq G}\exp\left[-\log D\left|\partial(\Sigma A)\right|-S_{\Sigma}^{(2)}(\rho_b)\right]
\end{align}
Here $s_x=\pm 1$ denotes the identity and swap operators, the two elements of permutation group $S_2=\mathbb{Z}_2$. $\Sigma$ is the spin down region, {\it i.e.} the region that is applied by a swap operator permuting the two copies. Only the EPR pairs crossing the boundary of region $\Sigma A$ has a nontrivial contribution to the entropy, which leads to the area law term $\log D\left|\partial(\Sigma A)\right|$. The second Renyi entropy $\log D$ of the link state becomes the coupling constant of the ferromagnetic Ising model. If different links have different second Renyi entropies, the coefficient $\log D$ should be replaced by $S^{(2)}_{xy}$ which is the second Renyi entropy of state $\ket{xy}$. In the Ising model language, the region $A$ translates to a boundary condition, with a fixed Ising spin on each boundary vertex, which is $-1$ in $A$ and $+1$ everywhere else. This Ising model picture is illustrated in Fig. \ref{fig:Ising}. 

In the limit of large $D$, the sum is dominated by a single term with minimal number of links connecting $\Sigma A$ with the complement, leading to the analog of the quantum extremal surface formula (in this case a minimal surface since there is only variation in spatial direction):
\begin{align}
    S_A^{(2)}\simeq \min_{\Sigma\subseteq G}\left(\log D\left|\partial(\Sigma A)\right|+S_{\Sigma A}^{(2)}(\rho_b)\right)\label{eq:RTNQES}
\end{align}

\begin{figure}[htbp]
    \centering
    \includegraphics[width=5.5in]{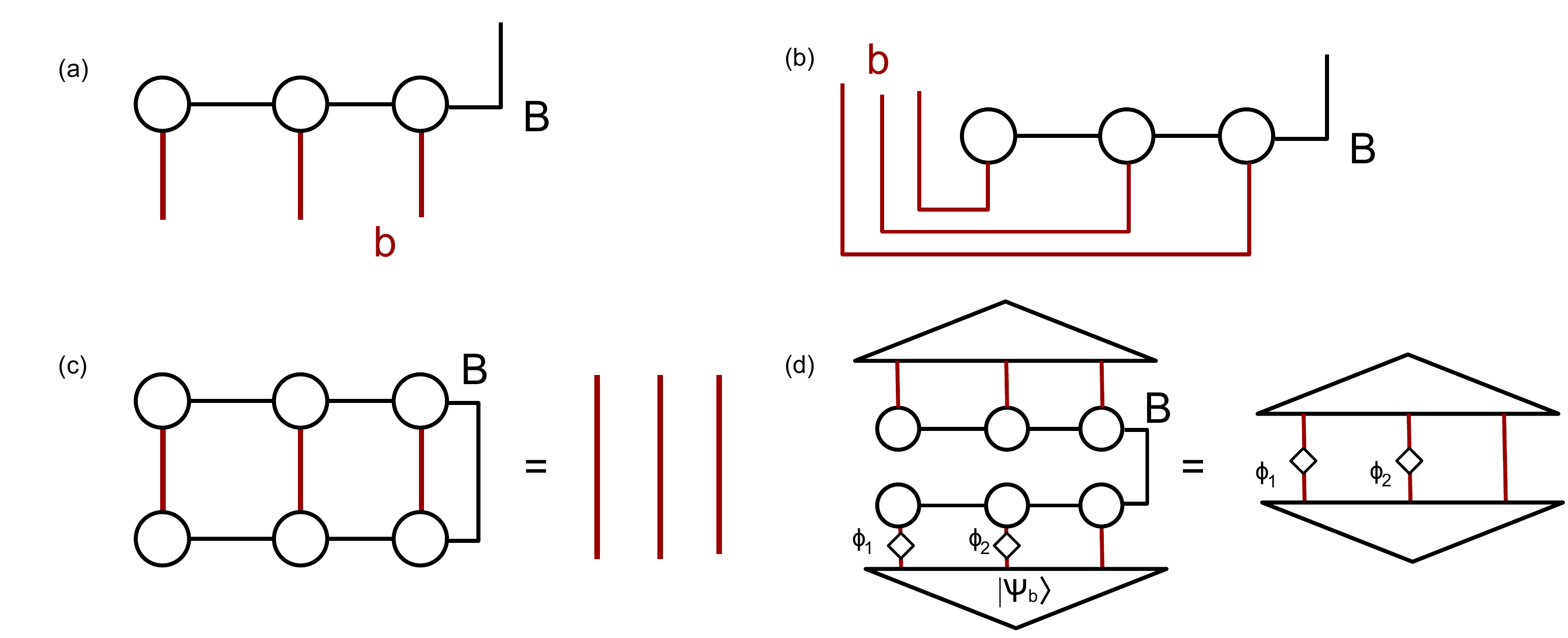}
    \caption{(a) A tensor network with bulk state $\rho_b=\ket{\psi_b}\bra{\psi_b}$. The red lines are bulk quantum field theory degrees of freedom, and the black lines are EPR pair (geometry) degrees of freedom. (b) By leaving the $b$ legs open, we can view the tensor network as a map from $b$ to $B$, or equivalently, an entangled state of $b$ and $B$. (c) The isometry condition is equivalent to the statement that in the state of $b$ and $B$, tracing over $B$ leads to a maximally mixed state of $b$. (d) The isometry condition implies that the bulk correlation functions are equal to those in QFT.}
    \label{fig:isometry}
\end{figure}

The minimal region $\Sigma$ becomes the (spatial slice of) entanglement wedge, and $\partial(\Sigma A)=\gamma$ is the quantum extremal surface. The analog of geometry is the graph geometry, defined by the links $\overline{xy}$. More generally, if we allow different states $\ket{xy}$ with different second Renyi entropy, the geometry will be given by a weighted graph with each link weighted by $S_{xy}^{(2)}$. 
One interesting feature of RTN model is that the geometry is not restricted to negative curvature. The setup is well-defined in arbitrary graph geometry. 
For the purpose of our current discussion, we want to study the entropy of a bulk region in the bulk quantum field theory state $\rho_b$. In the limit of large bond dimension $D$ for the EPR pairs, the tensor network defines an isometry from the bulk QFT degrees of freedom to the boundary. This can be proved by leaving the bulk indices open, and define a state of $bB$ after the random projection (Fig. \ref{fig:isometry}). The isometry condition is equivalent to the condition that the mutual information $I(b:B)=2\log D_b$, where $D_b$ is the Hilbert space dimension of $\mathbb{H}_b$. This isometry guarantees that the bulk QFT degrees of freedom are encoded faithfully in the boundary Hilbert space $\mathbb{H}_B$. 

With this isometry condition, correlation functions of bulk operators are preserved. For example, Fig. \ref{fig:isometry} (d) illustrates a two-point function in the bulk, which is defined by inserting operators $\phi_1,\phi_2$ into the bulk:
\begin{align}
    \langle\phi_1\phi_2\rangle_{RTN}\equiv \frac{{\rm tr}\left(\phi_1\phi_2\rho_P\ket{V}\bra{V}\right)}{{\rm tr}\left(\rho_P\ket{V}\bra{V}\right)}
\end{align}
with $\rho_P=\rho_b\otimes \prod_{\langle xy\rangle}\ket{xy}\bra{xy}$. This is equivalent to mapping the operators to the boundary Hilbert space by the isometry, and compute the correlation function there. The isometry condition guarantees that the correlator is the same as that in the QFT:
\begin{align}
    \langle\phi_1\phi_2\rangle_{RTN}={\rm tr}\left(\rho_b\phi_1\phi_2\right)=\langle \phi_1\phi_2\rangle_{QFT}\label{eq:correlatorRTN}
\end{align}

For a bulk region $A$, the quantity ${\rm tr}\left(\rho_A^n\right)$ can be expressed as a sum over correlation functions in this region. For example, ${\rm tr}\left(\rho_A^2\right)=\sum_a{\rm tr}\left(\rho_AO_a\right)^2$ with the sum runs over an orthonormal basis of Hermitian operators in $A$. We discuss the more general $n$-th Renyi entropy case in Appendix \ref{app:renyi}. Therefore Eq. (\ref{eq:correlatorRTN}) implies that the entropy of a bulk region is also equal to that in the QFT state, due to the isometry condition.

\subsection{General geometries and super-density operator}

For a generic spatial geometry, which may have a small boundary or even no boundary, the isometry condition may fail, but the tensor network state (\ref{eq:RTNdef}) is still well-defined. We will apply this definition to general geometry and discuss the physical consequences and interpretations. Without the isometry condition, we need to rethink about bulk correlation functions and their interpretation. Due to the projection to random tensor states $\ket{V}=\otimes_x\ket{V_x}$, the general bulk-boundary correlation function looks like
\begin{align}
    C_{ab\alpha}={\rm tr}_B\left(O_\alpha\bra{V}\phi_a\rho_P\phi_b\ket{V}\right)\label{eq:bulkboundarycorrelation}
\end{align}
with $\phi_a,\phi_b$ operators acting in the bulk, and $O_\alpha$ acting on the boundary. Compared to the discussion in previous subsection, the main difference is that due to the projection on $\ket{V}$, in general we cannot move $\phi_b$ to the left of $\phi_a$ if there is no isometry condition\footnote{More precisely, even if we have the isometry condition, when $O_\alpha$ is a nontrivial operator on the boundary, in general we still cannot move $\phi_b$ to the left. In that case we didn't discuss this problem since we can simply push all operators to the boundary and discuss correlation functions there. In general geometry without isometry condition, there is no such ``anchor Hilbert space" and we have to do the discussion directly with bulk operators.}. 

\begin{figure}
    \centering
    \includegraphics[width=5.5in]{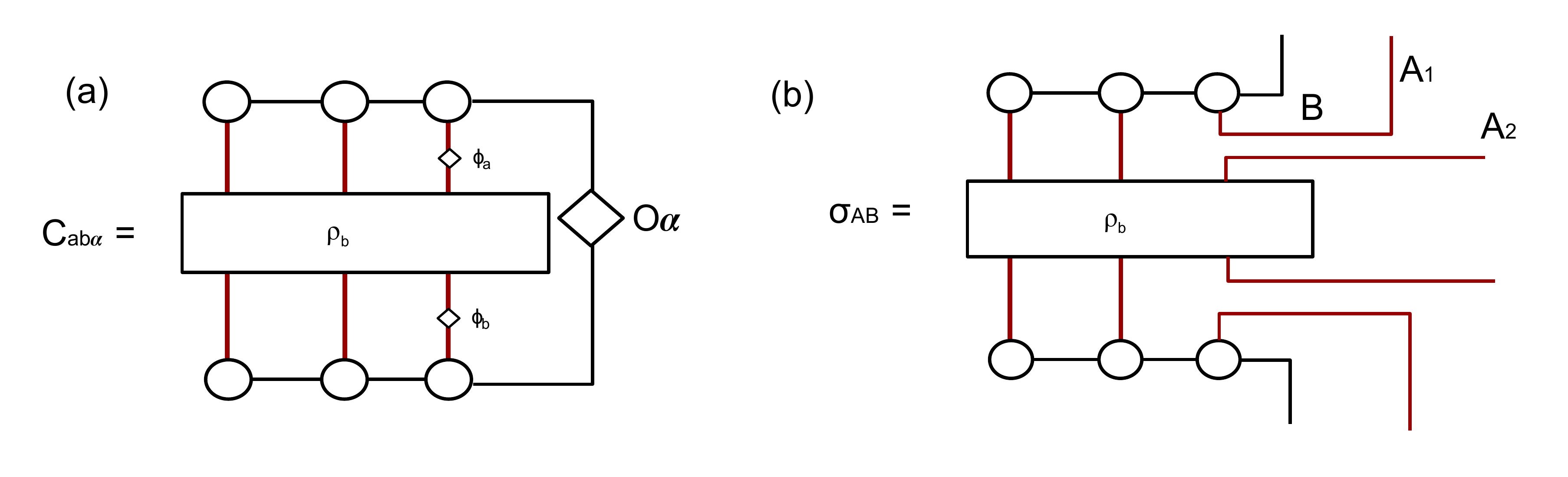}
    \caption{(a) Illustration of the bulk boundary correlator $C_{ab\alpha}$ in Eq.(\ref{eq:bulkboundarycorrelation}). (b) The super-density operator that determines all bulk-boundary correlators involving region $A$ and boundary $B$. Since we need to keep both $\phi_a,\phi_b$ on two side of $\rho_b$, the dimension of $\sigma_{AB}$ is $D_A^2D_B$ instead of $D_AD_B$ when the Hilbert space dimension of $A,B$ region are $D_A,D_B$ respectively.} 
    \label{fig:superdensity}
\end{figure}

Let us consider a bulk region $A$ and restrict operators $\phi_a,\phi_b$ to arbitrary QFT operators in $A$. In ordinary QFT, expectation values of the form $\langle \phi_a\rangle$ are all determined by the reduced density matrix $\rho_A$. Here the situation is different, because we need to keep a pair of operators $\phi_a,\phi_b$. The generalization of density operator is the linear map from operators $\phi_a,\phi_b,O_\alpha$ to $\mathbb{C}$ given by Eq. (\ref{eq:bulkboundarycorrelation}). More explicitly, we can take $\phi_a$ to be an orthonormal basis of the QFT operators of $A$, and introduce an auxiliary set of states $\ket{a},a=1,2,...,D_A^2$. Then we define the ``super-density operator" \cite{cotler2018superdensity}
\begin{align}
    \sigma_{AB}=\sum_{a,b}\bra{V}\phi_a\rho_P\phi_b\ket{V}\otimes \ket{a}\bra{b}
\end{align}
In the graphic representation of tensor networks, the super-density operator corresponds to opening up the bulk links in $A$, in addition to the boundary indices $B$, as is illustrated in Fig. \ref{fig:superdensity}. As has been discussed in Ref. \cite{cotler2018superdensity}, the superdensity operator is positive definite and satisfies all properties of the ordinary density operator. Physically, $\sigma_{AB}$ can be prepared by introducing an ancilla system that couples to the degrees of freedom in $A$ before the projections on state $\ket{V}$ is imposed. As is illustrated in Fig. \ref{fig:superdensity} (b), the ancilla system has a Hilbert space dimension of $D_A^2$, and was initialized in a maximally entangled state between two qudits each with dimension $D_A$. Then one of the qudit subsystem is coupled with $A$ by a swap gate. As a consequence, the state of $A$ is swapped to the ancilla and therefore survives from the projection by $\bra{V}$. The subsystem $A_1$ of the ancilla contains information about the QFT state of $A$, while the other subsystem $A_2$ is maximally entangled with the qudit that enters the random tensor network as bulk inputs. The entanglement structure of $\sigma_{AB}$ determines the quantum information flow in this model. 

To gain some physical understanding of the superdensity operator, let us first discuss what happens in the situation discussed in the previous subsection, when there is an isometry from bulk to boundary. In that case, it is easy to see that the mutual information between $A_2$ and $B$ is maximum:
\begin{align}
    S(A_2)&=\log D_A,~S(B)=S_{\rm QFT}(\overline{A})+\log D_A\\
    S(A_2B)&=S_{\rm QFT}(\overline{A})\\
    I_2(A_2:B)&=2\log D_A
\end{align}
which follows from the isometry condition in Fig. \ref{fig:isometry}. If we are interested in whether $A$ can be reconstructed locally in a boundary region $B_1\subset B$, we can also compute $I_2(A_2:B_1)$, which is required to be maximal in order for $A$ to be in the entanglement wedge of $B_1$. 

We can also compute $I_2(A_1:B)$ using the Ising model method. With the isometry condition, the Ising spin directions are all determined by the boundary condition at $B$. We obtain
\begin{align}
    S(A_1)&=S_{\rm QFT}(A),~S(B)=S_{\rm QFT}(\overline{A})+\log D_A\\
    S(A_1B)&=S_{\rm QFT}(\rho_b)+\log D_A\\
    I_2(A_1:B)&=I_{2\rm QFT}(A:\overline{A})
\end{align}
This equation shows that $A_1$ is only entangled with $B$ through the entanglement that is already in the QFT state $\rho_b$. Although in the ordinary RTN with a large boundary and isometry condition, we do not usually need to discuss the superdensity operator formalism, it is still helpful since it allows us to discuss the encoding map (from $A_2$ to $B$) and a particular bulk QFT state (saved in $A_1$) in a well-defined quantum state, rather than switching between two different tensor networks representing ``the holographic code" and ``the holographic state".

In the next two subsections, we will use the superdensity operator formalism to study generic RTN without isometry condition. We will show how an entanglement island could appear, which is the analog of the replica wormhole geometry in gravity calculation. We will analyze its physical interpretation in term of quantum information recovery and quantum error correction.

\subsection{Entanglement island}

\begin{figure}
    \centering
    \includegraphics[width=3in]{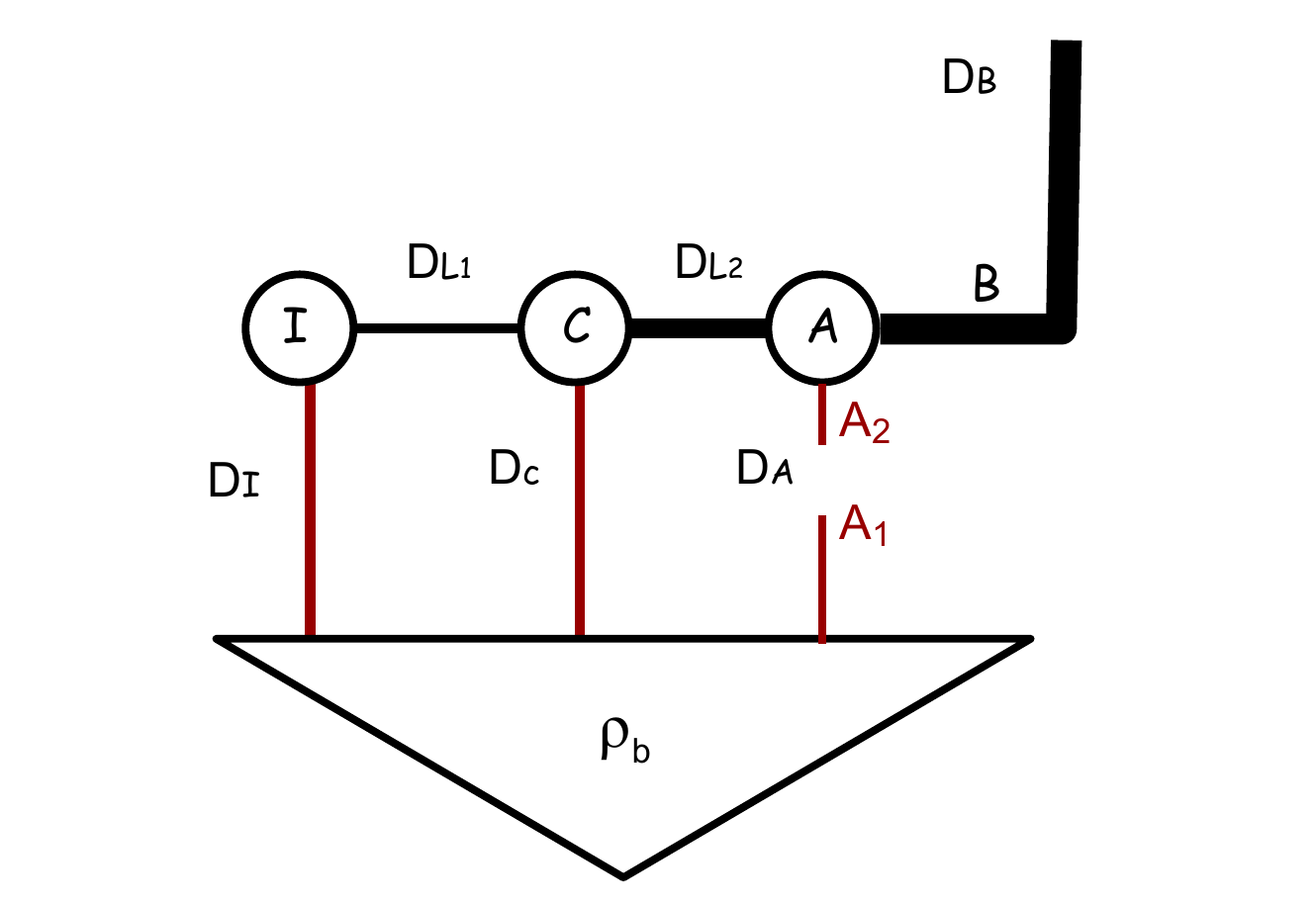}
    \caption{Illustration of the three-tensor model. The labels on each link $D_A$, $D_I$,  etc labels the dimension of each link, which are of the value $2^n$ with $n=|I|$, $|A|$, etc, i.e. the number of qubit EPR pairs. In general the bulk state can be a mixed state, but we draw the pure state case for simplicity. In the super density operator, $B$ and $A_1,A_2$ are external legs of the network.}
    \label{fig:threetensor}
\end{figure}

In this subsection we discuss different situations that may occur in the Renyi entropy calculation of the superdensity operator. For concreteness, we focus on the three-tensor model, which has been illustrated earlier in Fig. \ref{fig:superdensity}. All discussions can be generalized to more generic geometry straightforwardly. As is illustrated in Fig. \ref{fig:threetensor}, we denote the three sites by $I,C,A$, and the corresponding bulk dimensions are $D_I,D_C,D_A$. The links connecting different tensors have dimension $D_{L_1},D_{L_2}$ and $D_B$. For later convenience, we assume each link consists some integer number of qubit EPR pairs, such that $D_I=2^{|I|}$ and similarly for other links.
We denote the number of EPR pairs by $|I|$, $|A|$, etc.

The Renyi entropy of this state can be computed by spin model partition function, with three spins at the three tensors $I,C,A$. We are interested in considering the following situation:
\begin{align}
    |I|&>|L_1|\label{eq:sizeinequality1}\\
    |I|+|C|&<|L_2|\label{eq:sizeinequality2}\\
    |A|&<|L_2|\label{eq:sizeinequality3}
\end{align}
Physically, these inequalities indicate that the volume law entropy of each region is smaller than the area law entropy of its boundary, except the $I$ region. 

We first look at the second Renyi entropy of $A_1$. We can run over the different spin configurations and check their Ising action. In the following list, the three spin configurations are for the sites $ICA$ correspondingly. $+$ and $-$ represent identity and swap, respectively.
\begin{align}
    s_Is_Cs_A=\left\{\begin{array}{cc}+++,&S_{QFT}(A)\\
    -++,&S_{QFT}(IA)+|L_1|\log 2\\
    +-+,&S_{QFT}(CA)+(|L_2|+|L_1|)\log 2\\
    --+,&S_{QFT}(ICA)+|L_2|\log 2\\
    ++-,&S_{QFT}(A)+(|L_2|+|B|+|A|)\log 2\\
    -+-,&S_{QFT}(AI)+(|L_1|+|L_2|+|B|+|A|)\log 2\\
    +--,&S_{QFT}(AC)+(|L_1|+|B|+|A|)\log 2\\
    ---,&S_{QFT}(ICA)+(|B|+|A|)\log 2\end{array}\right.
\end{align}
In the region we are considering, the only two possible lowest action configurations are $+++$ and $-++$. Using the inequality (\ref{eq:sizeinequality1}-\ref{eq:sizeinequality3}) we can show that all other configurations are never preferred. Thus in the large bond dimension limit
\begin{align}
    S_{A_1}=\min\left\{S_{QFT}(A),S_{QFT}(IA)+|L_1|\log 2\right\}
\end{align}
If $S_{QFT}(IA)+|L_1|\log 2<S_{QFT}(A)$, the $-++$ configuration is dominant. The same analysis applies to higher Renyi entropy. The transition (switch of the dominant term in entropy contribution) for different Renyi entropies generically occur at different value of parameters in the system. Assuming that all Renyi entropies and also the von Neumann entropy are dominated by the $S_{\rm QFT}(IA)+|L_1|\log 2$ term, we recover the QES formula with $I$ the entanglement island of $A$.

The analysis here also applies to more general tensor network geometries. In general, we should minimize the entropy configuration over all regions $\Sigma$ that do not intersect $A$:
\begin{align}
    S_{A_1}=\min_{ I\cap A=\emptyset}\left(S_{QFT}(I A)+|\partial I|\log 2\right)
\end{align}
This is the analog of Eq. (\ref{eq:QES2}) in the gravitational calculation. It should be noted that the Ising spin is always $+$ in $A$ region due to the pinning field coming from tracing over $A_2$. This is similar to fixing the spatial geometry of $A$ in the gravity theory case. Note that the appearance of the island requires a necessary condition
\begin{align}
    S_{QFT}(I)\geq \left|S_{QFT}(A)-S_{QFT}(IA)\right|>|\partial\Sigma|\log 2
\end{align}
which means the QFT entropy of $I$ needs to exceed the area law ``entropy bound” $|\partial\Sigma|\log 2$.

There is another extreme limit: the ``entanglement island" could include the entire complement of $A$, which corresponds to the $--+$ configuration in the three-tensor model. In order for this configuration to be dominant, it is required that
\begin{align}
    S_{QFT}(A)>S_{QFT}(ICA)+|L_2|\log 2\geq |L_2|\log 2
\end{align}
In our model this will not occur due to inequality (\ref{eq:sizeinequality3}). In general, this will only occur if the QFT entropy of $A$ exceeds the area law bound $|\partial (AB)|\log 2$. (Note that what appears is the boundary of $AB$ rather than $A$, which means the boundary between $A$ and the boundary is excluded.) 

As we discussed at the beginning of this section, an RTN is defined by a parent state, which we take to be a product state of EPR pairs on links and remaining QFT state (Eq. (\ref{eq:geometryQFTseparation})). The separation of ``geometry"---link EPR pairs---and matter is the analog of choosing the UV cutoff of CFT. In the three-tensor model, we can move an EPR pair at the AC link from ``geometry" to matter, which means the entanglement in QFT between $A$ and $C$ will increase by $\log 2$, while $L_2$ will reduce by $1$. When we introduce ancilla, this separation between geometry and matter gets a nontrivial physical consequence. If this EPR pair is considered as part of QFT, the ancilla will couple to it, such that the Hilbert space dimension of $A_1,A_2$ will increase, while the area law bound for $A$ given by $|L_2|\log 2$ will decrease. This is similar to the effect of changing the UV cutoff of QFT in the gravity calculation. 

It should be noted that the discussion above does not require boundary $B$ to be large. When $|B|$ is sufficiently large, $A$ can be reconstructed in $B$, which corresponds to the fact $I\left(A_2:B\right)=2|A|\log 2$ in the superdensity operator. If $|B|<|A|$ this clearly will fail. The superdensity operator formalism makes the bulk region entropy quantities well-defined even when the isometry from boundary to bulk fails. In Sec. \ref{sec:closeduniverse} we will provide further discussion about the case of closed (or almost closed) universe. 


\subsection{Interpretation: recovery of quantum information}\label{sec:recovery}

\begin{figure}
    \centering
    \includegraphics[width=6in]{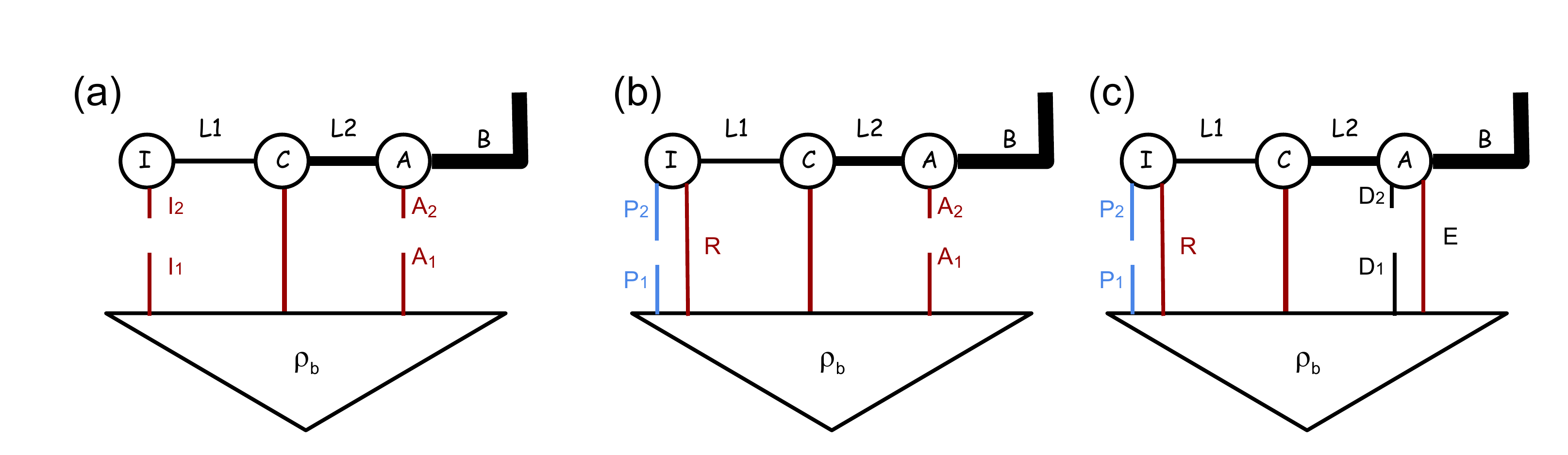}
    \caption{The super-density operator with (a) two ancilla coupled to $I$ and $A$ respectively; (b) two ancilla coupled to $A$ and a part of $I$, denoted as $P$; (c) two ancilla coupled to a part of $A$, denoted as $D$, and a part of $I$, denoted as $P$. }
    \label{fig:recovery}
\end{figure}

The next question is the physical interpretation of the entanglement island. From the discussion of AdS space black hole coupled with bath \cite{Penington:2019npb,Almheiri_2019}, it is natural to ask whether degrees of freedom in island $I$ can be reconstructed in region $A$ (which corresponds to the early radiation in the black hole case). In the superdensity operator formalism, one attempt to show this reconstruction is to introduce ancilla in $I$ in the same way as $A$, as is illustrated in Fig. \ref{fig:recovery} (a). However, one can verify that in the region we are considering, the mutual information
\begin{align}
    S(I_1I_2)&=|L_1|\log 2+S_{\rm QFT}(I),~~~(-++)\nonumber\\
    S(A_1A_2)&=S_{\rm QFT}(A)+|A|\log 2,~~~(+++)\nonumber\\
    S(I_1I_2A_1A_2)&=|L_1|\log 2+S_{\rm QFT}(IA)+|A|\log 2,~~~(-++)\nonumber\\
    I(I_1I_2:A_1A_2)&=I_{\rm QFT}(AI)\label{eq:ancillaAI}
\end{align}
Thus the mutual information is determined by the QFT state, and is not necessarily maximal. The mutual information is actually contributed by $I_1,A_1$ and there is no contribution from $I_2$ and $A_2$. This seems contradictory with the fact that $I$ is the entanglement island of $A$. Physically, this apparent contradiction is caused by an external field imposed by the ancilla of $I$: Now with the $I$ link broken into $I_1$ and $I_2$, there is a boundary condition at $I_1$ which is $+$ in the computation of $S(A)$. In the Ising model dynamics, the entanglement island is the region of which the spin is controlled by the boundary condition in $A$. Thus by introducing the ancilla in $I$, we have exclude this region from the possible location of the entanglement island. In other words, {\it entanglement island will only appear in region that is not accessible to an (arbitrarily powerful) observer.} 

One may worry that this means there is no physical way to observe the entanglement island. In fact, the information recovery from the entanglement island can be shown in a different ancilla setup, as is illustrated in Fig. \ref{fig:recovery} (b). Instead of introducing a ``complete probe" in $I$ region, we introduce a small ancilla that only couples to a small region $P$ in the island. In the tensor network it corresponds to opening up the legs in $P$ and leaving the remaining part of $I$ (denoted as $R$) uncoupled. This is the analog of the Hayden-Preskill setup \cite{hayden2007black} in the case of evaporating black hole. In this situation, the entropy calculation is different. If $|R|\gg |P|$, the spin configuration that determines $S(A_1)$ will still be $-++$ when there is an island. In this case we can study the mutual information between $P_2$ and $A_1$:
\begin{align}
    S(P_2)&=|P|\log 2,~~~(+++)\nonumber\\
    S(P_2A_1)&=|L_1|\log 2+S_{\rm QFT}(AR),~~~(-++)\nonumber\\
    S(A_1)&=|P|\log 2+S_{\rm QFT}(AR)+|L_1|\log 2,~~~(-++)\nonumber\\
    I(A_1:P_2)&=2|P|\log 2\label{eq:ancillaAP}
\end{align}
It should be noticed that $S(A_1)$ also depends on the choice of $P$ region, because introducing ancilla and tracing over them is different from having no ancilla. \footnote{In the super-desity operator formalism, removing ancilla corresponds to a post-selection on an EPR pair state of the ancilla. } The mutual information is maximal and equals to $2S(P_2)$. In the spin model language, this is a consequence that for small $|P|$, the spin at $I$ site is always controlled by the boundary condition at $A_1$. Due to this maximal mutual information, any operator applied to $P$ site can be mapped to $A$ by an isometry (which is defined by the channel that is dual to the state $\rho_{A_1P_2}$):
\begin{align}
    \mathcal{M}&: \mathbb{H}_P\otimes\overline{\mathbb{H}}_P\longrightarrow \mathbb{H}_A\otimes\overline{\mathbb{H}}_A
\end{align}
The reconstruction map is defined using Petz map. Technically, the reconstruction is similar to the holographic tensor network case studied in Ref. \cite{Jia:2020etj}, but in the current situation we need to consider a bulk-to-bulk reconstruction using superdensity operators. The detail of the map $\mathcal{M}$ is discussed in appendix \ref{app:isometry}. In term of the general bulk-boundary correlation function (\ref{eq:bulkboundarycorrelation}), the isometry condition means for any operators $\phi_P,\eta_P\in \mathbb{H}_P\otimes\overline{\mathbb{H}}_P$, $\phi_A,\eta_A\in\mathbb{H}_A\otimes\overline{\mathbb{H}}_A$ and $O_\alpha \in \mathbb{H}_B\otimes\overline{\mathbb{H}}_B$, the general correlation function
\begin{align}
    C_{PAB}={\rm tr}_B\left(O_\alpha \bra{V}\phi_A\phi_P\rho_P\eta_P\eta_A\ket{V}\right)
\end{align}
the operators $\phi_P,\eta_P$ can be mapped to $\mathcal{M}(\phi_P),~\mathcal{M}(\eta_P)\in \mathbb{H}_A\otimes\overline{\mathbb{H}}_A$ acting on $A$ region, such that
\begin{align}
    C_{PAB}={\rm tr}_B\left(O_\alpha \bra{V}\phi_A\mathcal{M}(\phi_P)\rho_P\mathcal{M}(\eta_P)\eta_A\ket{V}\right)\label{eq:correlation_reconstruction}
\end{align}
In other words, all correlation functions involving operators in $A, P$ and the boundary $B$ can be mapped to those only involving operators in $A$. 

The operators $\mathcal{M}(\phi_P),\mathcal{M}(\eta_P)$ from reconstruction of $P$ operators should always appear closer to $\rho_P$ than other operators acting on $A$. This ordering has an important consequence. Generically, a unitary operator in $P$, denoted $u_P$, is mapped to a unitary operator $\mathcal{M}(u_P)$ that acts on $A$. If we are allowed to apply an arbitrary measurement on $A$, we could see the change of measurement output induced by $\mathcal{M}(u_P)$. In other words, if we have control to the ancilla coupled to small probe region $P\subset I$, we can induce a physical response in $A$, although it is probably a complicated response that cannot be probed in simple operators (which we will discuss later in this section by considering operators acting on part of $A$). In contrast, we can consider the reverse situation by applying a unitary $u_A$ in $A$ and ask whether it could induce a nontrivial change of measurement result in ancilla system $P_1,P_2$. (In general, such an approach can be used to analyze causal structure of tensor networks and more general systems, which was proposed in Ref. \cite{cotler2019quantum}.) A subtlety is that not all unitaries in $A$ remains unitary after the random projections. In the superdensity operator defined in Fig. \ref{fig:recovery} (b), if $I(A_2:B)=2\log 2|A|$ is maximal, then the reduced density operator of $A_2$ is maximally mixed, and thus inserting an arbitrary unitary operator $u_A$ does not change the value of the network:
\begin{align}
    {\rm tr}_B\left(\bra{V}u_A\rho_Pu_A^\dagger\ket{V}\right)={\rm tr}_B\left(\bra{V}\rho_P\ket{V}\right)
\end{align}
More generally, this unitary may only apply to certain unitaries, such as those applied to a subsystem of $A_2$, if this subsystem has a maximally mixed density operator. Now if in addition to this unitary, we insert a measurement in $P$, the measurement result will be independent from $u_A$. For example we can consider a projective measurement given by operator $\phi_P$:
\begin{align}
    {\rm tr}_B\left(\bra{V}u_A\phi_P\rho_P\phi_Pu_A^\dagger\ket{V}\right)={\rm tr}_B\left(\bra{V}\phi_P\rho_P\phi_P\ket{V}\right)
\end{align}
This is because $P_2$ has maximal mutual information with $A_1$ and therefore has zero mutual information with the remainder of the system. Thus this discussion tells us that, in the sense of causal influence \cite{cotler2019quantum}, the island degrees of freedom lives in the ``causal past" of $A$, such that it is possible to influence $A$ by an unitary operation in the island (if it only applies to a small part $P$), but no influence occurs from $A$ to $P$. This property is similar to the causal structure of the black hole final state projection model \cite{horowitz2004black}, analyzed in Ref. \cite{cotler2019quantum}.

It is also interesting to comment that the map $\mathcal{M}$ preserves order of operators:
\begin{align}
    \mathcal{M}(\phi_P)\mathcal{M}(\eta_P)=\mathcal{M}(\phi_P\eta_P)
\end{align}
This is different from the ``mirror operators" \cite{papadodimas2013infalling} defined for an entangled state which involves a transpose operation. As will be discussed in more details in Appendix \ref{app:isometry}, the operator ordering is because the maximal entanglement occurs between $P_2$ and $A_1$. If we have an isometry from $P_1$ to $A_1$ (which is true if $P$ is maximally entangled with $A$ in the QFT state), that will define ``mirror operator" map that involves a transpose, such that $\mathcal{M}(\phi_P)\mathcal{M}(\eta_P)=\mathcal{M}(\eta_P\phi_P)$. More discussion about this will be presented in the closed universe case in Sec. \ref{sec:closeduniverse}.

If we gradually increase the size of the probed region $P$, some phase transition will occur in the Renyi entropy calculation. At a critical value 
\begin{align}
    (|P|+|L_1|)\log 2+S_{\rm QFT}(AR)=S_{\rm QFT}(A)\label{eq:criticialvalue1}
\end{align}
the dominant configuration for $S(A_1)$ will become $+++$. At a different (bigger) critical value
\begin{align}
    |P|\log 2=|L_1|\log 2+S_{\rm QFT}(R)\label{eq:criticalvalue2}
\end{align}
the dominant configuration for $S(P_2)$ becomes $-++$. If the size of $P$ exceeds both critical values, instead of Eq. (\ref{eq:ancillaAP}) we have 
\begin{align}
    I(A_1:P_2)=I_{\rm QFT}(A:R)
\end{align}
which gradually decreases to zero as $P_2$ approaches the entire $I$. The mutual information stops being maximal at critical value (\ref{eq:criticialvalue1}).

The calculation above can be directly generalized to generic tensor network geometries. As long as the prob $P$ is small enough, which means the Ising spin configuration in all entropy calculations are unchanged by introducing the ancilla at $P$, the mutual information $I(P_2:A_1)$ is maximal when $P$ is in the entanglement island of $A$. $I(P_2:A_1)=0$ for a small prob outside the entanglement island. 

The same analysis can also be used to discuss the situation when our access to $A$ is also incomplete. For example, we may be able to only measure few-point functions in $A$. In the three-tensor model, this is modeled by introducing the ancilla only for part of the degrees of freedom in $A$, as is illustrated in Fig. \ref{fig:recovery} (c). We denote the subsystem with probe as $D$ and its complement as $E$, such that $DE=A$. In the limit that the probe $P$ in the island is small, the transition of mutual information as a function of size of $D$ occurs when
\begin{align}
    S_{\rm QFT}(D_1R)+\log 2(|P|+|L_1|)=S_{\rm QFT}(D_1)
\end{align}
In the limit $|P|\ll |I|$, we can write $ S_{\rm QFT}(D_1I)+|L_1|\log 2=S_{\rm QFT}(D_1)$, or
\begin{align}
    S_{\rm QFT}(I|D_1)+|L_1|\log 2=0
\end{align}
This condition tells us that a small probe in the island appears independent from subsystem $D\subset A$, until $D$ is large enough to recover the message. If the entanglement between $I$ and $A$ are simple EPR pairs, the condition is simply that $D$ needs to include $|L_1|$ number of qubits that are maximally entangled with $I$. 

In more general geometry, the situation will be more complicated since there is the possibility of flipping spin in only part of $A$, but the general picture remains valid: small probes at different location of the tensor network appear independent from each other, while a small region may become reconstructable from a large region $A$. The location of such small regions outside $A$ defines its entanglement island.

\subsection{Further discussion on closed universe}\label{sec:closeduniverse}


As we have discussed earlier, the definition of bulk region entropy does not require a large boundary. In this subsection we discuss further the extreme situation of a (spatially) closed universe, which corresponds to $D_B=1$. In this case, the projection $\ket{V}\bra{V}$ is rank $1$. If we do not introduce ancilla, the tensor network only defines a non-negative real number $\bra{V}\rho_P\ket{V}$ rather than a quantum state. We can insert operators and define correlation function
\begin{align}
    C_{ab}=\frac{\bra{V}\phi_a\rho_P\phi_b\ket{V}}{\bra{V}\rho_P\ket{V}}
\end{align}
In the same way as the case with boundary, we can define the super-density operator $\sigma_A$ for any region $A$, which is a density operator of two ancillas $A_1,A_2$. In this way we have defined a well-defined quantum state for the region we are observing. When we observe different regions, we obtain different states in different Hilbert spaces, but they are all compatible to each other, in the sense that for two intersecting regions $A$, $C$, one can obtain $\sigma_{A\cap C}$ from $\sigma_A$ or $\sigma_C$--not by partial trace but by reconnecting the legs outside $A\cap C$--and the answer should agree with each other. 

\begin{figure}
    \centering
    \includegraphics[width=6in]{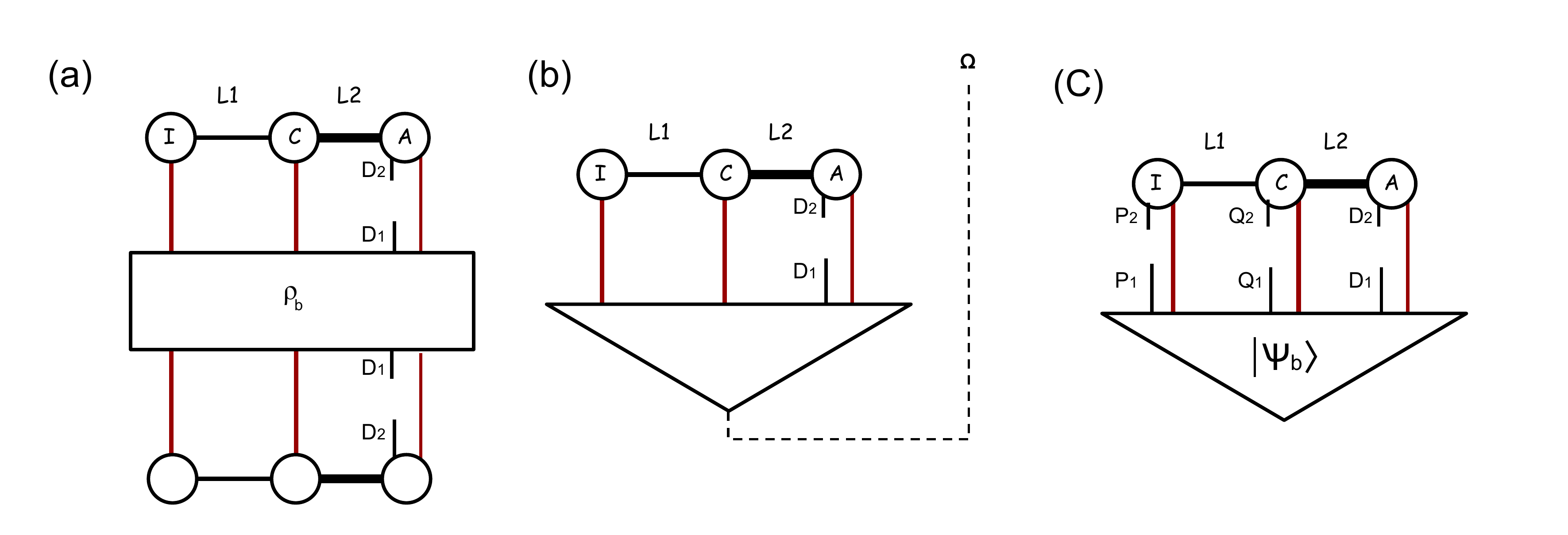}
    \caption{(a) A small probe $D$ in a closed universe without boundary. The super-density operator $\sigma_D$ consists of ancilla system $D_1,D_2$. (b) A pure state of $D_1,D_2,\Omega$ defined by purification of $\rho_b$. (c) A pure state closed universe with small probes $P,Q$ and a large probe in region $A$.}
    \label{fig:smallprobeclosed}
\end{figure}

It is clear that in the closed universe, degrees of freedom in the bulk are not independent with each other. However, to make the description meaningful, it is essential that degrees of freedom in a low energy long-wavelength region behave like an ordinary quantum field theory. To see how this works in the tensor network model, we can discuss two different situation. 

The first situation is when the bulk state $\rho_b$ before projection is a mixed state, and we consider a small probe $D$ which does not induce an entanglement island. For concreteness we can consider the three-tensor model as is illustrated in Fig. \ref{fig:smallprobeclosed} (a). It is important to remember the fact that the bulk state $\rho_b$ can be a mixed state. (This is also true in the previous discussion in this section, but we did not emphasize that since it was not essential.) We assume the probe is small enough, such that any Ising domain wall is not preferred since the area law entropy cost is too large. Then the only possible configurations left are $+++$ and $---$. The entropy becomes
\begin{align}
    S(D_1)&=S_{\rm QFT}(D),~~~(+++)\\
    S(D_2)&={\rm min}\left\{S_{\rm QFT}(\overline{D}),|D|\log 2\right\}\\
    S(D_1D_2)&={\rm min}\left\{S_{\rm QFT}(\rho_b),S_{\rm QFT}(D)+|D|\log 2\right\}
\end{align}
Here $S_{\rm QFT}(\rho_b)$ is the entropy of the entire state. If we have a large enough entorpy such that
\begin{align}
    S_{\rm QFT}(\rho_b)>S_{\rm QFT}(D)+|D|\log 2\label{eq:closedisometrycond}
\end{align}
that implies
\begin{align}
    S_{\rm QFT}(D)+S_{\rm QFT}(\overline{D})\geq S_{\rm QFT}(\rho_b)&>S_{\rm QFT}(D)+|D|\log 2\nonumber\\
    S_{\rm QFT}(\overline{D})&>|D|\log 2
\end{align}
which leads to $S(D_2)=|D|\log 2,~S(D_1D_2)=S_{\rm QFT}(D)+|D|\log 2$. In this limit there is no mutual information, and $D_2$ has maximal entropy. If we introduce an auxiliary $\Omega$ which purifies $\rho_b$, as is shown in Fig. \ref{fig:smallprobeclosed} (b), we define a pure state of $D_1D_2\Omega$, and under condition (\ref{eq:closedisometrycond}) we have shown that $D_2$ is maximally entangled with $\Omega$. This is an important condition since it shows that when we create a ``baby universe" with some degrees of freedom that are entangled with a reference $\Omega$, and then probe the ``baby universe" with a small probe, the information obtained by the probe can be reconstructed from $\Omega$. In this limit, the probe will not see any violation of unitarity due to the effect of the projection $\ket{V}$, since the condition
\begin{align}
    S(D_1)=S_{\rm QFT}(D),~S(D_2)=|D|\log 2
\end{align}
applies to all Renyi entropies, which requires
\begin{align}
    \sigma_{D_1D_2}=\rho_{\rm QFT}(D)\otimes 2^{-|D|}\mathbb{I}_{D}
\end{align}
This equation guarantees that any operator insertion in region $D$ satisfies
\begin{align}
    C_{ab}=\frac{\bra{V}\phi_a\rho_P\phi_b\ket{V}}{\bra{V}\rho_P\ket{V}}={\rm tr}_D\left(\phi_a\rho_D\phi_b\right)=\langle\phi_b\phi_a\rangle_{\rm QFT}\label{eq:correlationprobe}
\end{align}

On the contrary, if the amount of information one wants to retrieve is too large and Eq. (\ref{eq:closedisometrycond}) is violated,  $S_{D_2}=S_{\rm QFT}(\overline{D})<|D|\log 2$ in this region, and thus $D_2$ is not maximally mixed any more. Consequently, evidence of unitarity violation will start to be observed, since operator $\phi_b$ on the right side of $\rho_P$ cannot be pushed to the left any more. For example, we can consider $\phi_a,\phi_b$ as projectors to orthogonal measurement outputs labeled by $a$, then the different measurement outputs fail to decohere, which means $C_{ab}$ is not diagonal in $a,b$ indices.

The other region is when $\rho_b$ is pure, but in addition to the small probe we have a stronger probe to a larger region. This is exactly the entanglement island probe situation we discussed earlier in Fig. \ref{fig:recovery} (b), with $D_B=1$. Here we would like to consider a more general situation, when the small probe is not necessarily in the entanglement island of $A$. In Fig. \ref{fig:smallprobeclosed} (c), we show two probes $P$ and $Q$, one in the island of $A$ and the other outside it. When the probes are small, if we compute $S(P_1P_2Q_1Q_2)$, it is dominated by $+++$ configuration, and leads to
\begin{align}
    S(P_1P_2Q_2Q_2)=(|P|+|Q|)\log 2+S_{\rm QFT}(P)+S_{\rm QFT}(Q)
\end{align}
The $(|P|+|Q|)\log 2$ terms guarantee that correlation functions in $P,Q$ agree with the QFT expectation without the projection, just like the earlier discussion in Eq. (\ref{eq:correlationprobe}). We see that this is independent from whether the probe is in the entanglement island. 

\begin{figure}
    \centering
    \includegraphics[width=5.5in]{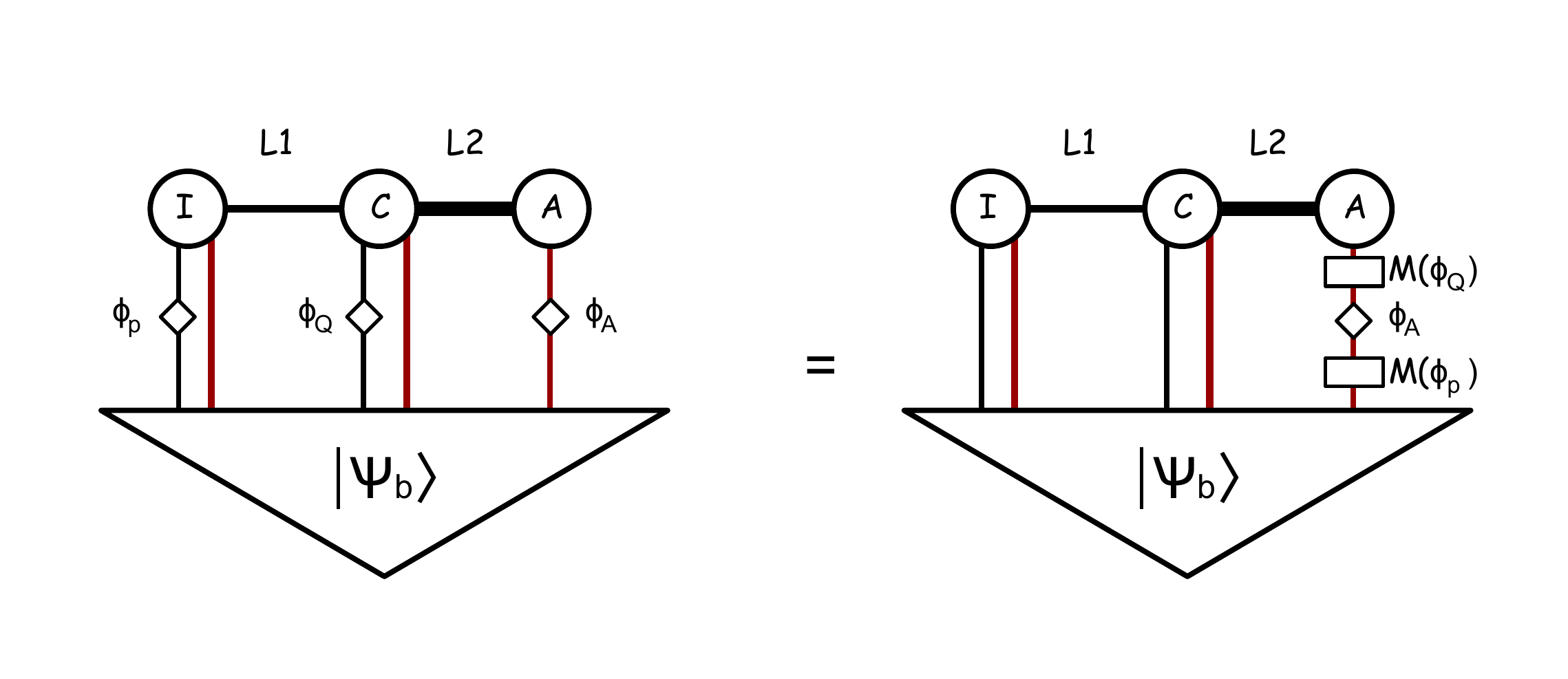}
    \caption{The reconstruction of operators in small probe regions $P,Q$ to a region $A$, when $P$ is part of the entanglement island $I$ of $A$ and $Q$ is outside $I$. Notice the different location of operator insertions on the right side of the equality. }
    \label{fig:operatormapclosed}
\end{figure}

If we also insert an operator in the large probe $A$, we can see the difference between $P$ and $Q$. Generalizing the discussion in subsection (\ref{sec:recovery}) we can obtain
\begin{align}
    I(P_2:A_1)=2|P|\log 2,~I(Q_2:A_2)=2|Q|\log 2
\end{align}
which is determined by the fact that the $I$ site spin is controlled by $A_1$ boundary condition, while $C$ site (everywhere outside the island) is controlled by $A_2$. This difference means that unitarity in $P$ can be violated by inserting operators in $A_1$, but is independent from operator insertion in $A_2$. For $Q$ the situation is opposite. In term of operator reconstruction, this difference means operators in $P$ and operators in $Q$ can both be reconstructed to region $A$, but the reconstruction map reverses the operator order for $Q$. For general correlator involving operators in $P,Q,A$, the reconstruction map satisfies
\begin{align}
    C_{PQA}&=\bra{V}\phi_A\phi_P\phi_Q\rho_P\eta_Q\eta_P\eta_A\ket{V}\nonumber\\
    &=\bra{V}\mathcal{M}(\phi_Q)\phi_A\mathcal{M}(\phi_P)\rho_P\mathcal{M}(\eta_P)\eta_A\mathcal{M}(\eta_Q)\ket{V} 
\end{align}
Notice the ordering of the operators is such that the reconstruction of $P$ operators are inserted next to $\rho_P$ while that of $Q$ operators are next to $\ket{V}$. Also the mapping satisfies
\begin{align}
    \mathcal{M}(\phi_P\eta_P)&=\mathcal{M}(\phi_P)\mathcal{M}(\eta_P)\\
    \mathcal{M}(\phi_Q\eta_Q)&=\mathcal{M}(\eta_Q)\mathcal{M}(\phi_Q)
\end{align}
The different order comes from an extra transpose operation when we map operators in $Q$ to $A$. This is illustrated in Fig. \ref{fig:operatormapclosed}

Interestingly, this discussion indicates that if $A$ is the only large probe in the universe, then sufficiently small probe operators in the entire complement of $A$ --- rather than just the entanglement island --- can be reconstructed in $A$. If there are multiple probes inserted in the universe which are not in the small limit, or if $\rho_b$ is not pure, this is not true any more, because the region controlled by $A_2$ boundary condition and that controlled by $A_1$ boundary condition are not complement to each other. This sounds counter-intuitive, but one should keep in mind that the operator reconstruction map depends on the pair of states $\ket{\psi_b},~\ket{V}$ of the entire universe. 

\subsection{Relation between closed universe and open universe}

\begin{figure}
    \centering
    \includegraphics[width=5.5in]{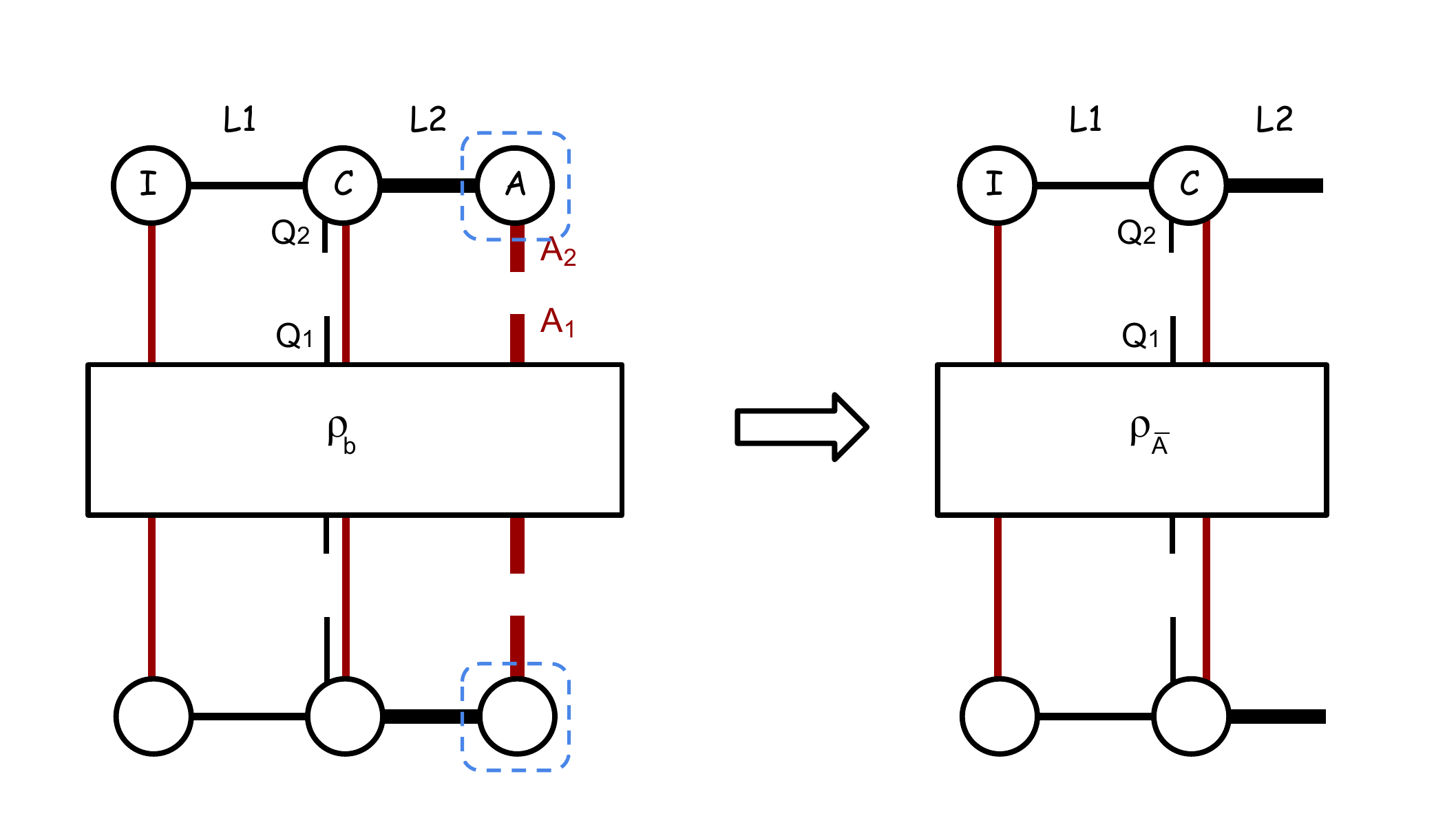}
    \caption{Illustration of a closed universe with a region $A$ satisfying $|A|>|L_2|$ and no entanglement island. By cutting the network at $L_2=\partial A$, the super-density operator of probe $Q$ and region $A$ can be decomposed to that of $Q,A_1,L_2$ conjugated by an isometry $K_A$, defined by the tensor(s) in the blue dashed region. The right panel illustrates the open universe obtained by removing $K_A$ part of the tensor network and tracing over $A_1$ to obtain $\rho_{\overline{A}}={\rm tr}_A(\rho_b)$.  }
    \label{fig:cutopen}
\end{figure}

It is also interesting to point out how the closed universe situation is related to the asymptotic AdS case with a large boundary. We consider a closed universe with a large region $A$, and consider the situation when the dimension of $A$ exceeds the area law bound. In the three-tensor model, this means $|A|>|L_2|$. We also assume $A$ has no entanglement island. In this case, following the discussion in the previous subsection, operators acting on a small probe can be reconstructed to $A_2$. Further more, in the area law phase we can cut the tensor network at the boundary of $A$ (which is $|L_2|$ in the three-tensor model) and denote the super-density operator of $A$ and the small probe $Q$ as
\begin{align}
    \sigma_{QA_1A_2}=K_A\sigma_{L_2A_2}K_A^\dagger\label{eq:composition}
\end{align}
This is illustrated in Fig. \ref{fig:cutopen}. Here $K_A$ is a linear map from $L_2$ to $A$ defined by cutting the tensor network open at $L_2$. In the three-tensor model, this is given by a single tensor. When $|A|>|L_2|$, this map is an isometry from the boundary $L_2$ to the bulk $A$. More generally, we can take the tensor network state for $A$ region, which is an entangled state of $A$ bulk indices and $\partial A$ boundary indices. The second Renyi entropy of $A$ is given by 
\begin{align}
    S(A)=\min_{\Sigma \subseteq A}\left(|\partial \Sigma|+|A\backslash \Sigma|\right)\log 2
\end{align}
If this minimization is given by $\Sigma=A$, the entropy is area law $S(A)=|\partial A|\log 2$, which means $K_A$ is an isometry from $\partial A$ to $A$. In this case, the operator reconstruction from a small probe region $Q$ to $A_2$ can be considered as the composition of two reconstruction maps, from $Q$ to $L_2$ and then to $A_2$. The first map from $Q$ to $L_2$ can be viewed as the ordinary bulk-to-boundary isometry in the holography case. The physical interpretation of this connection between open and closed universe in gravity theory is unknown, but we will make some speculation. This transition from closed universe to open universe is analogous to a black hole formation process. When energy in $A$ region is small enough, we have a closed universe, and the small probe $Q\subset \overline{A}$ can be reconstructed in $A$, as we discussed in the previous subsection. As we increase the matter energy in $A$, a black hole is eventually formed, and the physical degrees of freedom in $A$ are now defined on the horizon. A reconstruction map from $Q$ to the horizon $\partial A$ can be defined. It is interesting to explore the analog of this phenomenon in gravity theory.

\section{Conclusion and further discussion}
\label{sec:conclusion}

In conclusion, we have proposed a general framework for defining effective entropy in systems with dynamical gravity. Physically, the effective entropy of a region describes the entanglement entropy of the matter field in the region below certain UV cutoff. We have discussed how the quantum extremal surface formula and entanglement island appears as a consequence of path integral over replica geometries.

We applied our results to two example systems. One system is a 2d gravity path integral that defines a density matrix of a system on a 1d spatial interval in a closed universe. The other system is a Schwarzchild black hole with an external region including early Hawking radiation. 
In the first example, the density matrix is nonperturbatively defined using the no boundary proposal \cite{PhysRevD.34.2267,Hawking:1986vj,Barvinsky:2008vz,Maldacena:2019cbz,maldacenaStrings2019,Anous:2020lka}.
The closed universe is obtained by gluing the boundary of a Euclidean Anti-de Sitter space so that the original AdS boundary becomes a bulk spatial slice.
We treat the Euclidean partition function as defining a density matrix of the matter field which becomes effectively a Hartle-Hawking state.
In the presence of large number (order $N$) matter fields, the Von Neumann entropy has a phase transition with respect to the size of the spatial interval so that satisfies the Bekenstein bound.
And after phase transition, a disjointed spacetime region inside the closed universe becomes the entanglement wedge of the matter field on the original interval.
In the second example, we carried an approximate calculation of the entropy of the Hawking radiation in a Schwarzchild black hole in a specific state. The region is defined as the exterior of a sphere with fixed radius. 
We confirmed the late time Page transition due to a nontrivial quantum extremal surface whose location is slightly inside the horizon for our particular state.

 To obtain a more explicit understanding to the quantum information properties of such systems, we studied random tensor network models. By introducing ancilla and using the framework of superdensity operators, we see how quantum extremal surface and entanglement island appears in Renyi entropy calculation of a bulk region in the tensor network. We show that operators acting on a small region in the entanglement island of region $A$ can be reconstructed in $A$, but the reconstruction is a state-dependent mapping which relies on the knowledge that the remainder of the island is not probed. We discuss the case of closed universe, and show how quantum information in a local probe can still make sense if the bulk matter field is in a mixed state with large entropy, or if an ancilla is introduced to a bigger region than the probe. Interestingly, when the closed universe matter field is in a pure state, there is a complementary recovery, which means operators in entire complement of $A$ can be reconstructed on $A$, but the reconstruction map for operators in $A$'s entanglement island is different from those elsewhere. We also discuss how the closed universe case and the open universe case can be related. 

There are obviously many open questions for future research. The definition of effective entropy relies on gauge invariant ways to determine the location of the region. We proposed a few different ways to do that. It is not clear whether there is a unique way that is non-perturbatively defined and does not require to make a choice among different ways of defining the region. As we discussed in Sec. \ref{sec:gravity}, our effective entropy is cutoff dependent. We view this as not a problem but a feature, since the tensor factorization of matter field Hilbert space is only well-defined when we focus on certain states with semi-classical gravity description. In the extreme case of closed universe with pure state matter, the tensor network picture suggests that the entanglement entropy of a region can be viewed as the entanglement between matter and geometry. It is interesting to seek for a more nonperturbative framework for describing such entanglement. 

Another lesson we learn from tensor network models is that the quantum state definition depends on the choice of observers, which is the key difference from ordinary quantum many-body systems. In ordinary holographic duality, bulk degrees of freedom (in the code subspace) can be mapped to the boundary with an isometry, such that there is a single quantum many-body system with a given Hilbert space which provides the ``anchor point" for the bulk theory. In general geometries, there is no built-in isometry structure, such that correlation functions have to be described by superdensity operators. The superdensity operator formalism automatically require a more fluid structure of Hilbert space definition. Instead of a single quantum many-body system, we have a family of different quantum many-body systems, depending on which regions we introduce ancilla at, which satisfy certain consistency conditions with each other in overlapping regions. The quantum information reconstruction is also more observer-dependent than the AdS/CFT case. Whether a few qubits of quantum information can be reconstructed from a bigger region depends on the observer's access to that region. The entanglement island of a bulk region can be destroyed if the observer try to access the island and the original region simultaneously. There are a lot of new phenomena that we are observing in the tensor network models. Understanding their counterpart in the gravity theory seems to require a new mathematical framework beyond (unitary) many-body quantum mechanics. 

{\bf \noindent Acknowledgement.}
We would like to thank Ahmed Almheiri, Yiming Chen, Juan Maldacena, Geoff Penington, and Douglas Stanford for helpful discussion. This material is based upon work supported by the Air Force Office of Scientific Research under award number FA9550-19-1-0360 (XD) and also by funds from the University of California (XD).  This work is supported by the National Science Foundation Grant No. 1720504 (XLQ) and the Simons Foundation (XLQ). This work is also supported in part by the DOE Office of Science, Office of High Energy Physics, the grant de-sc0019380 (XLQ and ZS).  This work was developed in part at the Kavli Institute for Theoretical Physics during the workshop ``Gravitational Holography", which is supported in part by the National Science Foundation under Grant No.\ PHY-1748958.

\appendix

\section{Renyi entropy and correlation functions} \label{app:renyi}

To provide further physical understanding of the effective entropy, in this appendix we present some operator identity that relates Renyi entropy with correlation functions in the original (single copy) system. 


For a region $A$ with Hilbert space $\mathbb{H}_A$ in a quantum system, we consider an orthonormal basis of Hermitian operators $T_a,~a=1,2,...,D_A^2$. $T_a$ satisfies
\begin{align}
    {\rm tr}\left(T_aT_b\right)=\delta_{ab},~\sum_aT_a^{\alpha\beta}T_a^{\gamma\delta}=\delta^{\alpha\delta}\delta^{\beta\gamma}
\end{align}
The second equation can be rewritten as
\begin{align}
    \sum_aT_a\otimes T_a=X_A
\end{align}
with $X_A$ the swap operator acting on two copies of $A$. Therefore we can relate the second Renyi entropy to correlators:
\begin{align}
    {\rm tr}\left(\rho_A^2\right)={\rm tr}\left(\rho\otimes \rho X_A\right)=\sum_a{\rm tr}\left(\rho T_a\right)^2
\end{align}

\begin{figure}
    \centering
    \includegraphics[width=3.5in]{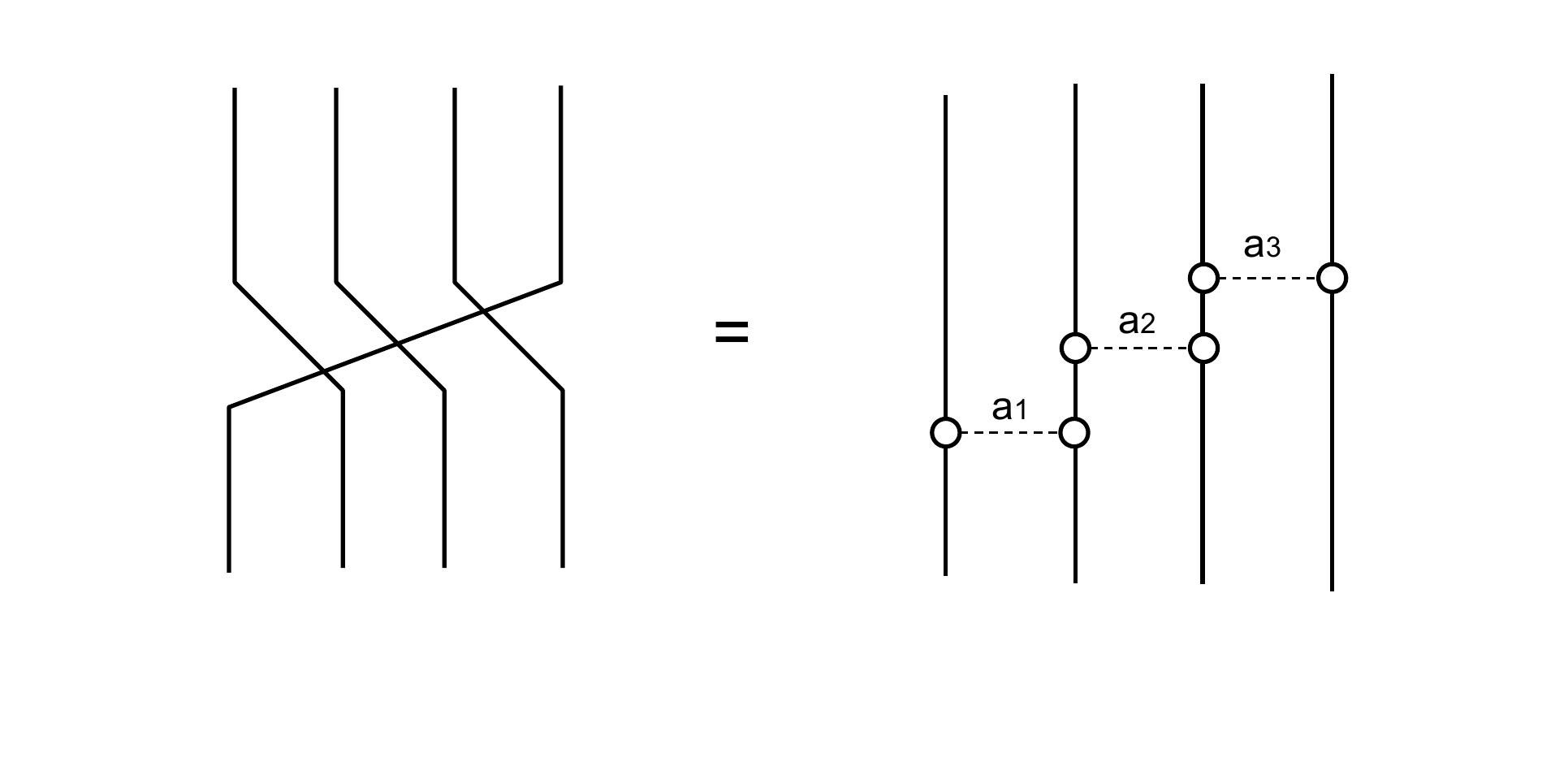}
    \caption{Illustration of the decomposition of cyclic permutation operator $X_{nA}$ into a sum of correlation functions.}
    \label{fig:swapdecomposition}
\end{figure}

This discussion can be generalized to higher Renyi entropies, since the cyclic permutation can be decomposed into pair swap operators:
\begin{align}
    X_{nA}=X_{A,n-1,n}...X_{A23}X_{A12}
\end{align}
as is illustrated in Fig. \ref{fig:swapdecomposition}. This decomposition thus enables us to obtain the relation
\begin{align}
    e^{-S^{(n)}_A}&\equiv\sum_{a_1,a_2,...,a_{n-1}}\langle T_{a_1}\rangle\langle T_{a_2}T_{a_1}\rangle...\langle T_{a_{n-1}}T_{a_{n-2}}\rangle\langle T_{a_{n-1}}\rangle\label{eq:effectiveentropydecomposition}
\end{align}
This relation tells us that in gravitational systems, if correlation functions on the right-hand-side of Eq. (\ref{eq:effectiveentropydecomposition}) are well-defined for an orthonormal basis of low energy QFT operators, the effective Renyi entropy $S^{(n)}_A$ is well-defined. (Note that $T_a$ only needs to form an orthonormal basis, and does not need to generate a closed algebra.)

\section{Operator reconstruction map}\label{app:isometry}

For completeness, in this appendix we present the details of the operator reconstruction map.\footnote{For recent discussion on the operator reconstruction map in the context of black hole system, see \cite{Cotler:2017erl,Hayden:2018khn,Chen:2019gbt,Penington:2019kki,Chen:2019iro,Jia:2020etj, Gilyen:2020gmg}} Consider a quantum state consists of three parts (tensor factors of Hilbert space) $P,A,E$. We assume the Hilbert space dimension satisfies $D_P<D_A$ and the mutual information between $P$ and $A$ is maximal: $I(P:A)=2\log D_P$. Without losing generality, we take the quantum state to be a pure state $\ket{\psi_{PAE}}$, since we can always purify it by enlarging $E$ if that is not the case.

For a pure state, we have the identity
\begin{align}
    I(P:A)+I(P:E)=I(P:AE)=2S(P)
\end{align}
Therefore when $I(P:A)$ is maximum $I(P:A)=2S(P)=2\log D_P$, we have $I(P:E)=0$. Consequently, tracing over $A$ leads to a direct product state:
\begin{equation}
    \rho_{PE}=\includegraphics[scale = .4,valign = c]{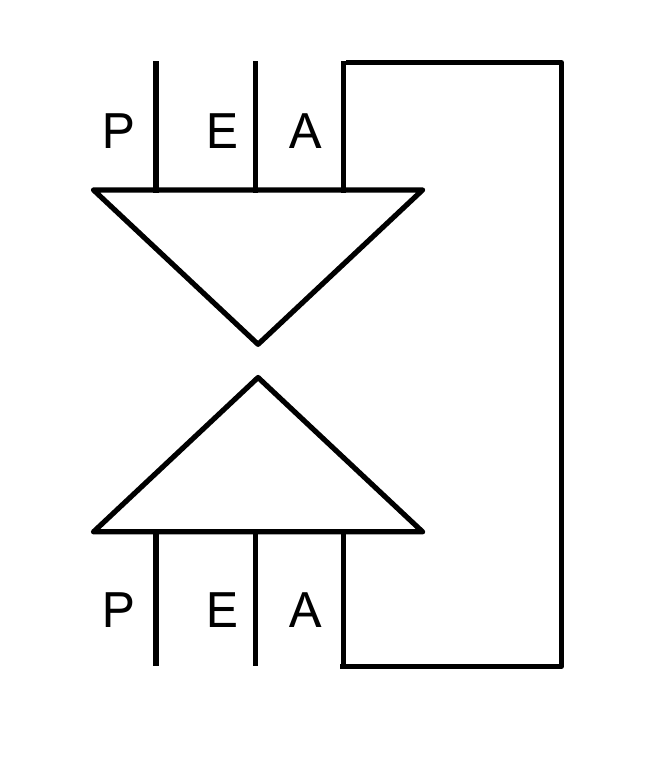}=\includegraphics[scale = .4,valign = c]{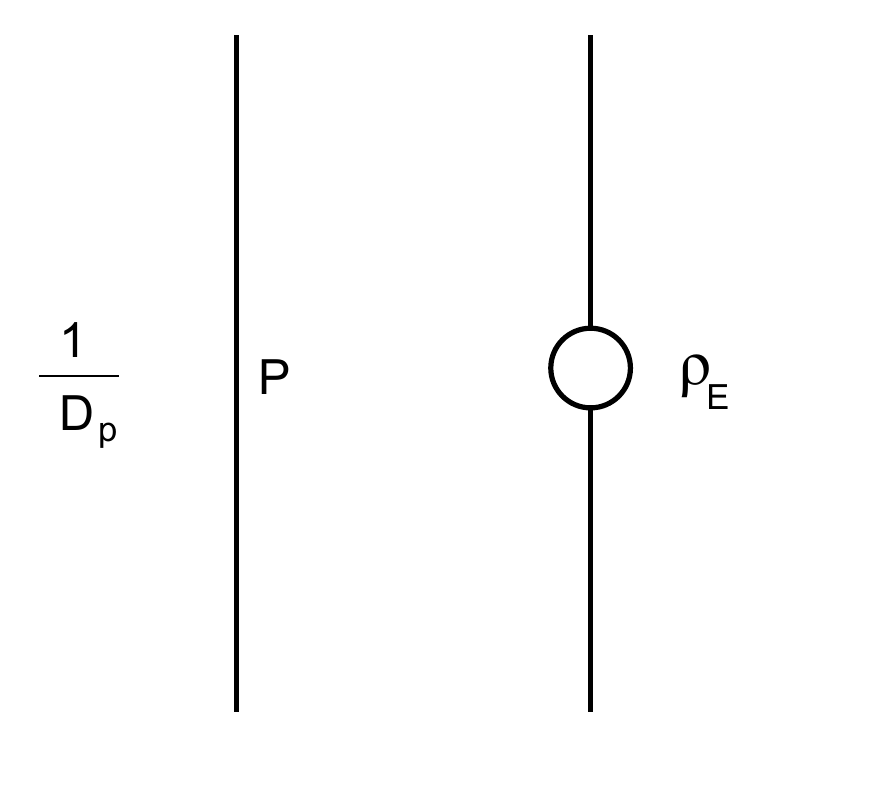}\label{eq:zeroIPE}
\end{equation}
Using this fact, we can use the reduced density matrix $\rho_{PA}$ and define the following reconstruction map $\mathcal{M}_0: \mathbb{H}_P\otimes \overline{\mathbb{H}}_P\rightarrow \mathbb{H}_A\otimes \overline{\mathbb{H}}_A$. Denote the spectral decomposition of the reduced density matrix $\rho_{PA}$ as
\begin{align}
    \rho_{PA}=\sum_n\lambda_n\ket{n}\bra{n}
\end{align}
The map is defined by
\begin{align}
    \mathcal{M}_0\left(\phi_P\right)&=D_P{\rm tr}_P\sum_{n,~\lambda_n>0}\left(\phi_P\ket{n}\bra{n}\right)\nonumber\\
    &\equiv D_P{\rm tr}_{PE}\left(\phi_P\rho_E^{-1}\ket{\psi_{PAE}}\bra{\psi_{PAE}}\right)=D_P\times \includegraphics[scale = .4,valign = c]{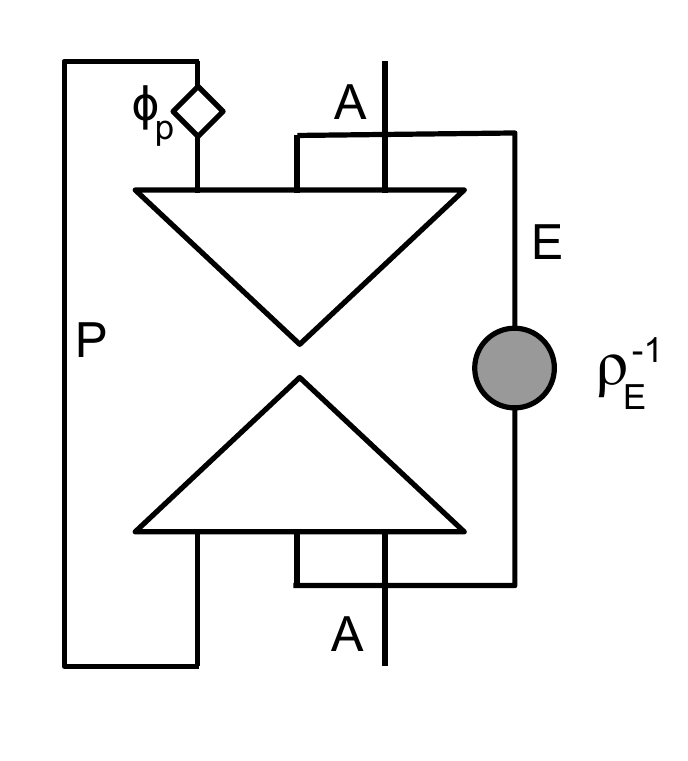}
\end{align}
Using Eq. (\ref{eq:zeroIPE}) one can prove that the map satisfies 
\begin{align}
    \mathcal{M}_0(\phi_P)\ket{\psi_{PAE}}=\phi_P\ket{\psi_{PAE}}\label{eq:stateaction1}
\end{align}
and
\begin{align}
    \mathcal{M}_0(\phi_P)\mathcal{M}_0(\eta_P)=\mathcal{M}_0(\eta_P\phi_P)\label{eq:convolution1}
\end{align}
More explicitly, this is proven pictorially in the following:
\begin{align}
    D_P\includegraphics[scale = .4,valign = c]{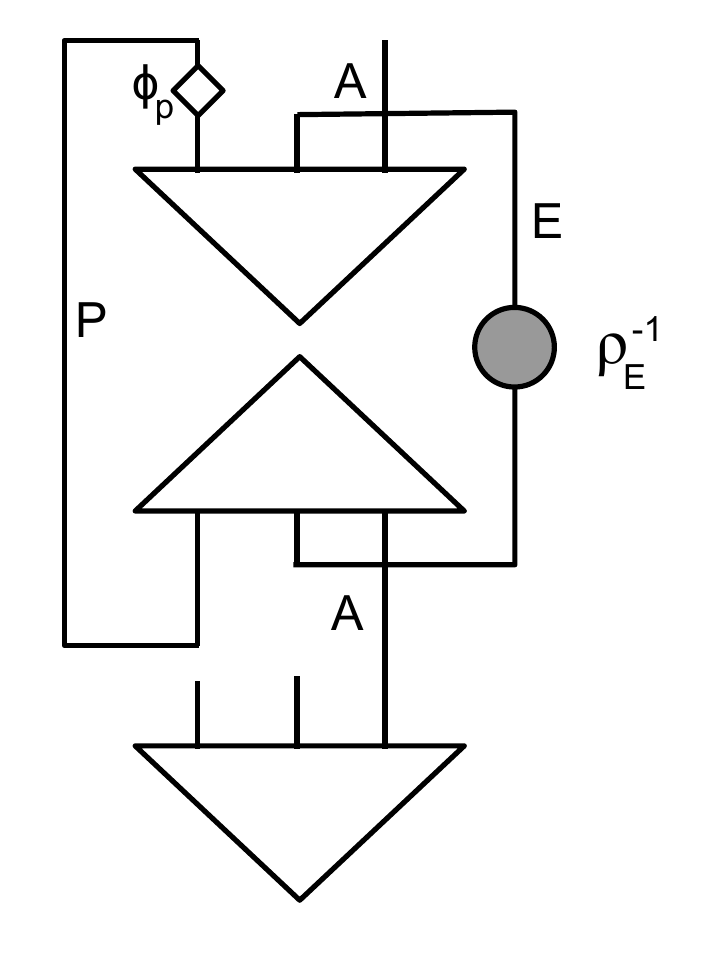}=\includegraphics[scale = .45,valign = c]{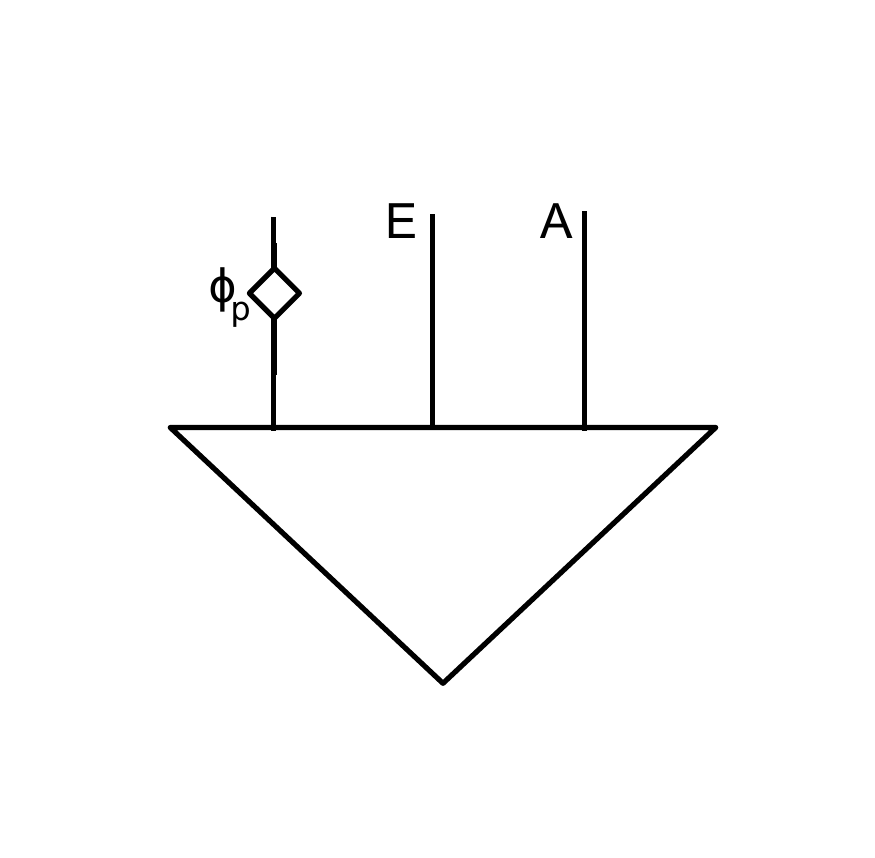}\label{eq:stateaction2}
\end{align}
\begin{align}
    \mathcal{M}_0(\phi_P)\mathcal{M}_0(\eta_P)&=D_P^2\includegraphics[scale = .4,valign = c]{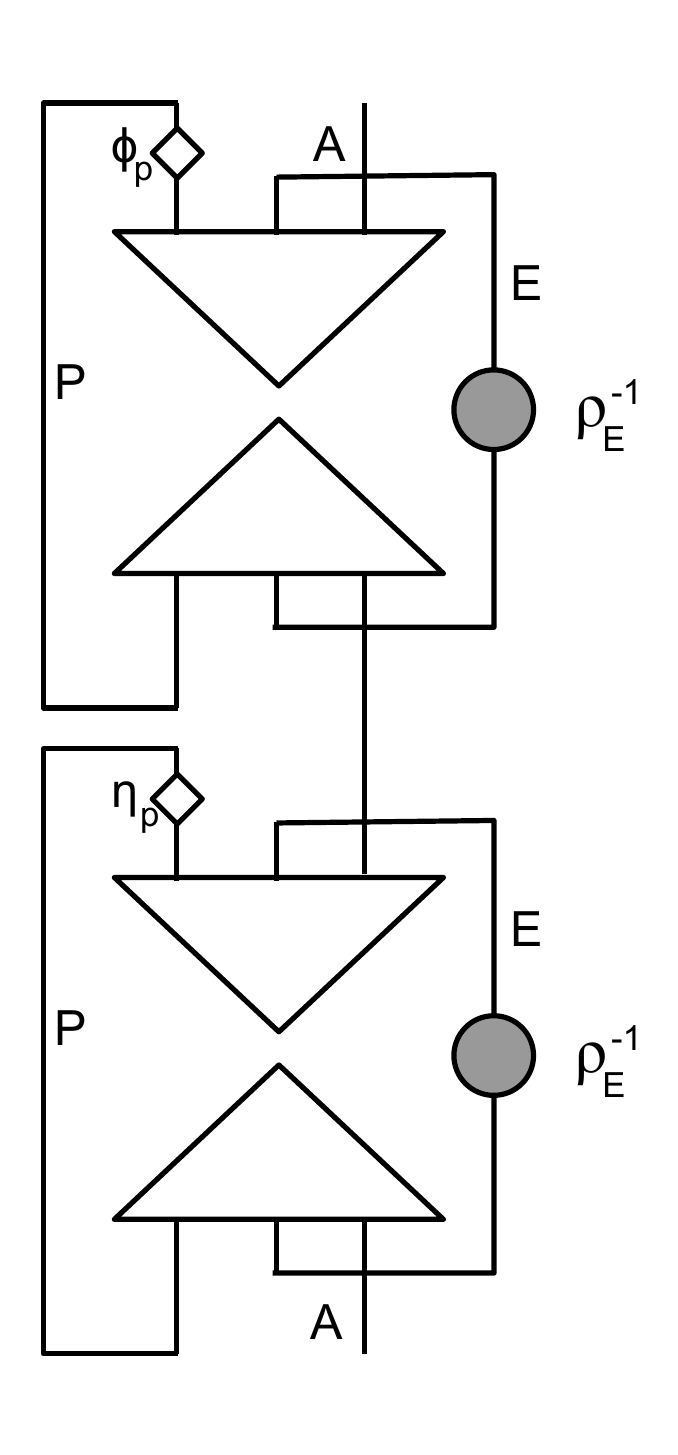}=D_P\includegraphics[scale = .4,valign = c]{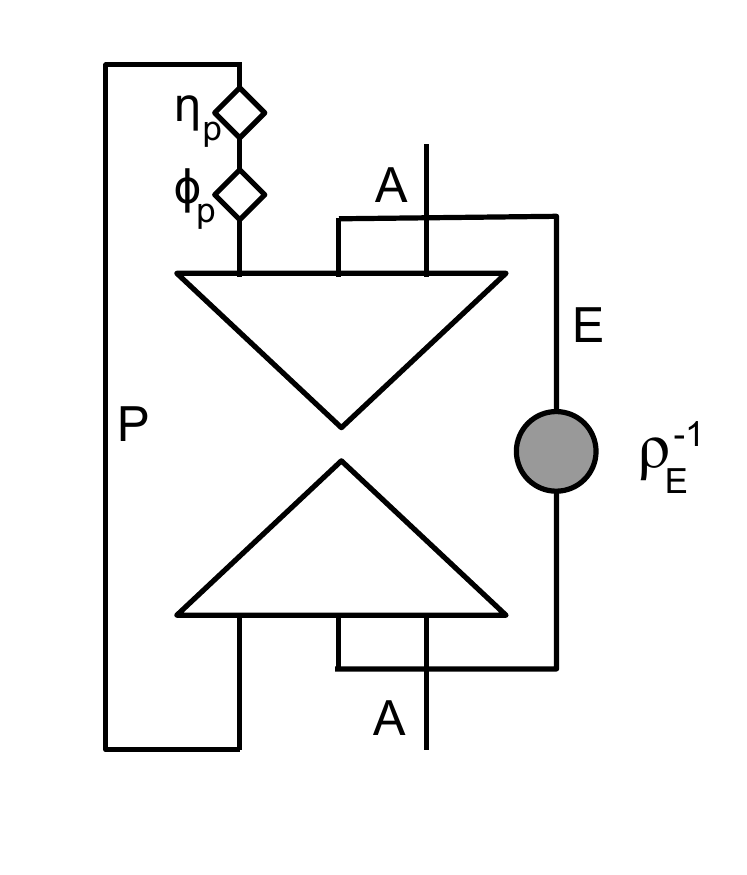}
\end{align}

Now we apply this recovery map to the superdensity operator in Fig. \ref{fig:recovery} (b). The reduced density matrix of $\rho_{A_1P_2}$ in the superdensity operator is 
\begin{align}
    \rho_{A_1P_2}=\includegraphics[scale = .4,valign = c]{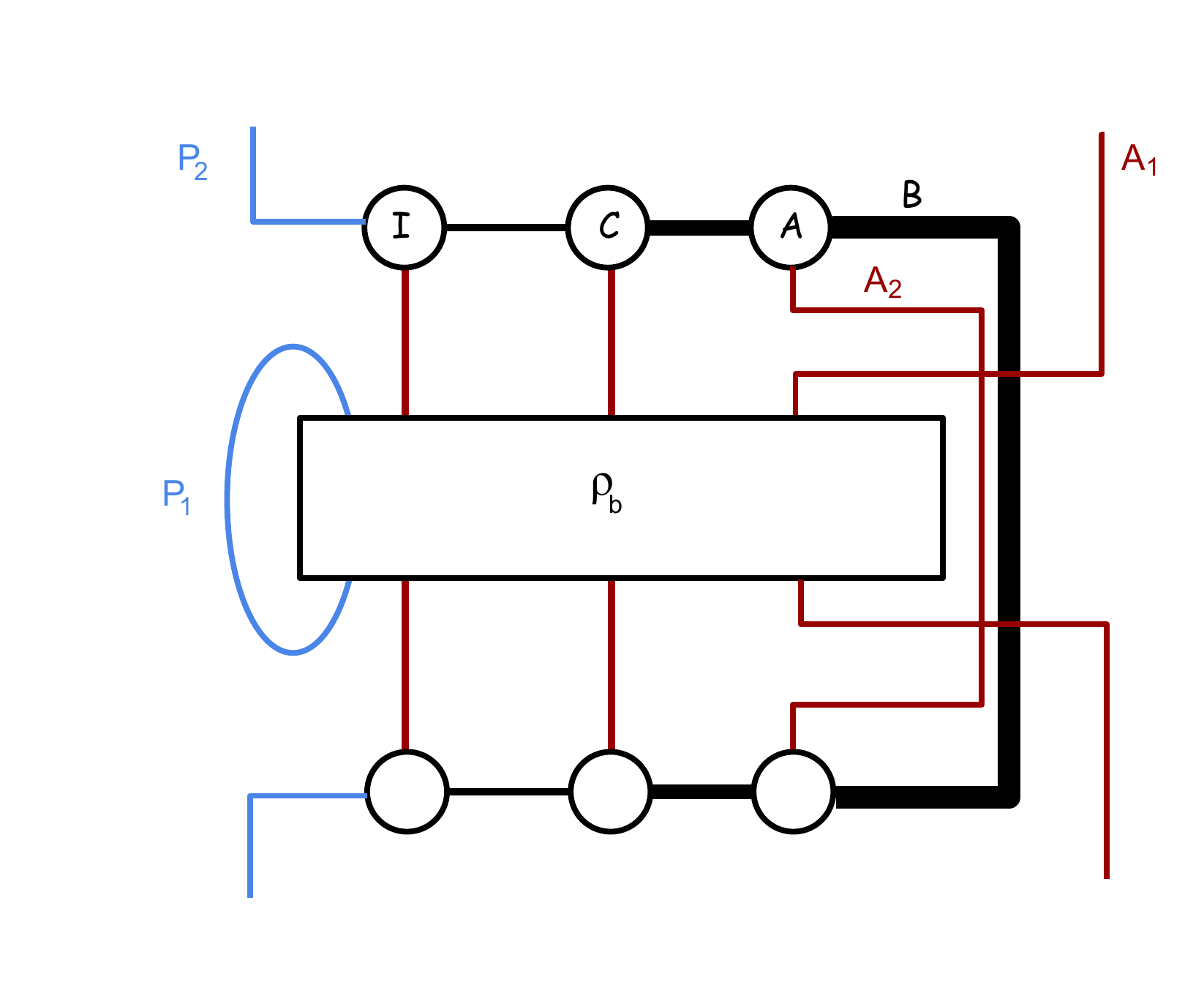}
\end{align}
When $P$ is a small probe inside entanglement island $I$, as we discussed in the draft, the mutual information $I(P_2:A_1)=2\log D_{P}$ is maximal. Therefore we can define the reconstruction map following the general prescription above. The remainder of the system $P_1$, $A_2$, $B$ and the purification of bulk QFT state $\rho_b$ (if it is a mixed state) together plays the role of $E$. However, one difference is that there is an extra transpose in mapping operators in ancilla $P_2$ to operators acting in the original $P$ system, as is illustrated in the following picture:
\begin{align}
    \includegraphics[scale = .4,valign = c]{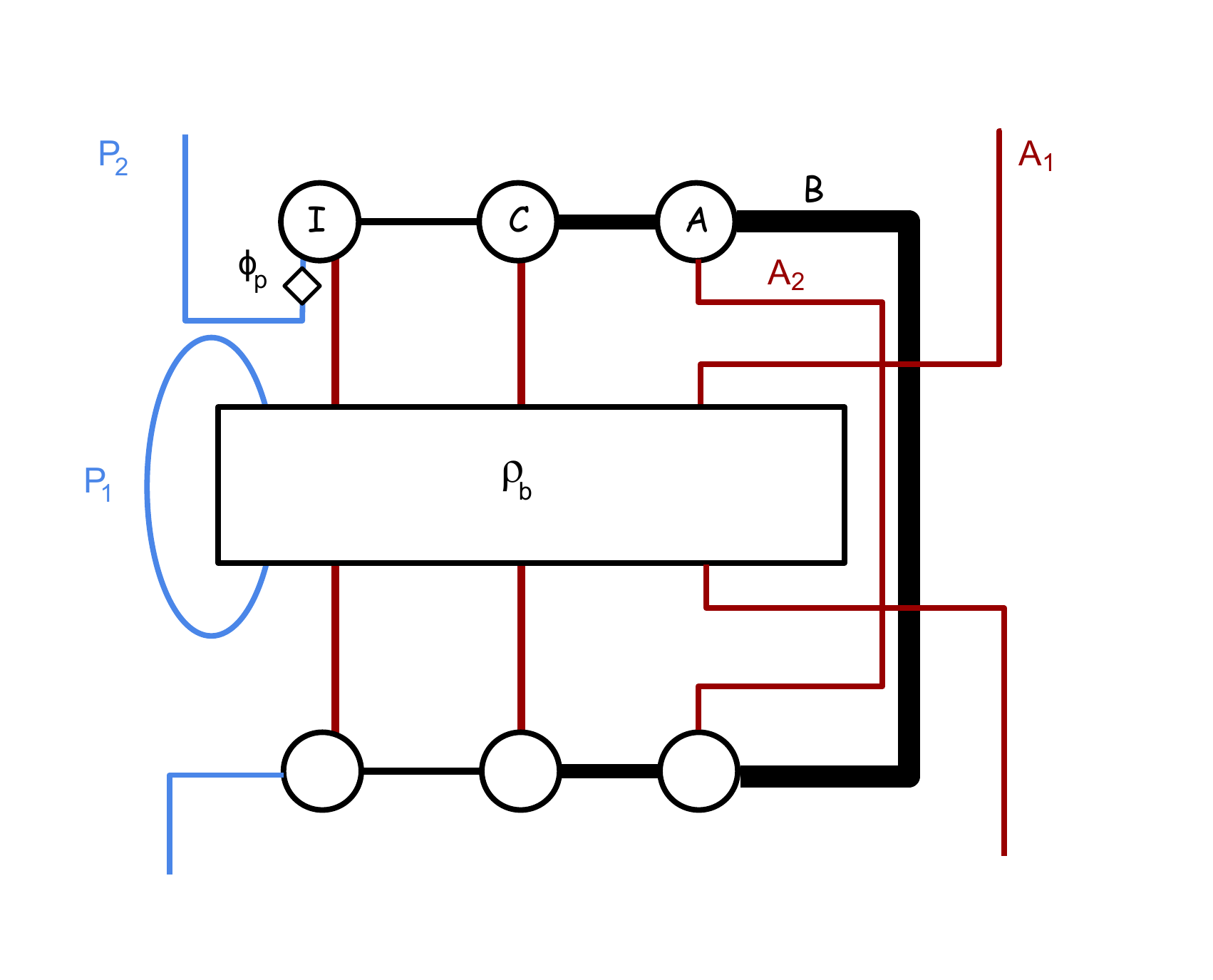}=\includegraphics[scale = .4,valign = c]{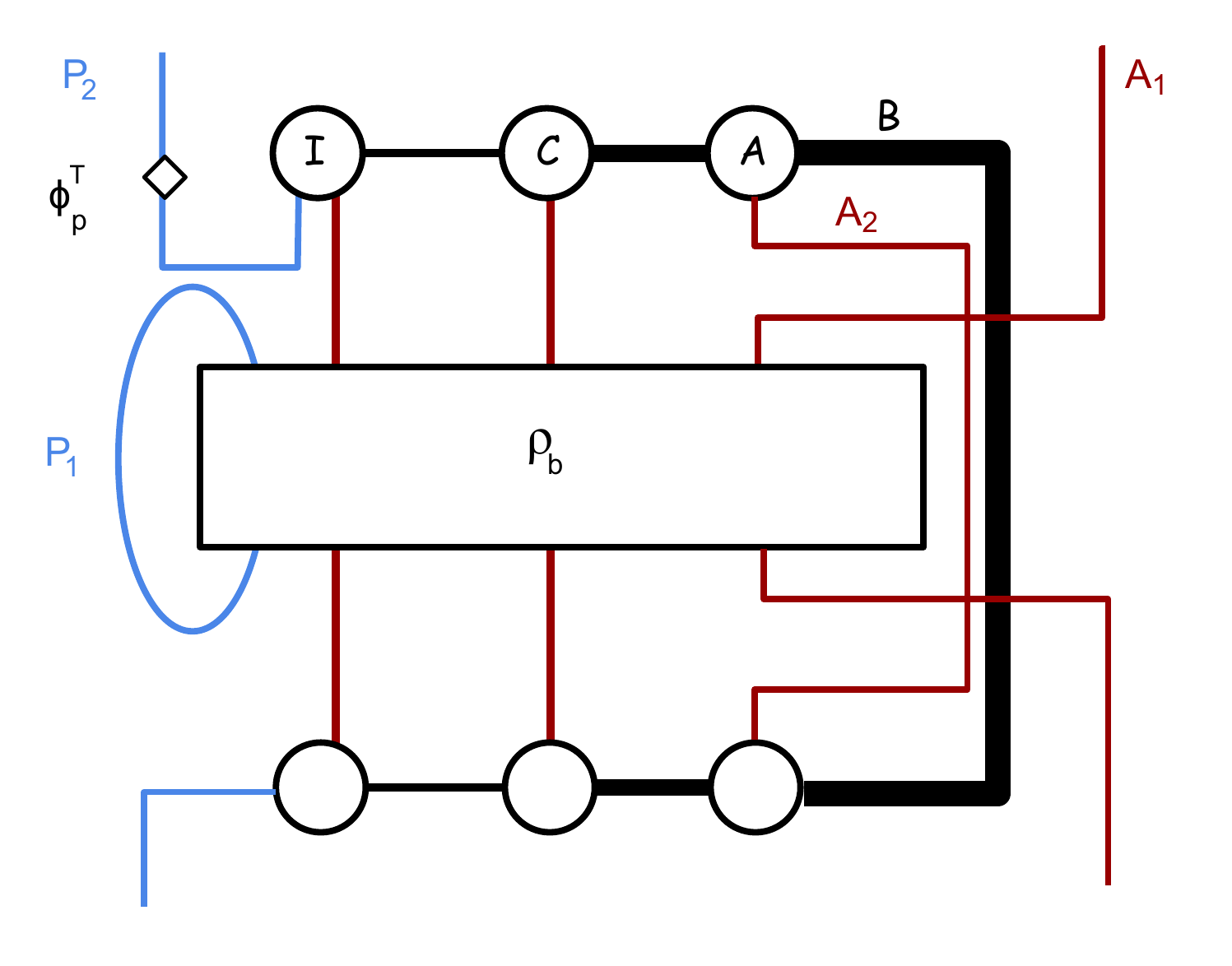}
\end{align}
Consequently, we can define the reconstruction map 
\begin{align}
    \mathcal{M}(\phi_P)=\mathcal{M}_0\left(\phi_P^T\right)
\end{align}
This changes the convolution rule (\ref{eq:convolution1}) to
\begin{align}
    \mathcal{M}(\phi_P)\mathcal{M}(\eta_P)=\mathcal{M}(\phi_P\eta_P)
\end{align}

It may seem that the two definitions $\mathcal{M}$ and $\mathcal{M}_0$ are just different conventions, but there is an important difference. If one is given the state $\rho_{P_2A_1}$, like in our discussion of $\mathcal{M}_0$, then it determines a map $\mathcal{M}_0$ with the convolution rule (\ref{eq:convolution1}). Of course one can still define $\mathcal{M}(\phi_P)=\mathcal{M}_0(\phi_P^T)$, but the transpose map $\phi_P\rightarrow \phi_P^T$ is basis dependent:
\begin{align}
    \phi_P^T=\sum_{n,m}\bra{m}\phi_P\ket{n}\ket{n}\bra{m}
\end{align}
so one has to make an arbitrary choice. In the superdensity operator discussion, the difference is that such a transpose map is provided by the ancilla coupled to $P$. The initial state of the ancilla is a maximally entangled state which can be written as
\begin{align}
    \ket{P_1P_2}=D_P^{-1/2}\sum_n\ket{n}_{P_1}\ket{n}_{P_2}
\end{align}
which defines the transpose map
\begin{align}
    \phi_P^T=\sum_{n,m}\bra{m}_{P_1}\phi_P\ket{n}_{P_1}\ket{n}_{P_2}\bra{m}_{P_2}
\end{align}
Acting operator $\phi_P^T$ on the ancilla $P_2$ is equivalent to acting $\phi_P$ on $P_1$, which is the input state of the random projection at $I$. Therefore in the case we are considering, $\mathcal{M}(\phi_P)$ is uniquely determined, which is independent from the choice of maximally entangled ancilla states at $I$ and $A$. 

The super-density operator is introduced as a tool, and we should apply it to the original system without ancilla. Applying Eq. (\ref{eq:stateaction1}) we obtain
\begin{align}
    \includegraphics[scale = .4,valign = c]{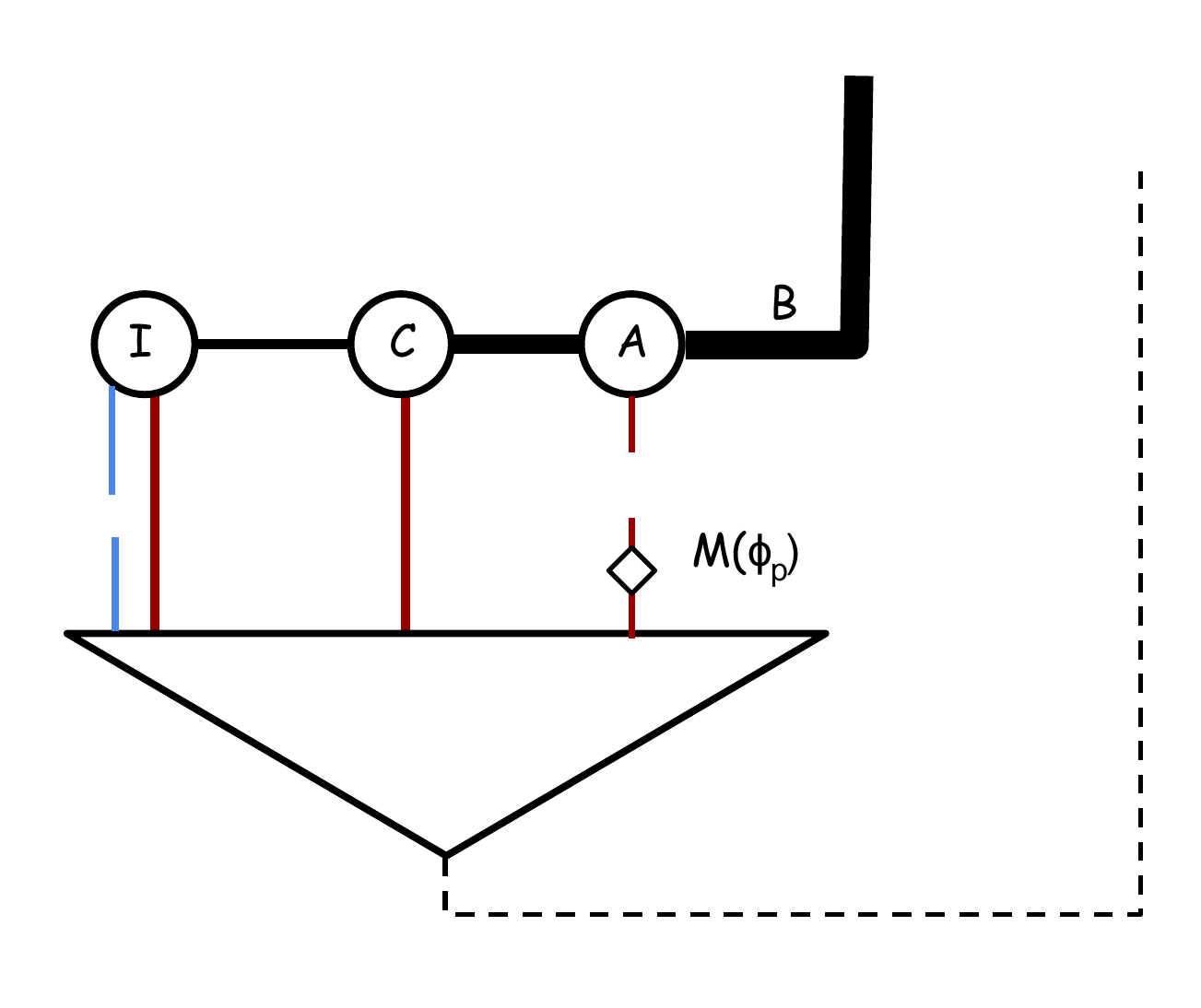}~=~~\includegraphics[scale = .4,valign = c]{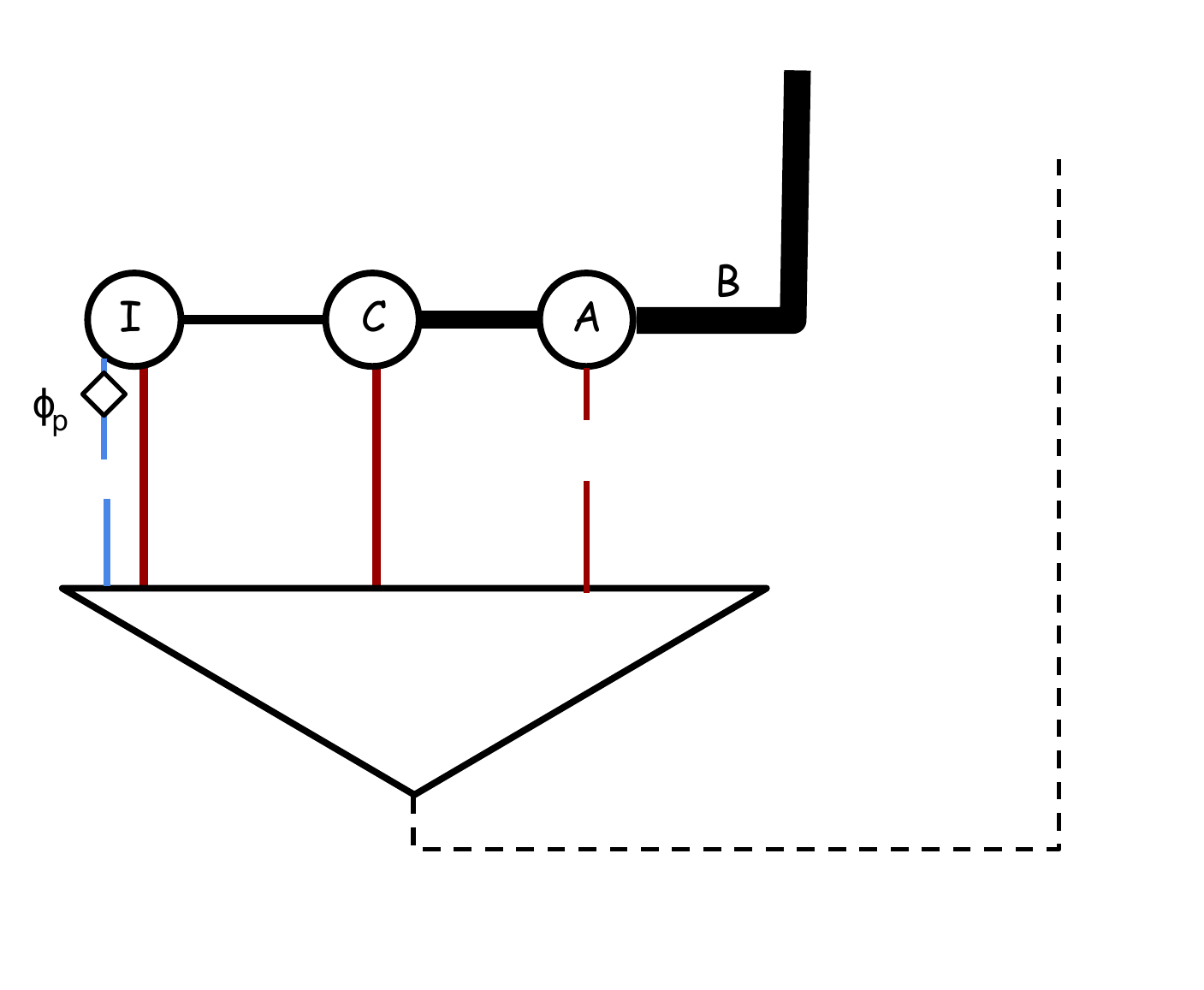}
\end{align}
Thus by reconnecting the lines we obtain the statement in the original tensor network correlation functions in Eq. (\ref{eq:correlation_reconstruction}), or equivalently
\begin{align}
\includegraphics[scale = .4,valign = c]{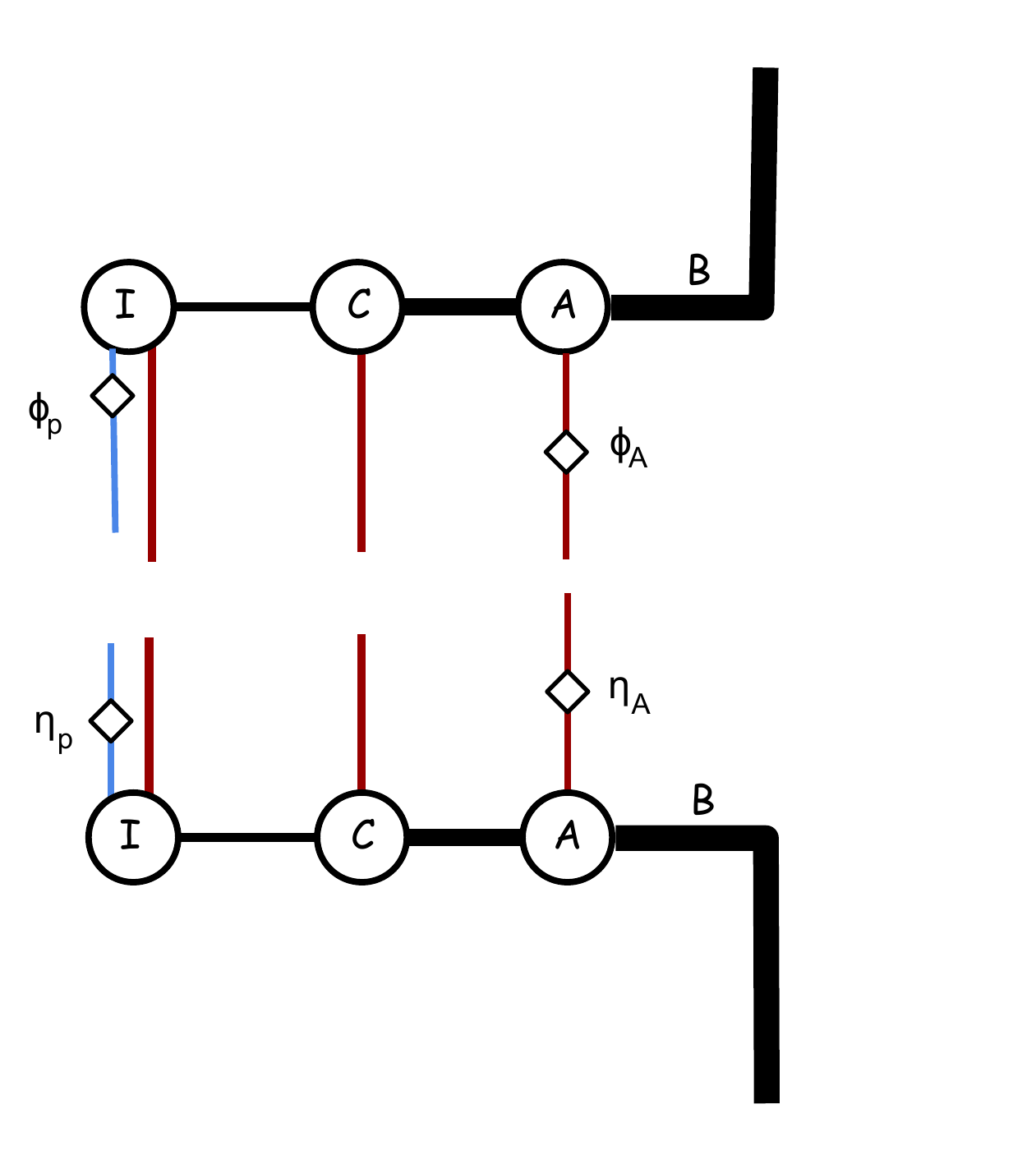}=\includegraphics[scale = .4,valign = c]{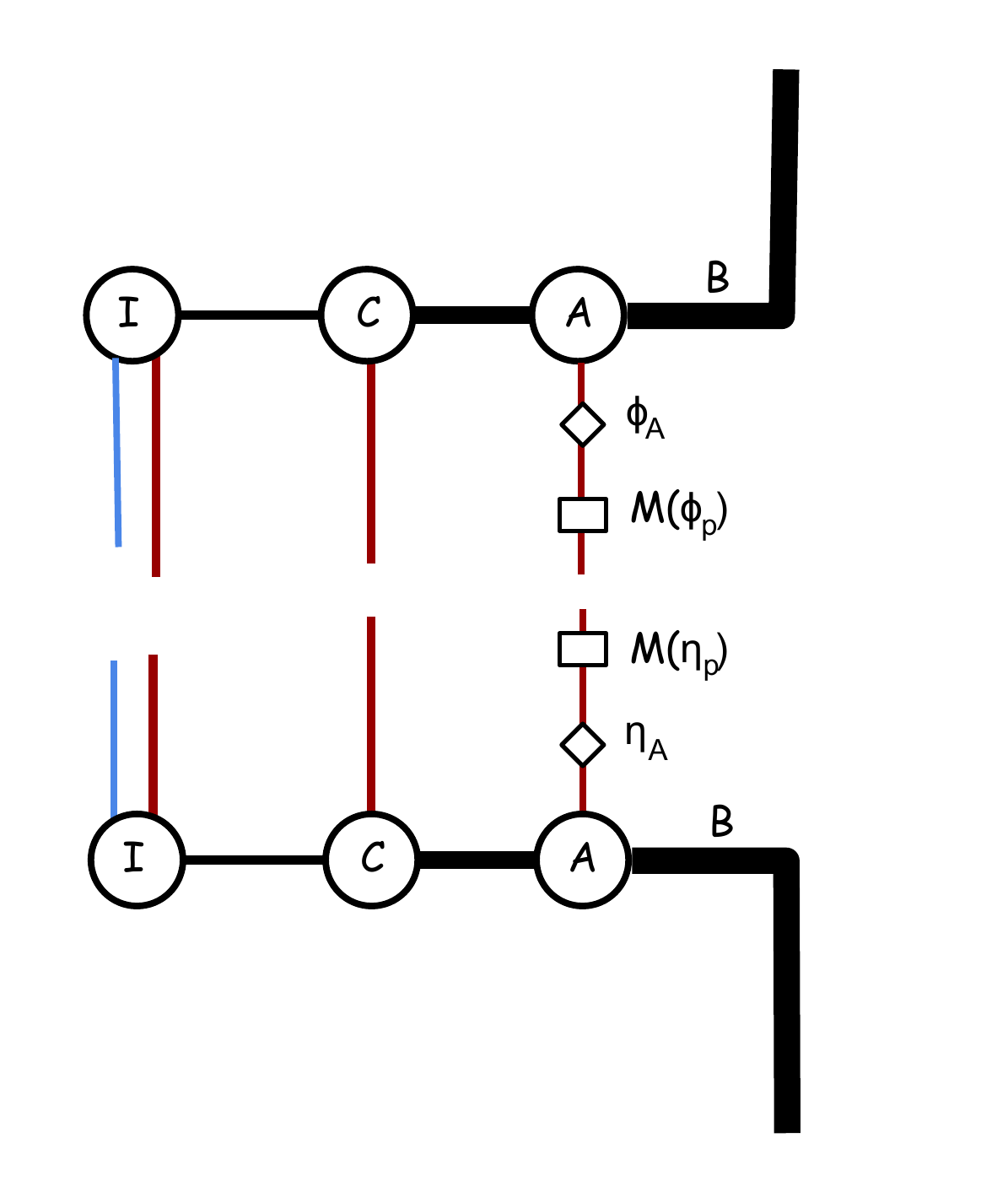}
\end{align}
The reconstructed operators $\mathcal{M}(\phi_P),\mathcal{M}(\eta_P)$ are always inserted closer to $\rho_b$ than all additional operators $\phi_A,\eta_A$ that act on $A$.

\bibliography{refs}
 \bibliographystyle{utphys}

\end{document}